\documentclass[DIV=15, bibliography=totoc]{scrartcl}  

\usepackage[a4paper,margin=2.25cm,bottom=2.5cm]{geometry}
\usepackage{amsfonts}
\usepackage{textcomp}
\usepackage{xurl}

\usepackage[table]{xcolor}
\definecolor{lightblue}{rgb}{0,0.5,1.0}
\definecolor{linkblue}{rgb}{0,0.1,0.6}
\definecolor{citegreen}{rgb}{0,0.4,0.0}
\definecolor{linkred}{rgb}{0.8,0,0.005}
\definecolor{mailviolet}{rgb}{0.3,0,0.35}
\definecolor{tumblue}{rgb}{0,0.396,0.741}
\definecolor{darkgreen}{rgb}{0,0.4,0} 
\definecolor{darkbrown}{rgb}{0.5, 0.396, 0.09}
\usepackage[hyperindex,colorlinks=true,linkcolor=linkblue,citecolor=citegreen,urlcolor=mailviolet,filecolor=linkred]{hyperref}

\usepackage{caption, subcaption}
\usepackage{fancyhdr}

\usepackage{algorithmicx}
\usepackage{algpseudocode} 
\usepackage{algorithm}
\usepackage{placeins}

\usepackage{siunitx}
\usepackage[square,comma,nonamebreak,compress,numbers]{natbib}

\usepackage{soulpos}
\usepackage{ulem}
\usepackage{authblk} 
\usepackage{float}

\usepackage[inkscapearea=page]{svg}
\usepackage{amsmath,amsfonts,amssymb,amsthm}
\usepackage[noabbrev]{cleveref}
\usepackage{upgreek}
\usepackage{adjustbox}
\usepackage{color}
\usepackage{enumitem}

\usepackage{pgfplots}
\usepackage{pgfplotstable}
\usepackage{filecontents}
\usepackage{tikz}
\usepackage{tikzscale}
\usepackage{tikz-3dplot}

\usetikzlibrary{spy}
\usetikzlibrary{intersections}
\usetikzlibrary{arrows,shapes}
\usetikzlibrary{spy}
\usetikzlibrary{backgrounds}  
\usetikzlibrary{decorations}
\usetikzlibrary{decorations.markings}
\usetikzlibrary{positioning}
\usetikzlibrary{patterns}
\usetikzlibrary{calc} 
\usetikzlibrary{quotes} 
\usetikzlibrary{external} 
\usetikzlibrary{matrix}

\pgfplotscreateplotcyclelist{customCycleList}
{%
  {blue, mark=o},
  {red, mark=square},
  {darkgreen, mark=diamond},
  {cyan, mark=diamond},
  {darkbrown, mark=triangle*},
  {black, mark=pentagon},
}

\pgfplotsset{every axis/.append style= {
    cycle list name=customCycleList,
}}

\usepackage{soul}

\title{A Memory Efficient Adjoint Method to Enable Billion Parameter Optimization on a Single GPU in Dynamic Problems
}

\author[1]{Leon Herrmann\thanks{\href{mailto:leon.herrmann@uni-weimar.de}{\texttt{leon.herrmann@uni-weimar.de}},
    Corresponding author}}
\author[2]{Tim Bürchner}
\author[1]{László Kudela}
\author[1,2]{Stefan Kollmannsberger}
\affil[1]{Chair of Data Engineering in Construction, Bauhaus-Universit\"at Weimar, Coudraystraße 13 b, 99423, Weimar, Germany}

\affil[2]{Chair of Computational Modeling and Simulation, Technical University of Munich, School of Engineering and Design, Arcisstraße 21, Munich, 80\,333, Germany}

\newcommand{\publicationDate}{\today}

\date{}
\fancypagestyle{plain}{%
\fancyhf{}%
\fancyfoot[L]{
\vspace{-0.0cm} 
\footnotesize{Preprint submitted to Structural and Multidisciplinary Optimization. \hfill 
\publicationDate
\\
}} }

\begin{document}     

\normalem \maketitle  
\normalfont\fontsize{11}{13}\selectfont

\newcommand{\tim}[1]{\textcolor{blue}{#1}}

\vspace{-1.5cm} \hrule 

\section*{Abstract}
Dynamic optimization is currently limited by sensitivity computations that require information from full forward and adjoint wave fields.
Since the forward and adjoint solutions are computed in opposing time directions, the forward solution must be stored. This requires a substantial amount of memory for large-scale problems even when using check pointing or data compression techniques. 
As a result, the problem size is memory bound rather than bound by wall clock time, when working with modern GPU-based implementations that have limited memory capacity.
To overcome this limitation, we introduce a new approach for approximate sensitivity computation based on the adjoint method (for self-adjoint problems) that relies on the principle of superposition. 
The approximation allows an iterative computation of the sensitivity, reducing the memory burden to that of the solution at a small number of time steps, i.e., to the number of degrees of freedom. 
This enables sensitivity computations for problems with billions of degrees of freedom on current GPUs, such as the A100 from NVIDIA$^{\text{\textregistered}}$ (from 2020). 
We demonstrate the approach on full waveform inversion and transient acoustic topology optimization problems, relying on a highly efficient finite difference forward solver implemented in CUDA$^{\text{\textregistered}}$. 
Phenomena such as damping cannot be considered, as the approximation technique is limited to self-adjoint problems.

\vspace{0.25cm}
\noindent \textit{Keywords:} 
adjoint optimization, GPU acceleration, finite difference method, dynamic optimization, full waveform inversion, acoustics
\vspace{0.25cm}


\section{Introduction}
\label{sec:introduction}

The key difference between graphics processing units (GPUs) and central processing units (CPUs), the latter of which are more commonly used in computational mechanics, stems from the number and design of their cores. 
GPUs have far more cores ($\approx10^3$ versus $\approx10^0-10^1$), but each core is less sophisticated.
This makes GPUs well-suited for highly parallel tasks, provided these tasks are repetitive and relatively simple (regular workloads and predictable memory accesses). 
In the context of the finite element method~(FEM) or finite difference method~(FDM), this can lead to superior performance on GPUs --- especially for problems with low complexity, e.g., FDM on uniform grids. 
By contrast, adaptive methods, irregular meshes, and sparse computations are more challenging to implement efficiently on GPU architectures. 
A further limitation of GPUs is their memory capacity ($\approx10^1$ GB versus $\approx10^2$ GB for CPUs), which can restrict their applicability to smaller problems.\\

Thus, GPUs have primarily been applied to small- to mid-sized problems in the literature.
Examples hereof are~\cite{michea_accelerating_2010,hamilton_room_2013,ye_accelerating_2022} for FDM~\cite{schweiger_gpu-accelerated_2011,traff_simple_2023} for FEM,~\cite{afanasiev_modular_2019} for the spectral element method~(SEM), and~\cite{hamilton_room_2013,xu_generalized_2016} for the finite volume method~(FVM).
Further attempts have been made using deep learning~\cite{goodfellow_deep_2016,bishop_deep_2024} and their corresponding frameworks~\cite{abadi_tensorflow_2016,bradbury_jax_2018,paszke_pytorch_2019} with the goal to exploit advances in software and hardware from the machine learning community~\cite{kollmannsberger_deep_2021,herrmann_deep_2024}. 
Examples\footnote{Not all mentioned approaches mention GPUs directly, but only imply the possibility of using massively parallel hardware architectures.} are available for both FEM~\cite{takeuchi_neural_1994,ramuhalli_finite-element_2005,yao_fea-net_2019,yao_fea-net_2020,park_convolution_2023} and FDM~\cite{mishra_nfdtd_2005,richardson_seismic_2018,sun_theory-guided_2020,herrmann_use_2023}.\\

Although GPUs have been successfully applied to forward problems, their use for inverse optimization problems is still limited. 
A notable exception in static elasticity is~\cite{traff_simple_2023}. 
Unlike in static elasticity, the memory limitation in dynamic optimization problems originates from the sensitivity computation. 
Computing the sensitivity with the adjoint method~\cite{plessix_review_2006, Fichtner2006a, Fichtner2006b, givoli_tutorial_2021} requires information about both the forward and adjoint wave fields at the same time.
Since the forward wave field is solved forward in time, while the adjoint wave field is solved backward in time, the forward wave field must be saved.
Storing the full forward solution requires enormous amounts of memory, as it involves the number of degrees of freedom multiplied by the number of time steps. 
Even worse, the excitation frequency $\omega$ dictates the spatial and temporal resolutions.
For a $d$-dimensional problem, the size of the full wave field scales with $\propto \omega^{d+1}$, i.e., doubling the frequency for a three dimensional problem increases the memory requirement by a factor of $16$.
This high memory demand is difficult to satisfy with current GPU memory capacities, and imposes a restrictive bound upon the size of computable problems.
State-of-the-art remedies are checkpointing~\cite{griewank_achieving_1992,griewank_algorithm_2000,symes_reverse_2007,anderson_time-reversal_2012},
data compression~\cite{Boehm2016, Silva2019}, and multi-GPU implementations~\cite{michea_accelerating_2010,ye_accelerating_2022,omlin_high-performance_2024,omlin_distributed_2024,aloisi_seimicwavesjl_2024}.
Checkpointing stores only selective forward states.
Since the remaining states need to be recomputed during the adjoint computation, checkpointing leads to an increase in the computational effort.
Data compression reduces the resolution of the stored forward solution, which may result in a degradation of the accuracy of the sensitivity.
Finally, multi-GPU usage requires additional (and expensive) hardware with greater communication overheads.
In~\cite{Clapp2009, Shen2015}, the authors re-simulate the forward wavefield along the adjoint wavefield backward in time, allowing memory-efficient gradient computations.
Additionally, they use a random boundary approach to implement absorbing boundary conditions. \\ 

We propose an alternative workaround that does not increase the computational effort.
This approach relies on the superposition of forward and adjoint wave fields.
Corresponding sensitivities are computed approximately with the adjoint method in an almost memory-free manner.
Only information of current time steps are required. 
This reduces the memory requirement from the full forward wave solution to the forward solution at a small number of time steps. 
Given 40 GB of memory, available in a single A100 GPU chip, renders it possible to compute sensitivities of problems with billions of degrees of freedom.
As the proposed approach relies on time reversibility, its application is limited to self-adjoint problems and time-reverse time integration schemes.
We study the efficiency and accuracy of the proposed approach on two inverse problems based on the wave equation: full waveform inversion~(FWI)~\cite{tarantola_inversion_1984,Fichtner2011} and transient acoustic topology optimization~(TATO)~\cite{wadbro_topology_2006,lee_rigid_2009,du_minimization_2007,yoon_topology_2007,duhring_acoustic_2008,kook_acoustical_2012}\footnote{which is directly connected to topology optimization of photonics; see~\cite{jensen_topology_2011,christiansen_inverse_2021,christiansen_compact_2021}.}. 
In both cases, a scalar wave equation without damping is considered. 
The equations are solved with a highly efficient, yet simple, finite difference solver, implemented in CUDA$^{\text{\textregistered}}$~\cite{nickolls_scalable_2008}. The code is available at~\cite{herrmann_memory_2025}. 
Due to its simplicity, the code can readily be adapted to other dynamic forward problems and optimization scenarios. \\

Before explaining the approximate memory-efficient sensitivity computation in \Cref{sec:MEM}, we introduce FWI and TATO in~\Cref{sec:optimization}, and give details on the GPU-based finite difference implementation in~\Cref{sec:forwardsolver}.
In~\Cref{sec:results}, we demonstrate the efficiency and robustness of the introduced method on various optimization problems.
Finally, \Cref{sec:conclusion} concludes our findings.


\section{Dynamic Optimization Problems}
\label{sec:optimization}

\subsection{Underlying Problem}

In FWI and TATO, we aim to minimize a cost functional $C$ that provides a measure of how close a particular model is to a measured (FWI) or desired (TATO) state. 
In the sequel, $\gamma$ is a discrete realization of the corresponding continuous field that represents spatially distributed but time invariant material parameters. The minimization of $C$ is carried out with respect to the variables $\gamma$,
\begin{equation}
    \gamma^* = \arg \min_{\gamma} C(\gamma),
\end{equation}
where $\gamma^*$ minimizes the cost function $C$.
The cost function $C$ is minimized in an iterative procedure using a gradient based approach. To this end, a sensitivity analysis $\mathrm{d}C / \mathrm{d} \gamma = \nabla_\gamma C$ yields the direction of the steepest ascent. 
Thus, an update in the negative direction
\begin{equation}
    \gamma^{(j+1)}=\gamma^{(j)}-\alpha \frac{\mathrm{d} C\left(u(\gamma^{(j)});\gamma^{(j)}\right)}{\mathrm{d} \gamma^{(j)}}\label{eq:gradientdescent}
\end{equation}
with the step length $\alpha$ intends to reduce the objective function $C$\footnote{In the paper at hand, we rely on the gradient-based optimizer Adam~\cite{kingma_adam_2017}, which in PyTorch~\cite{paszke_pytorch_2019} is well-suited to large-scale optimization problems. 
Better choices are, however, L-BFGS-B~\cite{liu_limited_1989} or the method of moving asymptotes~\cite{svanberg_method_1987}, which were not used due to memory constraints by the implementations of SciPy~\cite{virtanen_scipy_2020}.}. 
The procedure is repeated iteratively until convergence is reached. The detailed problem formulations, including governing equations, objective function formulation, and sensitivity analysis with the adjoint method, are provided in~\Cref{ssec:FWI} for FWI and in \Cref{ssec:acousticTopOpt} for TATO.





\subsection{Full Waveform Inversion}\label{ssec:FWI}

\subsubsection*{FWI in a nutshell}

Originally developed for its use in geophysics, FWI can also be applied to ultrasonic testing.
The aim is to infer the internal state $\gamma$ of a sample by measuring wave signals $u$ at sensor locations $\boldsymbol{x}_i$. FWI exploits that waves provide information about the media they traverse. Measured wave signals are the image of a defined signal injected into the sample by a source at a defined position. 
These wave signals, thus, carry information about possible local material anomalies they traversed on their way to the receiver positions which allows to detect local perturbations in the material. 

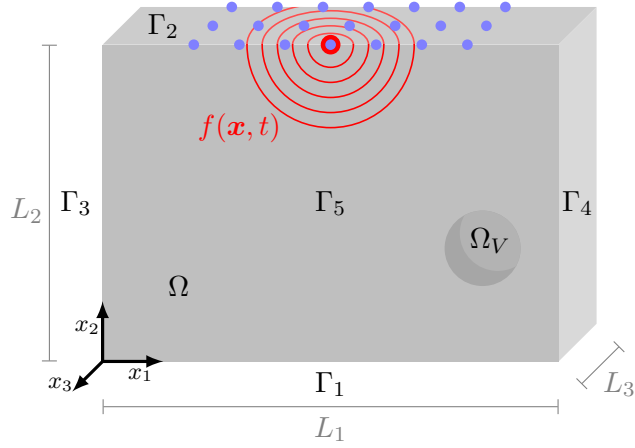
\begin{figure}[htbp]	
	\centering
	\begin{tikzpicture}

    \fill [lightgray!80] (0,4.2) -- (6,4.2) -- (6.5,4.7) -- (0.5,4.7) -- (0,4.2); 
    \fill [lightgray!60] (6,4.2) -- (6.5,4.7) -- (6.5,0.5) -- (6,0) -- (6,4.2);

	\fill [lightgray] (0,0) rectangle (6,4.2);
	
    \fill [lightgray!140] (5,1.5) circle (0.5cm);
    \begin{scope}
        \clip (5,1.5) circle (0.5cm); 
        \fill [lightgray!120] (5.2,1.7) circle (0.5cm); 
    \end{scope} 
    \node at (5.1,1.6) {$\Omega_V$};

    \begin{scope}
	\clip (0,4.2) -- (6,4.2) -- (6.5,4.7) -- (0.5,4.7) -- (0,4.2);
	\foreach {\r} in {0.3,0.5,...,1.2} {
      \draw [semithick, red!70] (3,4.2) ellipse ({\r cm} and {0.5*\r cm});
	}	
	\end{scope}

	\begin{scope}
	\clip (0,0) rectangle (6,4.2);
	\foreach {\r} in {0.3,0.5,...,1.2} {
		\draw [semithick, red] (3,4.2) circle (\r cm);
	}	
	\end{scope}

 	\draw[very thick, latex-latex] (0,0.8) -- (0,0) -- (0.8,0);
	\draw[very thick, -latex] (0,0) -- (-0.4,-0.4);  
    \node at (-0.2,0.45) {\footnotesize{$x_2$}};
	\node at (0.5,-0.18) {\footnotesize{$x_1$}};
    \node at (-0.55,-0.3) {\footnotesize{$x_3$}};

	\fill [red] (3,4.2) circle (0.13cm);
	\foreach {\x} in {1.2,1.8,...,5.1} {
		\foreach {\y} in {4.2} {
			\fill [blue!50] (\x,\y) circle (0.07cm);
            \fill [blue!50] (\x+0.25,\y+0.25) circle (0.07cm);
            \fill [blue!50] (\x+0.5,\y+0.5) circle (0.07cm);
        }
	}

    \node [red] at (1.8,3.1) {$f(\boldsymbol{x},t)$};

    \draw [thin, |-|, gray] (0,-0.6) -- (6,-0.6);
	\node [gray] at (3,-0.9) {$L_1$};
	\draw [thin, |-|, gray] (-0.7,0) -- (-0.7,4.2);
	\node [gray] at (-1,2) {$L_2$};
    \draw [thin, |-|, gray] (6.3,-0.3) -- (6.8,0.2);
    \node [gray] at (6.8,-0.3) {$L_3$};
 
	\node at (-0.35,2.1) {$\Gamma_{3}$};
	\node at (6.25,2.1) {$\Gamma_{4}$};
	\node at (3,-0.3) {$\Gamma_{1}$};
	\node at (0.8,4.45) {$\Gamma_{2}$}; 
    \node at (1,1) {$\Omega$};
    \node at (3,2.1) {$\Gamma_5$};
	\end{tikzpicture}
	\caption{Ultrasonic testing on a domain $\Omega$ ($L_1\times L_2\times L_3$) with a void $\Omega_V\subset \Omega$. The boundaries are defined as $\Gamma=\bigcup_{i=1}^{6} \Gamma_i$. The signal emitted by the source (red circle) is recorded at the sensors (blue circles) over time $\mathcal{T}$. Modified from~\cite{herrmann_use_2023}.}
	\label{fig:ndt}
\end{figure}

In this work we investigate the identification of voids using density-scaling FWI~\cite{burchner_immersed_2023}. 
\Cref{fig:ndt} provides a conceptual illustration of FWI. The source $f(\boldsymbol{x},t)$ (red circle) excites a wave $u$ at a position $\boldsymbol{x}$. After interacting with the domain $\Omega$ and the void $\Omega_V$, the wave $u$ is recorded at the sensor positions (blue circles). 

\subsubsection*{Physical Model}
In solid structures, wave propagation phenomena can be modeled by the elastic wave equation.
However, the scalar wave equation can be used to investigate FWI for method development~\cite{fichtner_adjoint_2006-1, fichtner_adjoint_2006} (see the Appendix of~\cite{herrmann_use_2023} for a brief but sufficient explanation). 
Thus, for the sake of simplicity, we consider the scalar wave equation.
Furthermore, we consider $\rho$-scaling to parameterize the scalar wave equation to detect void-like defects such as cavities or cracks~\cite{burchner_immersed_2023}.
The wave field $u: \Omega \rightarrow \mathbb{R}$ fulfills the scalar wave equation in the domain $\Omega \subset \mathbb{R}^d$ with dimension $d$ for time $\mathcal{T} \subset \mathbb{R}$ excited by a volumetric force $f$:
\begin{equation}
    \gamma(\boldsymbol{x}) \, \rho_0 \, \ddot{u}(\boldsymbol{x},t) - \nabla\cdot \left(\gamma(\boldsymbol{x}) \,\rho_0 \, c_0^2 \, \nabla u(\boldsymbol{x},t)\right)=f(\boldsymbol{x},t) \qquad \text{on }\Omega\times \mathcal{T},\label{eq:scalarwaveequation}
\end{equation}
where $\rho_0$ and $c_0$ are the density and wave speed of the undamaged material also called background material.
Thus, $\rho$-scaling can be interpreted as scaling the background density $\rho_0$ with a spatially distributed scaling function $\gamma\in[\epsilon,1]$, where $0<\epsilon\ll 1$ is a non-zero lower bound that ensures numerical stability of the simulations. 
In this work, we set $\epsilon=10^{-5}$. 
The material parameters used in this work are $\rho_0=2\,700$ $\text{kg}/\text{m}^3$, and $c_0=6\,000$ $\text{m}/\text{s}$.\\

We consider free reflecting boundaries, modeled by homogeneous Neumann boundary conditions
\begin{equation}
    \boldsymbol{n}\cdot \nabla u = 0 \qquad \text{on } \Gamma,\label{eq:BC}
\end{equation}
where $\boldsymbol{n}$ is the unit normal vector on the boundary $\Gamma$ of the domain. 
In addition, homogeneous initial conditions are considered.
\begin{equation}
    u(\boldsymbol{x},0)=\dot{u}(\boldsymbol{x},0)=0\label{eq:IC} \qquad \text{on }\Omega 
\end{equation}
The domain is excited by a point source at position $\boldsymbol{x}_s$ which is defined using the Kronecker delta function $\delta$
\begin{equation}        
    f(\boldsymbol{x},t)=\psi(t) \, \delta(\boldsymbol{x}-\boldsymbol{x}_s). \label{eq:source1}
\end{equation}
In the sequel, the time signal $\psi$, is modeled by the sine burst with $n_c$ cycles, a frequency $\omega$, and the amplitude $\psi_0$ such that
\begin{equation}
    \psi(t)=\begin{cases}
        \psi_0 \, \sin(\omega \, t) \, \sin^2\left(\frac{\omega \, t}{2 n_c}\right) &\quad \text{for } 0\leq t \leq \frac{2\pi \, n_c}{\omega}, \\
        0 &\quad \text{for } \frac{2\pi \, n_c}{\omega}< t.
    \end{cases}\label{eq:source2}
\end{equation}

\subsubsection*{Objective Function}
\label{ssec:fwioptimization}

For the sample under investigation, a set of measured wave signals $u^{\mathcal{M}}$ at the sensor locations $\boldsymbol{x}_i$ (see~\Cref{fig:ndt}) is compared with simulated wave signals $u$ modeled by \Cref{eq:scalarwaveequation}. 
The $L^2$ misfit between the measured and simulated wave signals is a suitable choice for the objective function to detect voids:
\begin{equation}
    C(\gamma) = \frac{1}{2} \int_\mathcal{T} \int_\Omega \sum_{i=1}^{\text{N}_r} \left( u(\boldsymbol{x},t;\gamma) - u^{\mathcal{M}}(\boldsymbol{x}_i,t) \right)^2 \delta(\boldsymbol{x}-\boldsymbol{x}_i) \, \mathrm{d}\Omega \, \mathrm{d}\mathcal{T}.\label{eq:cost1}
\end{equation}
The simulations are based on the material field $\gamma$. 
After evaluating the cost function, the internal material distribution $\gamma^{(j)}$ is updated according to \Cref{eq:gradientdescent}. 
This update relies on the derivative of the cost function with respect to $\gamma$, i.e., $\mathrm{d}C / \mathrm{d}\gamma$ that is obtained using sensitivity analysis.

\subsubsection*{Sensitivity Analysis}
\label{ssec:fwisensitivity}
Following the derivation of the adjoint method in~\cite{burchner_immersed_2023} (for which we consider the material $\gamma$ as continuous field), the sensitivity of the cost function with respect to a perturbation $\delta \gamma$ is
\begin{equation}
    \nabla_\gamma C \, \delta \gamma = \int_\Omega K_\gamma(u, u^\dagger) \, \delta \gamma \, \mathrm{d}\Omega,\label{eq:sensitivitygamma}
\end{equation}
where $K_\gamma$ is a bilinear Fréchet kernel, given as
\begin{equation}
    K_\gamma\left(u(\boldsymbol{x}, t\right), u^\dagger(\boldsymbol{x}, t)) = -\int_{\mathcal{T}}\left(\rho_0 \, \dot{u}^\dagger \, \dot{u} + \rho_0 \, c_0^2 \nabla u^\dagger \cdot \nabla u\right) \, \mathrm{d}\mathcal{T}.
\label{eq:kernel1}
\end{equation}
Here, $u^\dagger$ is the adjoint wave field underlying the adjoint wave equation
\begin{equation}
    \gamma(\boldsymbol{x}) \, \rho_0 \, \ddot{u}^\dagger(\boldsymbol{x}, t) - \nabla \cdot (\gamma(\boldsymbol{x}) \, \rho_0 \, c_0^2 \nabla u^\dagger(\boldsymbol{x}, t)) = f^\dagger(\boldsymbol{x}, t) \qquad \text{on } \Omega\times \mathcal{T} \label{eq:adjointequation1}
\end{equation}
with the adjoint source
\begin{equation}
    f^\dagger(\boldsymbol{x},t)=-\sum_{i=1}^{N_r}\left(u(\boldsymbol{x},t)-u^{\mathcal{M}}(\boldsymbol{x}_i,t)\right)\delta(\boldsymbol{x}-\boldsymbol{x}_i) .
    \label{eq:adjointSource1}
\end{equation}
and temporal end conditions at $t_\mathrm{end}$
\begin{equation}
    u^\dagger(\boldsymbol{x}, t_\mathrm{end})=\dot{u}^\dagger(\boldsymbol{x}, t_\mathrm{end}) = 0 \qquad \text{on }\Omega .
    \label{eq:ICadjoint}
\end{equation}
In practice, the end conditions are introduced as initial conditions by solving the adjoint equation (\Cref{eq:adjointequation1}) backwards in time.
As in the forward problem, homogeneous Neumann boundary conditions are enforced
\begin{equation}
    \boldsymbol{n} \cdot \nabla u^\dagger = 0 \qquad \text{on } \Gamma.
    \label{eq:BCadjoint}
\end{equation}
Solving the forward and adjoint wave equation with FDM provides discretized wave fields.
Finally, the derivative of the cost function with respect to the discretized material field at each grid point indicated by the subscript $i$ of the FDM is
\begin{equation}
    \frac{\mathrm{d} C}{\mathrm{d} \gamma_i}=\int_\Omega K_\gamma(u(\boldsymbol{x}, t), u^{\dagger}(\boldsymbol{x}, t)) \, \delta(\boldsymbol{x}_{\gamma_i} - \boldsymbol{x}) \, \mathrm{d}\Omega,\label{eq:gradientgamma}
\end{equation}
where $\gamma_i$ and $\boldsymbol{x}_{\gamma_i}$ are the material's value and position of each grid point $i$. 

\subsubsection*{Constraints}
As the indicator function $\gamma$ is defined in the range $[\epsilon,1]$, its values are clipped to $[\epsilon,1]$ after each gradient-based update performed by the Adam optimizer (\Cref{eq:gradientdescent}). 
Furthermore, as we are interested in detecting voids in an otherwise homogeneous material, the resulting material distribution should be binary, i.e., $\epsilon$ or $1$. This constraint is, however, not imposed as it is not strictly necessary, as empirically observed in~\cite{burchner_immersed_2023,burchner_isogeometric_2023,herrmann_use_2023}. 
Regularization techniques such as neural network parametrizations~\cite{herrmann_use_2023}, total variation~\cite{arridge_solving_2019}, projection~\cite{guest_achieving_2004,wang_projection_2011,li_volume_2015}, or even SIMP\footnote{solid isotropic material with penalization}-like penalizations~\cite{bendsoe_optimal_1989,bendsoe_material_1999,bendsoe_topology_2003}, may be beneficial but are not applied in this work.

\subsection{Transient Acoustic Topology Optimization}
\label{ssec:acousticTopOpt}

\subsubsection*{TATO in a nutshell}

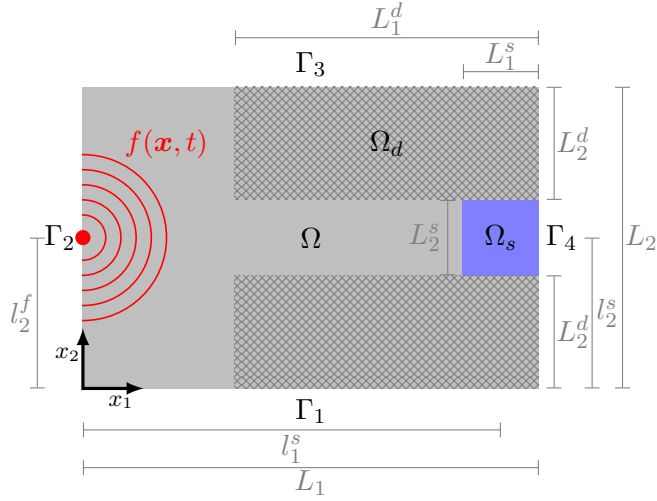
\begin{figure}[htbp]
	\centering
	\begin{tikzpicture}
	\node at (3,-0.3) {$\Gamma_{1}$};
	\node at (3,4.3) {$\Gamma_{3}$};
	\node at (-0.3,2) {$\Gamma_{2}$};
	\node at (6.3,2) {$\Gamma_{4}$};
	
	\fill [thin, fill=lightgray] (0,0) rectangle (6,4);
	
	\fill [thin, fill=blue!50] (5,1.5) rectangle (6,2.5); 
	\fill [thin, pattern=crosshatch, pattern color=gray] (2,2.5) rectangle (6,4);
	\fill [thin, pattern=crosshatch, pattern color=gray] (2,0) rectangle (6,1.5);
	
	\fill [red] (0,2) circle (0.1cm);
	\begin{scope}
	
    \clip (0,0) rectangle (6,4.2);
	\foreach {\r} in {0.3,0.5,...,1.1} {
		\draw [semithick, red] (0,2) circle (\r cm);
	}
	\end{scope}

    \node at (5.5,2) {$\Omega_s$};
	\node at (4,3.25) {$\Omega_d$};
	\node at (3,2) {$\Omega$};

	\draw[thin, |-|, gray] (-0.6,0) -- (-0.6,2);
	\node [gray] at (-0.8,1) {$l_2^f$};

	\draw[thin, |-|, gray] (5,4.2) -- (6,4.2);
	\draw[thin, |-|, gray] (4.8,1.5) -- (4.8,2.5);
	\node [gray] at (5.5,4.4) {$L_1^s$};
	\node [gray] at (4.5,2) {$L_2^s$};

	\draw [thin, |-|, gray] (0,-0.55) -- (5.5,-0.55);
	\draw [thin, |-|, gray] (6.7,0) -- (6.7,2);
	\node [gray] at (2.75,-0.8) {$l_1^s$};
	\node [gray] at (6.9,1) {$l_2^s$};

    \draw[thin, |-|, gray] (2,4.65) -- (6,4.65);
    \draw[thin, |-|, gray] (6.2,2.5) -- (6.2,4);
    \draw[thin, |-|, gray] (6.2,1.5) -- (6.2,0);

    \node [gray] at (4,4.9) {$L_1^d$};
    \node [gray] at (6.45,0.75) {$L_2^d$};
    \node [gray] at (6.45,3.25) {$L_2^d$};

	\draw[thin, |-|, gray] (0,-1.05) -- (6,-1.05);
	\node [gray] at (3,-1.25) {$L_1$};
	\draw[thin, |-|, gray] (7.1,0) -- (7.1,4);
	\node [gray] at (7.35,2) {$L_2$};
	
	\node [red] at (1.1,3.25) {$f(\boldsymbol{x},t)$};
	
	\draw[very thick, latex-latex] (0,0.8) -- (0,0) -- (0.8,0);
	\node at (-0.2,0.45) {\footnotesize{$x_2$}};
	\node at (0.5,-0.18) {\footnotesize{$x_1$}};
	\end{tikzpicture}
 	\caption{Topology optimization of an acoustic black hole $\Omega_d$ ($L_1^d \times L_2^d$) with the aim of suppressing the sound pressure in the domain $\Omega_s$ ($L_1^s\times L_2^s$) at ($l_1^s,l_2^s)$ induced by the source $f$ at ($l_1^f,l_2^f$). The physics of the domain $\Omega$ with dimensions $L_1\times L_2$, including $\Omega_d$ and $\Omega_s$, is governed by the acoustic wave equation. The boundaries are defined as $\Gamma=\bigcup_{i=1}^{4} \Gamma_i$. Modified from~\cite{herrmann_neural_2024}.}\label{fig:setupTopOpt}
  \end{figure}

TATO~\cite{hyun_transient_2021,dilgen_topology_2024} aims to find the optimal design of a domain $\Omega_d$ which is embedded in a larger domain $\Omega$.
The performance of the design is evaluated by an objective function. 
The optimal material distribution within $\Omega_d$ is computed such that acoustic waves minimize the objective. 
The acoustic waves are excited by a point source $f$. In the paper at hand, we consider the design of an acoustic black hole (see~\cite{pelat_acoustic_2020} for a general review and~\cite{mousavi_topology_2024} for a concrete topology optimization example). 
The problem setup is schematically illustrated in \Cref{fig:setupTopOpt}. 
Acoustic waves $u$ excited by the source $f$ travel through the domain $\Omega$ (including $\Omega_d$ and $\Omega_s$).
To achieve an optimal design, the waves are supposed to arrive in the domain $\Omega_s$ with a minimum amplitude.

\subsubsection*{Physical Model} 
The acoustic waves are governed by the acoustic wave equation, where we rely on the modified parametrization from~\cite{duhring_acoustic_2008}\footnote{modified from the Helmholtz equation for transient wave simulation, which employs an additional unit normalization}.
\begin{equation} 
    \kappa^{-1} \, (\boldsymbol{x}) \ddot{u}(\boldsymbol{x}, t) - \nabla \cdot \left(\rho^{-1}(\boldsymbol{x}) \nabla u(\boldsymbol{x},t) \right) = f(\boldsymbol{x},t) \qquad \text{on } \Omega\times \mathcal{T}\label{eq:wavetopopt}
\end{equation}
The material is specified in terms of the inverse bulk modulus $\kappa^{-1}$ and the inverse mass density $\rho^{-1}$, which are parametrized as
\begin{align}
    \rho^{-1}(\gamma(\boldsymbol{x}))&=\frac{1}{\rho_1}+\gamma(\boldsymbol{x})\left(\frac{1}{\rho_2}-\frac{1}{\rho_1}\right),\\
    \kappa^{-1}(\gamma(\boldsymbol{x}))&=\frac{1}{\kappa_1}+\gamma(\boldsymbol{x})\left(\frac{1}{\kappa_2}-\frac{1}{\kappa_1}\right),
\end{align}
with the spatially distributed indicator function $\gamma \in [0,1]$. 
The material parameters are $\rho_1=1.204$ $\text{kg}/\text{m}^3$ and $\kappa_1=1.419\cdot 10^5$ $\text{N}/\text{m}^2$ for air, while for the solid material\footnote{We use $\kappa_2=6.87\cdot 10^8$ $\text{N}/\text{m}^2$ instead of $\kappa_2=6.87\cdot 10^{10}$ $\text{N}/\text{m}^2$, as in~\cite{duhring_acoustic_2008} to be able to use larger time steps sizes, and thus a smaller number of time steps. Note, however, that the proposed memory-efficient sensitivity computation is better suited to larger time step numbers in comparison to standard adjoint sensitivity computation.} the values are $\rho_2=2\,643$ $\text{kg}/\text{m}^3$ and $\kappa_2=6.87\cdot 10^8$ $\text{N}/\text{m}^2$. 
For $\gamma=0$ the material parameters of air are recovered, while $\gamma=1$ mimicks the solid material. \\

Again, reflecting boundary conditions in terms of homogeneous boundary conditions are employed according to \Cref{eq:BC}, and homogeneous initial conditions are imposed as defined in \Cref{eq:IC}. 
The specimen is excited by a sine burst, \Cref{eq:source1,eq:source2}. Note that the source term $f$ in \Cref{eq:wavetopopt} is normalized by the inverse density $\rho^{-1}$, thus carrying the unit $\text{s}^{-2}$ instead of $\mathrm{N}/\mathrm{m}^2$.

\subsubsection*{Objective Function}
Based on the simulated wave pressures of a given material distribution $\gamma$, the design is assessed with the following objective function
\begin{equation}
    C(\gamma) = \frac{1}{A} \int_\mathcal{T}  \int_{\Omega_s}u(\boldsymbol{x},t;\gamma)^2 \, \mathrm{d} \Omega_s \,\mathrm{d} \mathcal{T}.
    \label{eq:cost2}
\end{equation}
where $A = \int_{\Omega_s}\mathrm{d}\Omega_s$ is the area of the domain of interest.
Analogously to \Cref{ssec:fwioptimization}, the current guess $\gamma^{(j)}$ is updated with the update rule from \Cref{eq:gradientdescent} using the derivative of the objective $\mathrm{d}C / \mathrm{d}\gamma$.

\subsubsection*{Sensitivity Analysis}\label{ssec:topoptsensitivity}

With the adjoint method, the sensitivity is computed equivalently to \Cref{ssec:fwisensitivity} using \Cref{eq:sensitivitygamma}. The Fréchet kernel corresponding to the acoustic model defined in \Cref{eq:wavetopopt} is 
\begin{align}
    K_{\gamma}(u(\boldsymbol{x},t), u^\dagger(\boldsymbol{x},t)) &=-\int_\mathcal{T} \Biggl( \left( \frac{1}{\kappa_2} -\frac{1}{\kappa_1} \right)\dot{u}^\dagger \, \dot{u}-\left(\frac{1}{\rho_2}-\frac{1}{\rho_1}\right)\nabla u^\dagger \cdot \nabla u\Biggr) \mathrm{d}\mathcal{T}.\label{eq:kernel2}
\end{align}

\noindent The adjoint wave field $u^\dagger$ obeys the adjoint wave equation
\begin{equation}
    \kappa^{-1}(\boldsymbol{x}) \ddot{u}^\dagger(\boldsymbol{x}, t) 
    - \nabla \cdot (\rho^{-1}(\boldsymbol{x}) \nabla u^\dagger(\boldsymbol{x}, t)) = f^\dagger(\boldsymbol{x}, t) \qquad \text{on } \Omega \times \mathcal{T}
\end{equation}
with the adjoint source
\begin{equation}
    f^\dagger(\boldsymbol{x}, t)=\begin{cases}
        -\frac{2 u(\boldsymbol{x}, t)}{A}\quad &\text{if } x\in\Omega_s,\\
        0\quad &\text{else}
    \end{cases}.\label{eq:adjointSource2}
\end{equation}
Again, homogeneous Neumann boundary conditions and homogeneous temporal end conditions must be employed, as in \Cref{eq:BCadjoint,eq:ICadjoint}.
After solving the forward and adjoint wave fields with the FDM, the gradient with respect to $\gamma_i$ considering each grid point $i$ can be performed following \Cref{eq:gradientgamma} with \Cref{eq:kernel2}.


\subsubsection*{Constraints}
To prevent mesh-dependent designs that include excessively small, and thus unmanufacturable, features, a filtering approach is utilized~\cite{sigmund_numerical_1998}. 
In particular, we apply a density filter with linear decay~\cite{bruns_topology_2001,bourdin_filters_2001}. The filter yields the following modified indicator function
\begin{equation}
    \tilde{\gamma}_i = \frac{\sum_{k\in N_i} w(\boldsymbol{x}_i - \boldsymbol{x}_k) \, \gamma_k}{\sum_{k\in N_i}w(\boldsymbol{x}_i-\boldsymbol{x}_k)},
\end{equation}
where $N_i$ is the set of grid points in the neighborhood of the grid point $i$ defined by the radius $r_f$, i.e., $N_i=\{k|\text{ }||\boldsymbol{x}_k-\boldsymbol{x}_i|| < r_f\}$. 
The linear decay is incorporated via the weighting function $w(\boldsymbol{x})$, given as
\begin{equation}
    w(\boldsymbol{x})=\begin{cases}
        r_f-||\boldsymbol{x}||\quad &\text{if } ||\boldsymbol{x}||\leq r_f, \\
        0 \quad &\text{else}.
    \end{cases}
\end{equation}
In this work, a filter radius of 1.5 voxels (corresponding to the same filter neighborhood as in the giga-voxel optimization in~\cite{aage_giga-voxel_2017}) is applied. \\

An unwanted side effect of filtering is that the designs are blurred, although sharp $0 / 1$ designs are desired. 
A subsequent projection recovers an almost binary design~\cite{guest_achieving_2004,wang_projection_2011,li_volume_2015}. 
Specifically, a smooth approximation of the Heaviside function $H(\tilde{\gamma})$ achieves the projection of the filtered indicator $\tilde{\gamma}$ to a smooth binary-like indicator $\bar{\tilde{\gamma}}$:
\begin{equation}
    \bar{\tilde{\gamma}}=H(\tilde{\gamma})=\frac{\tanh(\beta \, \eta)+\tanh(\beta \, (\tilde{\gamma} - \eta))}{\tanh(\beta \, \eta)+\tanh(\beta \, (1-\eta))}.
\end{equation}
The parameter $\eta$ shifts the thresholding, while $\beta$ adjusts the sharpness. 
Initially, $\beta$ must be kept small to allow a proper gradient flow from the projected indicator values $\bar{\tilde{\gamma}}$ to the indicator values $\gamma$. 
After initializing $\beta=1$, we follow a $\beta$-continuation scheme~\cite{guest_achieving_2004,wang_projection_2011,christiansen_creating_2015}, which increases $\beta$ by $10\%$ every five iterations. 
The threshold is set as $\eta=0.5$. 
Note that for smaller $\beta$ values, the projection does not guarantee $\bar{\tilde{\gamma}}\in[0,1]$, which is, however, needed to ensure numerical stability (specifically the upper bound). 
Therefore, the filtered and projected indicator values $\bar{\tilde{\gamma}}$ are clipped to the interval $[0,1]$ at every iteration. \\

Modifying the indicator by filtering and projection needs to be considered during the sensitivity computation. 
This yields the following chain rule
\begin{equation}
    \frac{\mathrm{d} C}{\mathrm{d} \gamma}= \frac{\mathrm{d} C}{\mathrm{d} \bar{\tilde{\gamma}}} \, \frac{\mathrm{d}\bar{\tilde{\gamma}}}{\mathrm{d}\tilde{\gamma}} \, \frac{\partial\tilde{\gamma}}{\partial\gamma},
\end{equation}
where the vector $\mathrm{d}C / \mathrm{d}\bar{\tilde{\gamma}}$ is obtained by the sensitivity analysis described in \Cref{ssec:topoptsensitivity}, and the matrices $\mathrm{d}\bar{\tilde{\gamma}} / \mathrm{d}\tilde{\gamma}$ and  $\partial \tilde{\gamma}/\partial\gamma$ are obtained as follows
\begin{align}
    \frac{\mathrm{d}\bar{\tilde{\gamma}}}{\mathrm{d} \tilde{\gamma}} &= \frac{\beta}{[\tanh(\beta \,\eta)+\tanh(\beta \, (1-\eta))]\cosh^2(\beta \, (\tilde{\gamma}-\eta))}, \\
    \frac{\mathrm{d} \tilde{\gamma}}{\mathrm{d} \gamma_k} &= \sum_{i} \frac{\partial \tilde{\gamma}_i}{\partial \gamma_k}, \\ 
    \frac{\partial \tilde{\gamma_i}}{\partial \gamma_k} &= \begin{cases}
        \frac{w(\boldsymbol{x}_i-\boldsymbol{x}_k)}{\sum_{k\in N_i}w(\boldsymbol{x}_i-\boldsymbol{x}_k)} &\text{ if } k \in N_i, \\
        0 &\text{ else}.
    \end{cases}
\end{align}

Note that designs obtained through acoustic topology optimization are typically not robust~\cite{christiansen_creating_2015}. 
Approaches discussed in~\cite{sigmund_morphology-based_2007,sigmund_manufacturing_2009,christiansen_creating_2015} can amend the robustness issue, which is, however, not considered in this work.






\section{Efficient Finite Difference Discretization on GPUs}\label{sec:forwardsolver}

\subsection{Second-order Finite Differences}
The wave equations in \Cref{eq:scalarwaveequation,eq:wavetopopt} are discretized with central second-order finite differences in space and time.
In one dimension, we obtain the following explicit schemes for
\begin{itemize}
    \item the scalar wave equation (\Cref{eq:scalarwaveequation}):
\begin{equation}
    \begin{split}
        u_i^{n+1}=&-u_i^{n-1}+2u_i^n \\
        &+\underbrace{\frac{2}{\gamma_i}\left(\frac{c_0 \, \Delta t}{\Delta x}\right)^2 \Biggl(\left(\frac{1}{\gamma_i+\gamma_{i+1}} \right)(u_{i+1}^n-u_i^n)-\left(\frac{1}{\gamma_{i-1}+\gamma_i}\right) (u_i^n-u_{i-1}^n) \Biggr)}_{\text{spatial stencil}}\\
        &+\frac{\Delta t^2}{\rho_0\gamma_i}f_i^n,\label{eq:FDstencil1}
    \end{split}
\end{equation}
    \item and the acoustic wave equation (\Cref{eq:wavetopopt}):
\begin{equation}
    \begin{split}
        u_i^{n+1}=& -u_i^{n-1}+2u_i^n\\
        &+\underbrace{2 \kappa_i\left(\frac{\Delta t}{\Delta x}\right)^2\Biggl( \left( \frac{1}{\rho_i + \rho_{i+1}}\right) (u_{i+1}^n-u_i^n) - \left(\frac{1}{\rho_{i-1} + \rho_i} \right) (u_i^n-u_{i-1}^n) \Biggr)}_{\text{spatial stencil}}\\
        &+ \kappa_i \, \Delta t^2 \, f_i^n.\label{eq:FDstencil2}
    \end{split}
\end{equation}

\end{itemize}
The are straightforward to extend to two and three dimensions by expanding the spatial stencil. 
The spatial grid spacing is denoted as $\Delta x$ and the time step size as $\Delta t$. 
Note that material points are interpolated via a harmonic mean, which is preferred for materials exhibiting discontinuities~\cite{langtangen_finite_2017}. 
The homogeneous Neumann boundary conditions from \Cref{eq:BC} are incorporated via ghost cells. The material at the ghost cells is set equal to the material at the boundary of the domain, ensuring the correct averages of $\gamma_i$ in \Cref{eq:FDstencil1} and $\rho_i$ in \Cref{eq:FDstencil2}.


\subsection{Implementational Aspects}

From a computational standpoint, the big advantage of the field update schemes of \Cref{eq:FDstencil1,eq:FDstencil2} is that the computation of each grid point $i$ is a local operation that only involves points in its close proximity but is independent of the computation of all other locations. 
Such schemes benefit from the massively parallel data processing capabilities provided by modern general purpose GPU computing. 
To achieve maximum computational efficiency in this setting, the presented examples are computed using a CUDA$^{\text{\textregistered}}$ implementation. Therein, updating each grid point is handled by simultaneously executing the rule from \Cref{eq:FDstencil1,eq:FDstencil2} in the form of \textit{CUDA$^{\text{\textregistered}}$ kernels}. 
To minimize the implementational effort, only the kernels that represent a computational bottleneck are implemented in CUDA$^{\text{\textregistered}}$, while the rest of the implementation uses CuPy\footnote{\href{https://cupy.dev}{https://cupy.dev}}, a convenient Python wrapper layer on top of the standard CUDA$^{\text{\textregistered}}$ functionality. The corresponding implementation can be found at~\cite{herrmann_memory_2025}.\\

Alternatively, a simpler implementation can be achieved by viewing the stencils (\Cref{eq:FDstencil1,eq:FDstencil2}) as $3\times 3$ convolutional operations. 
Thereby, the stencil can be implemented as a hard-coded recurrent convolutional neural network (CNN), as proposed in~\cite{richardson_seismic_2018,herrmann_use_2023}; see Appendix~\ref{appendix:conv} for technical details. 
This leads to straightforward and surprisingly efficient finite difference implementations, which we will consider as a baseline for the computation times in the sequel (relying on the implementation in PyTorch~\cite{paszke_pytorch_2019} from~\cite{herrmann_use_2024}).

\subsection{Computation Time}

\begin{figure}[htbp]
    \centering
    \begin{subfigure}[t]{0.49\textwidth}
        \includegraphics[width=\textwidth]{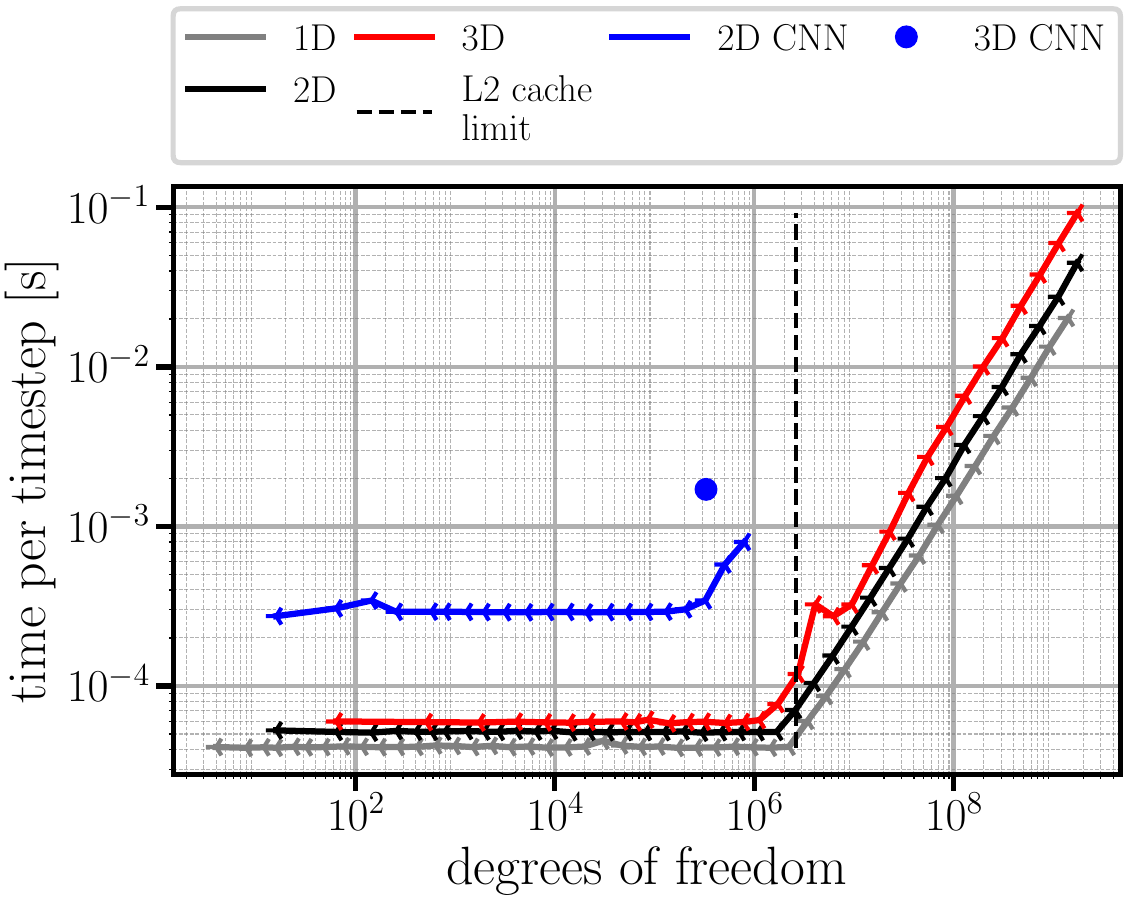}
    \end{subfigure}
    \caption{Computation time of the forward simulation versus number of degrees of freedom in one, two, and three dimensions. The timings rely on the CUDA$^{\text{\textregistered}}$ implementation and are compared to the timings from~\cite{herrmann_use_2023}, which relied on the CNN implementation. Here, the scalar wave equation from \Cref{eq:scalarwaveequation} is considered.}\label{fig:computationtime1}
\end{figure}

Considering the scalar wave equation and single precision floating-points, the computation time per time step of the finite difference forward solver versus the degrees of freedom is shown in \Cref{fig:computationtime1} for one, two, and three dimensions. 
All computations in this and upcoming studies were performed on an NVIDIA$^{\text{\textregistered}}$ A100 40 GB GPU. 
The CUDA$^{\text{\textregistered}}$ implementation (gray, black, and red lines in~\Cref{fig:computationtime1}) is compared to the PyTorch version based on CNNs from~\cite{herrmann_use_2023} (blue line), which is about one to two orders of magnitude slower. The two-dimensional timings were recomputed, while the three-dimensional data point is taken directly from~\cite{herrmann_use_2023}. 
Both represent the upper limit in terms of problem sizes, due to PyTorch's significant memory overhead. 
Thus, the CUDA$^{\text{\textregistered}}$ implementation is clearly superior as a forward solver in both memory efficiency and speed. The graph for the CUDA$^{\text{\textregistered}}$ implementation consists of two parts. 
The computational time is independent of the number of degrees of freedom until about $2\cdot 10^6$ after which it increases with a slope of three in the double logarithmic plot. The point of increase corresponds to the maximum capacity of the L2 cache of 40 MB\footnote{A cache of 40 MB can store 4 times 2.621.440
 short variables with 4 B ($u_i^{n-1}, u_i^n, u_i^{n+1},\gamma^j$ ). A corresponding behavior was also observed on the RTX500 grapics card with 16 MB L2 cache as well as the RTX4060 graphics card with 32 MB L2 cache. All cards clearly indicating a strong decrease of the cache hit rate beyond this point.} (indicated by the vertical dashed line). 

\section{Memory-Efficient Adjoint Sensitivity Computation}
\label{sec:MEM}

To compute the sensitivity with the adjoint method in a memory-efficient manner, we consider the superposed wave field $u^s$, where $u^s$ is the linear combination of the forward wave field $u$ and $k$ times the adjoint wave field $u^\dagger$
\begin{equation}
    u^s(\boldsymbol{x},t)=u(\boldsymbol{x},t) + k \, u^\dagger(\boldsymbol{x},t) .
\end{equation}
Due to linearity of the forward and adjoint equations in $u$ and $u^\dagger$, the governing equation for the superposed solution $u^s$ can be written as
\begin{equation}
    \gamma(\boldsymbol{x}) \, \rho_0 \, \ddot{u}^s(\boldsymbol{x},t) - \nabla \cdot (\gamma \, \rho_0 \, c_0^2 \, \nabla u^s(\boldsymbol{x},t)) = f^s(\boldsymbol{x},t)
    \label{eq:superposedPDE}
\end{equation}
in FWI, while it is
\begin{equation}
    \kappa^{-1} \, \ddot{u}^s(\boldsymbol{x},t) - \nabla \cdot(\rho^{-1} \, \nabla u^s(\boldsymbol{x},t)) = f^s(\boldsymbol{x},t)\label{eq:superposedPDE2}
\end{equation}
in TATO. 
The superposed force $f^s$ is defined as
\begin{equation}
    f^s(\boldsymbol{x},t) = f(\boldsymbol{x},t) + k \, f^\dagger(\boldsymbol{x},t),\label{eq:superposedforce}
\end{equation}
where $f^\dagger$ is computed according to \Cref{eq:adjointSource1} or \Cref{eq:adjointSource2}. Furthermore, superposed end conditions can be defined as
\begin{align}
    u^s(\boldsymbol{x}, t_{\text{end}}) = u(\boldsymbol{x},t_{\text{end}}) + k \, u^\dagger(\boldsymbol{x},t_{\text{end}}) = u(\boldsymbol{x},t_{\text{end}}),\\
    \dot{u}^s(\boldsymbol{x},t_{\text{end}}) = \dot{u}(\boldsymbol{x},t_{\text{end}}) + k\, \dot{u}^\dagger(\boldsymbol{x},t_{\text{end}}) = \dot{u}(\boldsymbol{x},t_{\text{end}}),
\end{align}
because of the homogeneous adjoint end conditions (\Cref{eq:ICadjoint}). 
Equivalently to the adjoint equation, the superposed equation can be solved backwards in time. \\

Inserting the superposed solution $u^s$ into the sensitivity kernel expression (\Cref{eq:kernel1} or \Cref{eq:kernel2}), yields
\begin{align}
    \begin{split}
    K_\gamma(u^s,u^s)&=K_\gamma(u+ku^\dagger,u+ku^\dagger)\\
    &=\underbrace{K_\gamma(u,u)}_{\text{to be subtracted}}+ 2k \, \underbrace{K_\gamma(u,u^\dagger)}_{\text{gradient } \mathrm{d} C/\mathrm{d} \gamma_i}+k^2 \,\underbrace{K_\gamma(u^\dagger,u^\dagger)}_{\text{small}}\label{eq:superposedkernel1}
    \end{split}
\end{align}
for bilinear kernels, i.e., $K_\gamma(u,u^\dagger)=K_\gamma(u^\dagger,u)$. 
As $u^\dagger$ is typically multiple orders of magnitude smaller than $u$, the third term $K_\gamma(u^\dagger,u^\dagger)$ in \Cref{eq:superposedkernel1} is negligible for small enough $k$ values. 
This enables an approximation of the mixed kernel $K(u,u^\dagger)$ needed for the sensitivity computation in \Cref{eq:sensitivitygamma}. 
With a suitable choice of $k$, we can approximate the sensitivity $ K_\gamma(u,u^\dagger)$, by introducing the mixed kernel $\tilde{K}_\gamma$, which is given as
\begin{equation}
    K_\gamma(u,u^\dagger) \approx \tilde{K}_\gamma= \frac{1}{2k}\left(K_\gamma(u^s,u^s)-K_\gamma(u,u)\right).\label{eq:superposedkernel}
\end{equation}
Note that \Cref{eq:superposedkernel} is only applicable in practice for time-reversible time integrators such as the second-order central differences for self-adjoint problems.
Since $K_\gamma(u, u)$ can be integrated alongside the forward simulation, and since $K_\gamma(u^s,u^s)$ is independent of $u$, the approximated sensitivity can be computed without saving the forward wave field $u$. 
The pseudo-code illustrated in \Cref{alg:memoryTrick} computes the cost and gradient for a material distribution. \\


\begin{algorithm*}[htbp]
	\caption{Memory-efficient mixed kernel computation $\tilde{K}_\gamma(u,u^\dagger)\approx K_\gamma(u,u^\dagger)$}\label{alg:memoryTrick}
	\begin{algorithmic}[1]
		\Require measurement $u^{\mathcal{M}}$, indicator $\gamma$, source $f^n$, factor $k$, number of time steps $N$
        \State initialize cost $C=0$ kernel $\tilde{K}_\gamma=0$
        \State initialize $u^0 = 0, u^1 = 0$
        \Statex \textbf{solve forward problem, compute adjoint force \& increment kernel} (forward in time)
        \For{time step $n = 1, ..., N-1$}
            \State $u^{n+1}\leftarrow$ \texttt{propagateStep}$(u^{n-1}, u^{n}, f^{n}, \gamma)$ \Comment(cf. \Cref{eq:FDstencil1,eq:FDstencil2} including BCs)
            \State $f^{\dagger^{n+1}}\leftarrow$ \texttt{computeAdjointSource}$(u^{n+1},u^{\mathcal{M}})$ \Comment(cf. \Cref{eq:adjointSource1,eq:adjointSource2})
            \State $C\leftarrow C +$ \texttt{incrementCost}$(u^{n+1},u^{\mathcal{M}})$ \Comment(cf. \Cref{eq:cost1,eq:cost2})
            \State $\tilde{K}_\gamma\leftarrow \tilde{K}_\gamma-$ \texttt{incrementKernel}$(u^{n+1})$ \Comment{cf. \Cref{eq:kernel1,eq:kernel2,eq:superposedkernel}($-K_\gamma(u,u))$ }
            \State delete $u^{n-1}$
        \EndFor
        \State scale adjoint force $f^{\dagger^{n+1}} \leftarrow k f^{\dagger^{n+1}}$ \Comment{cf. \Cref{eq:superposedforce}}
        \State initialize $u^{s^{N}}=u^{{N}}, u^{s^{N-1}}=u^{N-1}$
        \Statex \textbf{solve superposed problem \& increment kernel} (backward in time)
        \For{time step $n = N-1, ..., 1$}
            \State $u^{s^{n-1}}\leftarrow$ \texttt{propagateStep}$(u^{s^{n+1}}, u^{s^{n}}, f^{\dagger^{n}}+f^{n}, \gamma)$ \Comment(cf. \Cref{eq:FDstencil1,eq:FDstencil2} including BCs)
            \State $\tilde{K}_\gamma\leftarrow \tilde{K}_\gamma+$\texttt{incrementKernel}$(u^{s^{n-1}})$ \Comment{cf. \Cref{eq:kernel1,eq:kernel2,eq:superposedkernel}($K_\gamma(u^s,u^s)$)}
            \State delete $u^{s^{n+1}}$
        \EndFor
       \State \Return $C$, $\tilde{K}_\gamma$ 
    \end{algorithmic}
\end{algorithm*}

Given the current indicator $\gamma$, the forward problem is solved (line 4), yielding $u$ and $f^\dagger$ (line 5). 
Simultaneously, the cost is incremented (line 6), and the contribution of $K_\gamma(u,u)$ is computed and added to $\tilde{K}_\gamma$ (line 7). 
After scaling the adjoint forces, according to \Cref{eq:superposedforce}, the superposed problem (\Cref{eq:superposedPDE} or \Cref{eq:superposedPDE2}) is solved for $u^s$ (line 13). 
During the time integration, the contribution of $K_\gamma(u^s,u^s)$  is computed and added to $\tilde{K}_\gamma$ (line 14). 
After all time steps, the approximated mixed kernel $\tilde{K}_\gamma$ from \Cref{eq:superposedkernel} is retrieved, which approximates the sensitivity from \Cref{eq:sensitivitygamma}. 
Importantly, note, that after each incremented time step correspondingly $u^{n-1}$ (line 8) and $u^{s^{n+1}}$ (line 15) are deleted, meaning that at most four solution vectors are needed ($u^{n-1}/u^{s^{n+1}}, u^{n}/u^{s^{n}}, u^{n+1}/u^{s^{n+1}}, \tilde{K}_\gamma$), which all have the size of the number of degrees of freedom. \\

\subsection{Approximation Quality}

\begin{figure}[htbp]
    \centering
    \begin{subfigure}[t]{0.175\textwidth}
        \includegraphics[width=\textwidth]{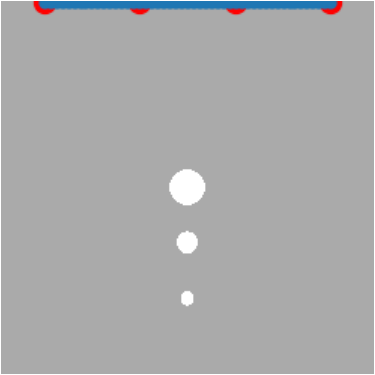}
        \caption{$\gamma(\boldsymbol{x})$}\label{fig:2Dcases1}
    \end{subfigure}
    \begin{subfigure}[t]{0.175\textwidth}
        \includegraphics[width=\textwidth]{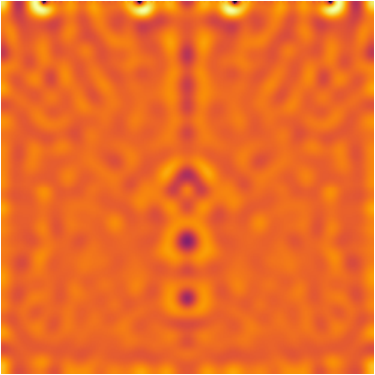}
        \caption{$\nabla C(\boldsymbol{x})$}\label{fig:2Dcases2}
    \end{subfigure}
    \begin{subfigure}[t]{0.07\textwidth}
        \raisebox{-0.11cm}{\includegraphics[width=\textwidth]{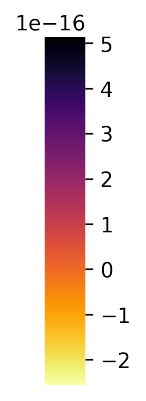}}
    \end{subfigure}
    \begin{subfigure}[t]{0.175\textwidth}
        \includegraphics[width=\textwidth]{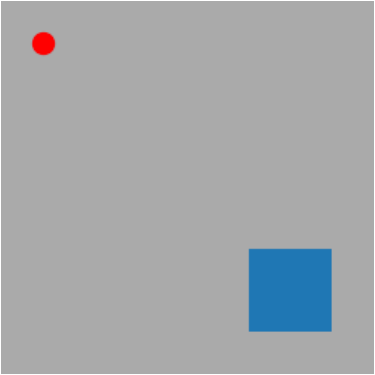}
        \caption{$\gamma(\boldsymbol{x})$}\label{fig:2Dcases3}
    \end{subfigure}
    \begin{subfigure}[t]{0.175\textwidth}
    \includegraphics[width=\textwidth]{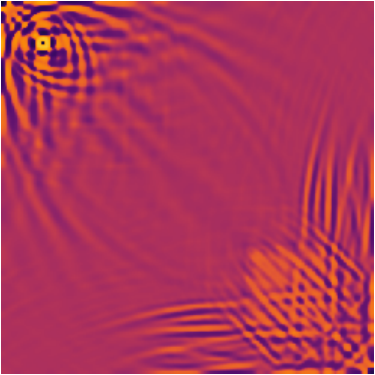}
    \caption{$\nabla C(\boldsymbol{x})$}\label{fig:2Dcases4}
    \end{subfigure}
    \begin{subfigure}[t]{0.07\textwidth}
        \raisebox{-0.11cm}{\includegraphics[width=\textwidth]{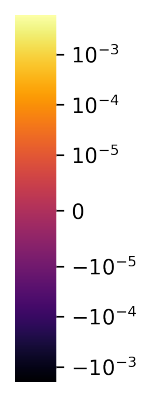}}
    \end{subfigure}
    \caption{Reference gradients for FWI (left) and TATO (right). During the gradient computation homogeneous material distributions of correspondingly $\gamma(\boldsymbol{x})=1$ and $\gamma(\boldsymbol{x})=0$.}\label{fig:2Dcases}
\end{figure}

The quality of the approximation depends on the choice of the parameter $k$. 
However, it is easy to show that the range of permissible values of $k$ is relatively large for FWI and TATO. 
To demonstrate this, we consider the FWI example illustrated in \Cref{fig:2Dcases1}, and the TATO problem illustrated in \Cref{fig:2Dcases3}. 
For details, see \Cref{tab:setups2D}\footnote{For the FWI example, the sensor measurements with the true material field are generated on a finite difference grid with twice as many points in each dimension and double the number of time steps. This is to avoid an inverse crime~\cite{wirgin_inverse_2004}.}. 
Assuming homogeneous material distributions of $\gamma(\boldsymbol{x})=1$ for FWI and $\gamma(\boldsymbol{x})=0$ for TATO, reference sensitivities were computed with double floating-point precision using the standard adjoint sensitivity method, described in \Cref{ssec:fwisensitivity,ssec:topoptsensitivity}.
The sensitivity fields are depicted in \Cref{fig:2Dcases2} and \Cref{fig:2Dcases4}. \\

\begin{table}[htbp]
    \centering
    \caption{Problem parameters of the two-dimensional FWI and TATO defined in \Cref{fig:2Dcases}, including the domain size $L_1\times L_2$, grid points $n_1\times n_2$, number of time steps $N$, time step size $\Delta t$, and source parameters $f, n_c, \psi_0$ defined according to \Cref{eq:source1,eq:source2}. For the detailed sensor and source arrangement, see~\cite{herrmann_memory_2025}.}\label{tab:setups2D}
    \begin{tabular}{cccccccccc}
    \hline
    $L_1$ & $L_2$ & $n_1$ & $n_2$ & $N$ & $\Delta t$ & $f$ & $n_c$ & $\psi_0$ \\
    \hline
    \hline
    \multicolumn{9}{l}{\textbf{FWI} (\Cref{fig:2Dcases1})} \\
    \hline
    0.02 m & 0.02 m & 251 & 251 & 3\,200 & $7.5\cdot 10^{-9}$ s & $10^6$ Hz & 2 & $10^{12}$ $\text{N}/\text{m}^2$\\
    \hline
    \hline
    \multicolumn{9}{l}{\textbf{TATO} (\Cref{fig:2Dcases3})} \\ 
    \hline
    9 m & 9 m & 363 & 363 & 1\,800 & $3.4\cdot 10^{-2}$ s & 650 Hz & 2 & $10^2 \text{ s}^{-2}$ \\
    \hline
    \end{tabular}
\end{table}

\begin{figure}[htbp]
    \centering
    \begin{subfigure}[t]{0.49\textwidth}
        \includegraphics[width=\textwidth]{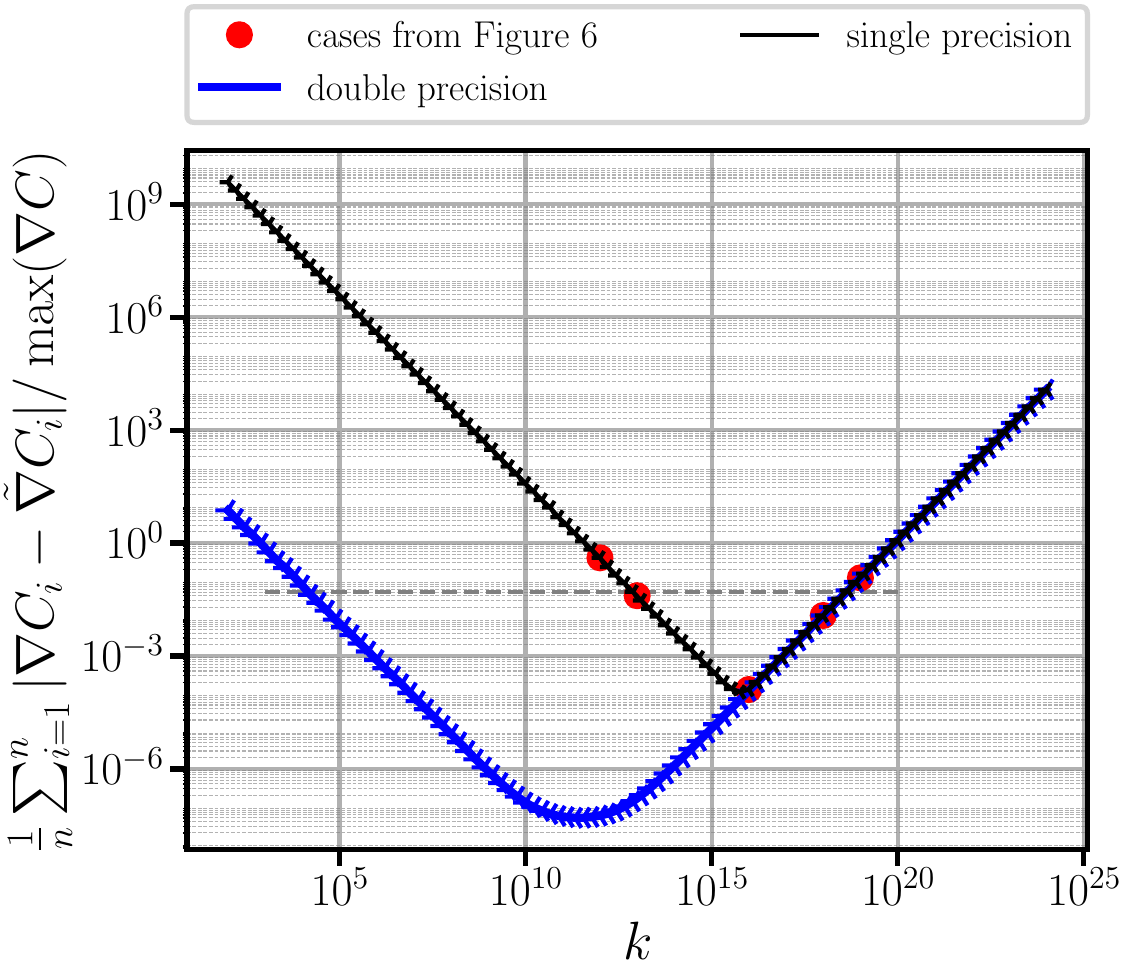}
        \caption{FWI}\label{fig:mseversusk1}
    \end{subfigure}
    \hfill
    \begin{subfigure}[t]{0.49\textwidth}
        \includegraphics[width=\textwidth]{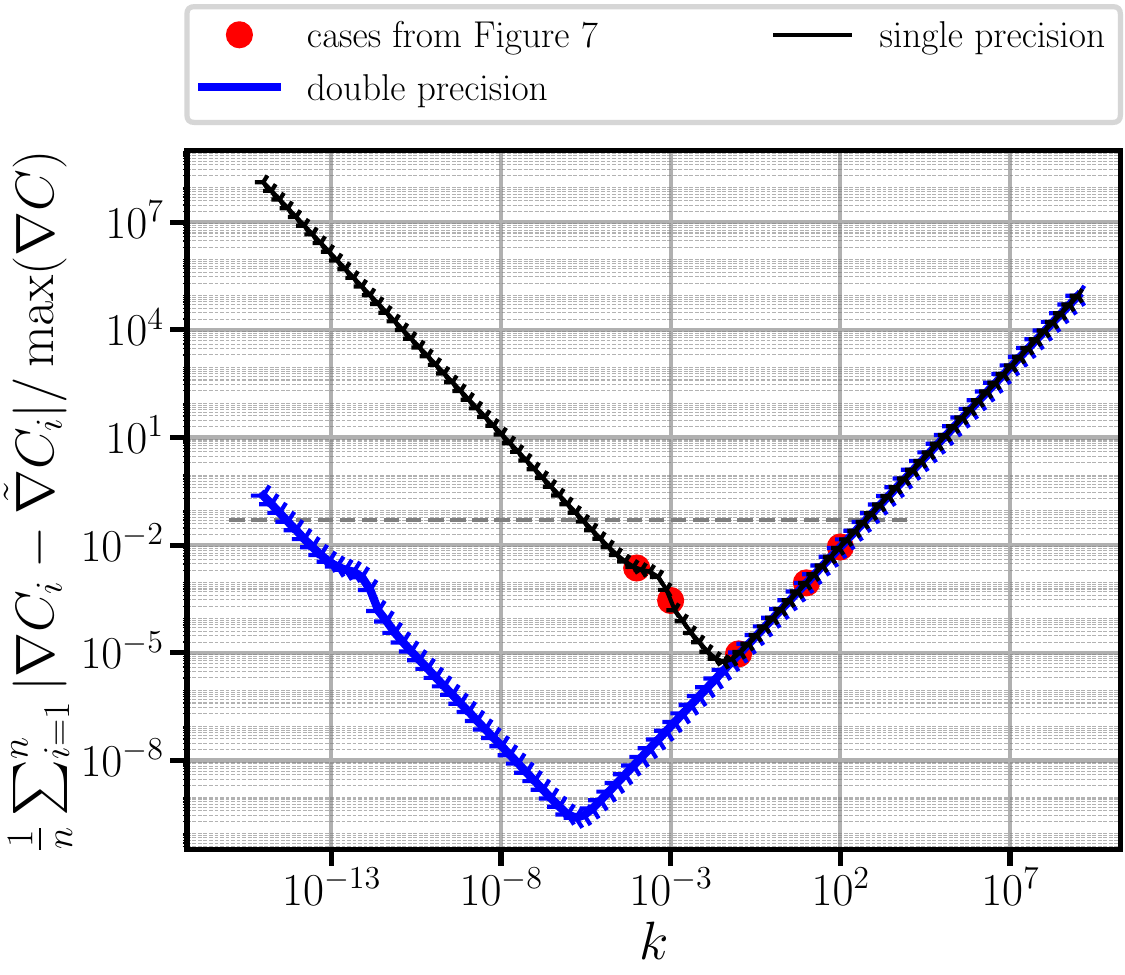}
        \caption{TATO}\label{fig:mseversusk2}
    \end{subfigure}
    \caption{Relative error magnitude versus the parameter $k$ for floating and double precision. Both graphs span 22 orders of magnitude of $k$.}\label{fig:mseversusk}
\end{figure}

Performing the gradient computation with the memory-efficient adjoint sensitivity approximation, summarized by \Cref{alg:memoryTrick}, for multiple $k$ values and using the reference gradients from \Cref{fig:2Dcases}, the relative error with respect to $k$ is illustrated in \Cref{fig:mseversusk}. 
The relative absolute mean squared error is shown for both single floating-point precision (black) and double floating-point precision (blue). 
The dashed line indicates a $5\%$ threshold, leading to a range of about seven orders of magnitude, resulting in a relative error of less than $5\%$. 
Double precision provides an even greater range, as the errors for smaller $k$ values are slower to rise, due to smaller overflow/underflow problems. 
The increasing error for larger $k$ values is caused by neglecting the term $k^2 \, K_\gamma(u^\dagger,u^\dagger)$ in \Cref{eq:superposedkernel}. 

\begin{figure}[htbp]
    \centering
    \begin{subfigure}[t]{0.175\textwidth}
        \includegraphics[width=\textwidth]{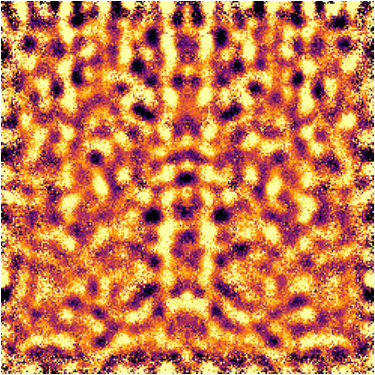}
        \caption{$k=10^{12}$}
    \end{subfigure}
    \hfill
    \begin{subfigure}[t]{0.175\textwidth}
        \includegraphics[width=\textwidth]{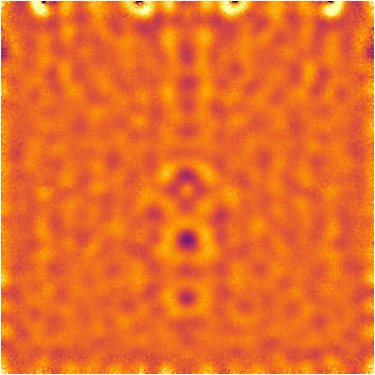}
        \caption{$k=10^{13}$}
    \end{subfigure}
    \hfill
    \begin{subfigure}[t]{0.175\textwidth}
        \includegraphics[width=\textwidth]{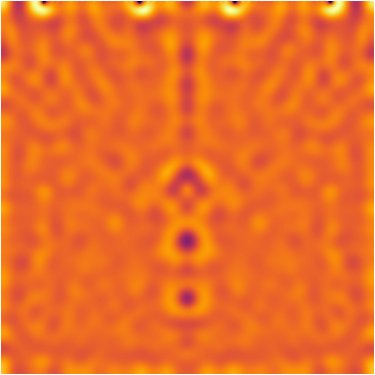}
        \caption{$k=10^{16}$}
    \end{subfigure}
    \hfill 
    \begin{subfigure}[t]{0.175\textwidth}
        \includegraphics[width=\textwidth]{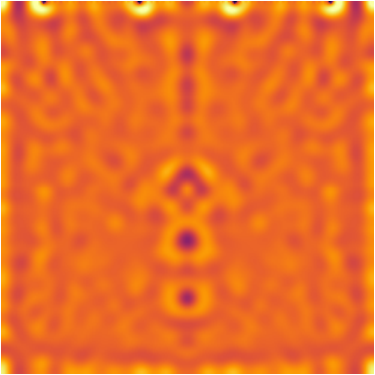}
        \caption{$k=10^{18}$}
    \end{subfigure}
    \hfill 
    \begin{subfigure}[t]{0.175\textwidth}
        \includegraphics[width=\textwidth]{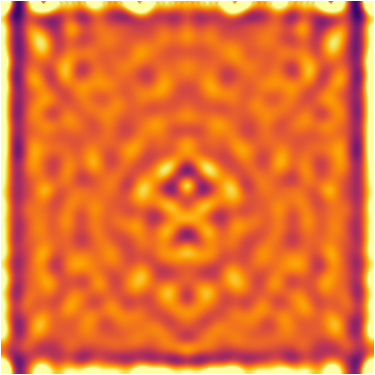}
        \caption{$k=10^{19}$}
    \end{subfigure}
    \hfill 
    \begin{subfigure}[t]{0.07\textwidth}
        \raisebox{-0.11cm}{\includegraphics[width=\textwidth]{figures/kfactorColorbar.pdf}}
    \end{subfigure}
    \caption{FWI: approximate sensitivity with different $k$ values (reference sensitivity given in \Cref{fig:2Dcases2})}
    \label{fig:FWI2D}
\end{figure}

\begin{figure}[htbp]
    \centering
    \begin{subfigure}[t]{0.175\textwidth}
        \includegraphics[width=\textwidth]{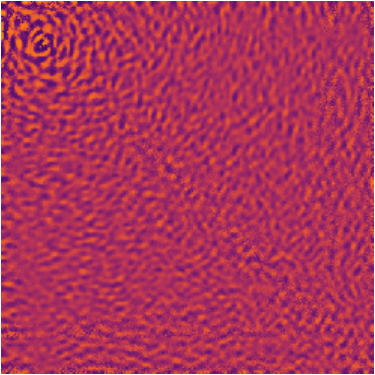}
        \caption{$\tilde{K}(u,u^\dagger)$ with $k=10^{-4}$}
    \end{subfigure}
    \hfill
    \begin{subfigure}[t]{0.175\textwidth}
        \includegraphics[width=\textwidth]{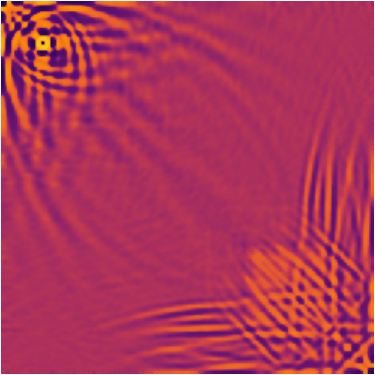}
        \caption{$\tilde{K}(u,u^\dagger)$ with $k=10^{-3}$}
    \end{subfigure}
    \hfill
    \begin{subfigure}[t]{0.175\textwidth}
        \includegraphics[width=\textwidth]{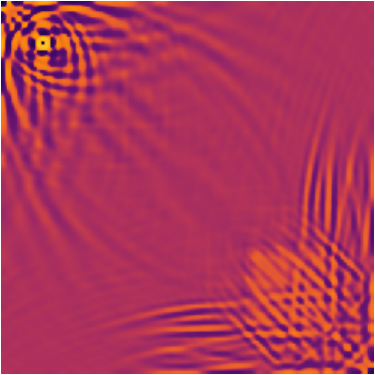}
        \caption{$\tilde{K}(u,u^\dagger)$ with $k=10^{-1}$}
    \end{subfigure}
    \hfill 
    \begin{subfigure}[t]{0.175\textwidth}
        \includegraphics[width=\textwidth]{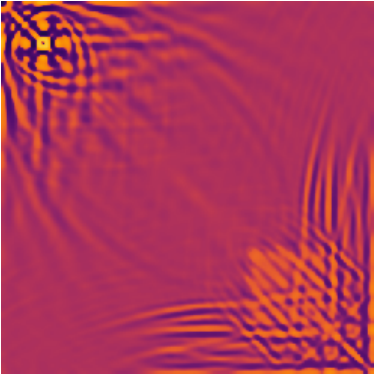}
        \caption{$K(u,u^\dagger)$ with $k=10^1$}
    \end{subfigure}
    \hfill
    \begin{subfigure}[t]{0.175\textwidth}
        \includegraphics[width=\textwidth]{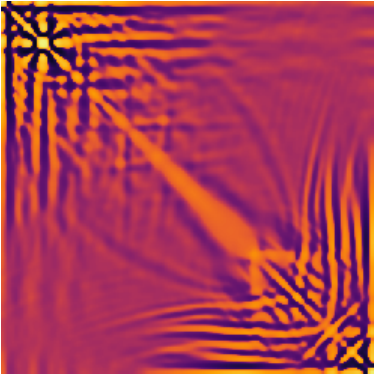}
        \caption{$K(u,u^\dagger)$ with $k=10^2$}
    \end{subfigure}
    \hfill 
    \begin{subfigure}[t]{0.07\textwidth}
        \raisebox{-0.11cm}{\includegraphics[width=\textwidth]{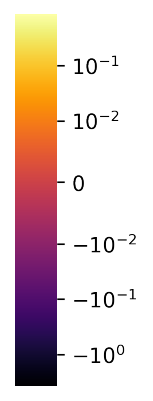}}
    \end{subfigure}
    \caption{TATO: approximate sensitivity with different $k$ values (reference sensitivity given in \Cref{fig:2Dcases4})}
    \label{fig:topopt2Dsmall}
\end{figure}

The approximate sensitivities for five different $k$ values --- ($10^{12}, 10^{13}, 10^{16}, 10^{18}, 10^{19}$) for FWI and ($10^{-4}, 10^{-3}, 10^{-1}, 10^{1}, 10^{2}$) for TATO --- are illustrated in \Cref{fig:FWI2D,fig:topopt2Dsmall}. 
The qualitative comparison to the reference gradients from \Cref{fig:2Dcases2,fig:2Dcases4} highlights the effect of the overflow/underflow error for small $k$ and the approximation error for large $k$ values\footnote{Similar error curves, known as hockey stick plots, can be observed when the sensitivity is computed using finite differences; see~\cite{Fichtner2021}. The head of the hockey stick corresponds to floating-point errors, while the handle is caused by increasing finite difference approximation errors.}. 
Yet, still, the robustness of the approximation over a large range of $k$ values is emphasized. 
Finding an appropriate $k$ value can be achieved during the first sensitivity computation of the optimization, in which large $k$ values are chosen and decreased until overflow/underflow errors become apparent. 
An overflow/underflow error can unambiguously be identified by comparing the single and double precision computations, which would lead to different results (unlike when the approximation errors for large $k$ values arise). 
The smallest possible $k$ value before overflow/underflow is an appropriate choice. 
Note that this calibration does not require the true gradient as ground truth. \\

\begin{figure}[htbp]
    \centering
    \begin{subfigure}[t]{0.49\textwidth}
        \includegraphics[width=\textwidth]{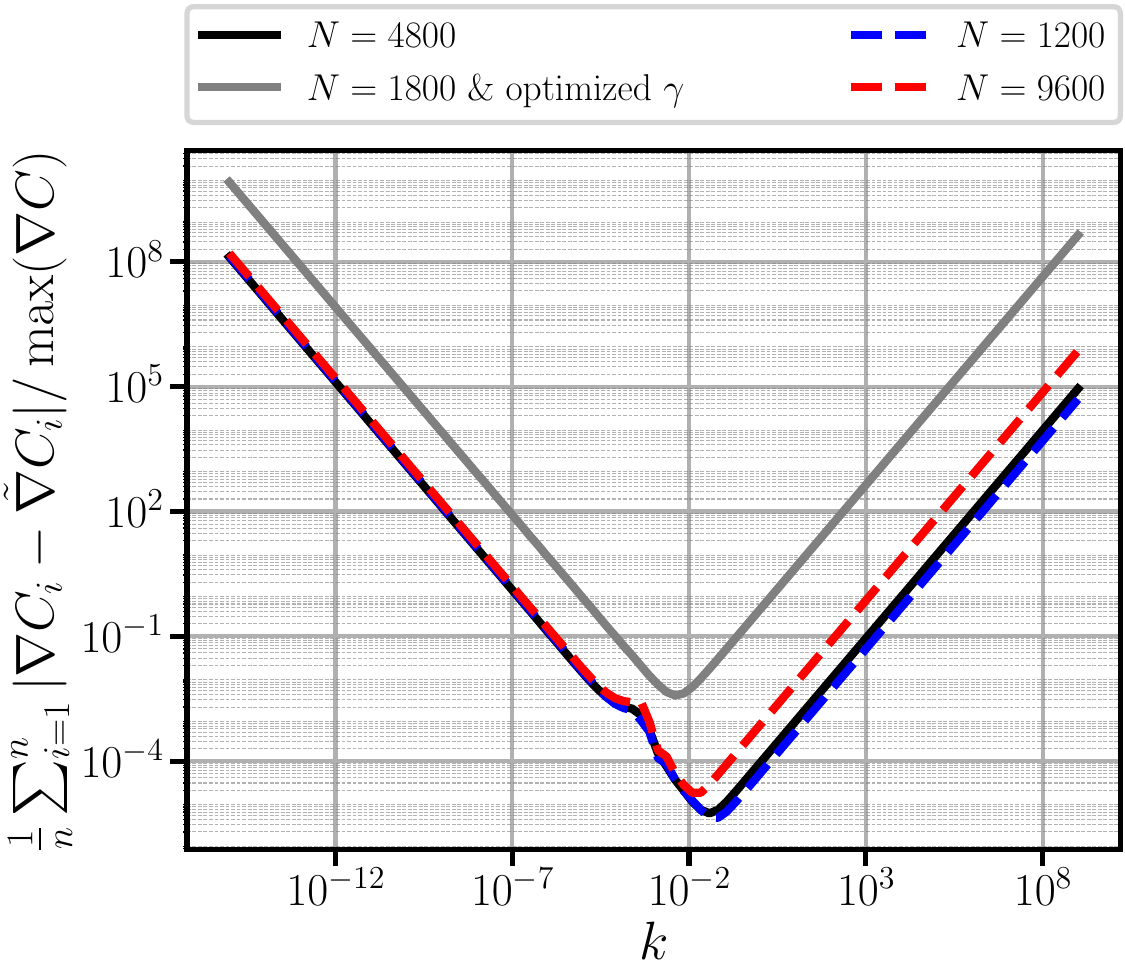}
        \caption{Effects on approximation quality}\label{fig:effectsonapproximation}
    \end{subfigure}
    \hfill
    \begin{subfigure}[t]{0.42\textwidth}
        \includegraphics[width=\textwidth]{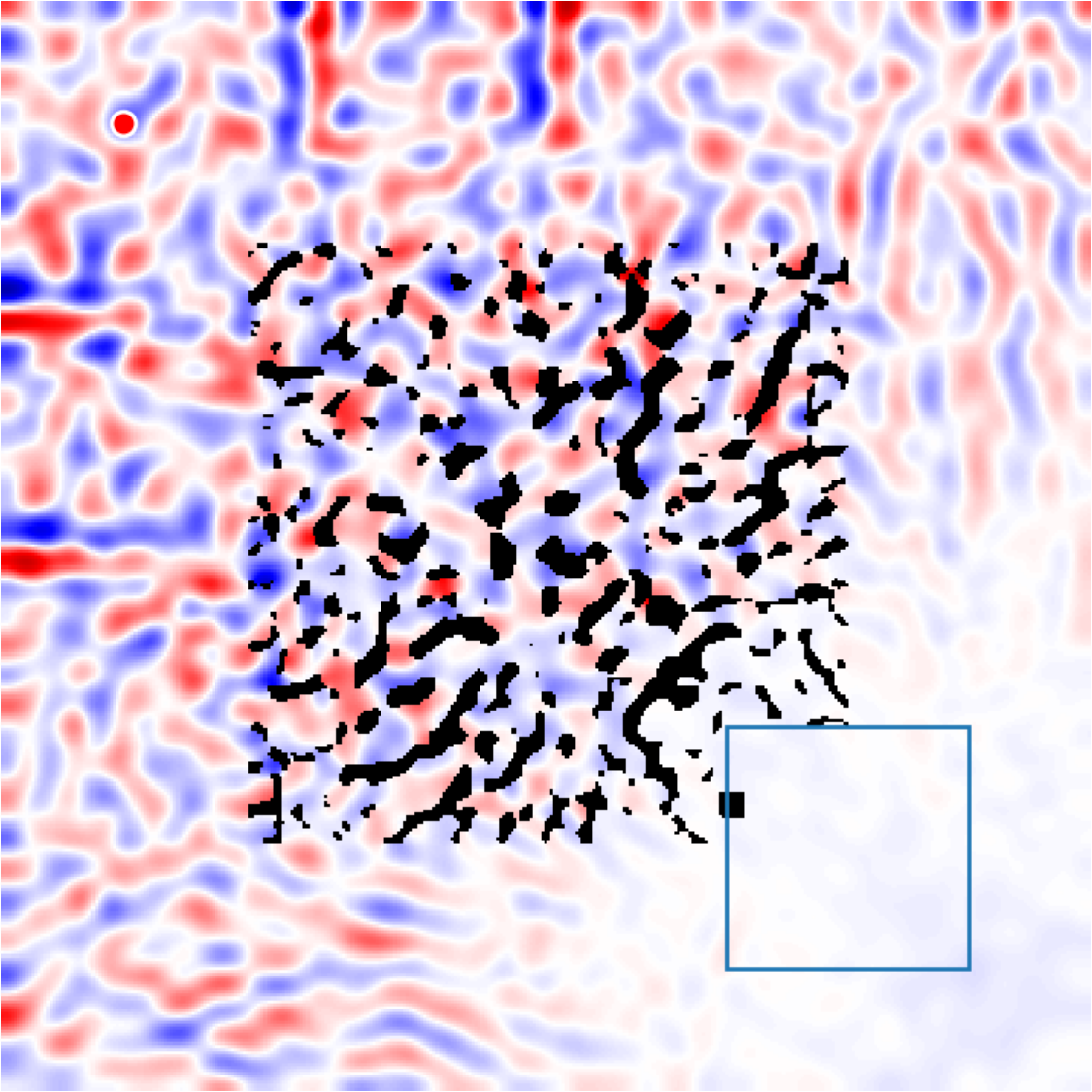}
        \caption{Optimized structure given by \Cref{fig:2Dcases3} with the corresponding wave field at $t=4.42\cdot 10^{-2}$ s}\label{fig:optimized2DTopOpt}
    \end{subfigure}
    \caption{Number of time steps and indicator function can influence quality of the approximation negatively. Results were all computed with single floating-point precision.}
\end{figure}

Furthermore, we have empirically observed that the $k$ value selected for the first optimization iteration is equally applicable to later stages of the optimization procedure. 
To illustrate this, we consider an optimized structure (\Cref{fig:optimized2DTopOpt}) for the TATO problem defined by \Cref{fig:2Dcases3}. Computing the gradient at the optimum and comparing it with a reference gradient using the standard adjoint sensitivity method yields a slightly worse approximation, as seen by \Cref{fig:effectsonapproximation} (gray line). Nevertheless, the range of appropriate $k$ is only slightly reduced and not shifted drastically, thus retaining the validity of a $k$ tuned at the first iteration.\\

Lastly, the number of time steps also influences the quality of the approximation. 
This is illustrated for $N=1\,200, N=4\,800$ and $N=9\,600$ time steps in \Cref{fig:effectsonapproximation}. 
The effect is, however, limited. 
More importantly, for the problem at hand, $N=9\,600$ time steps was the maximum of time steps possible with the standard adjoint sensitivity method, as the memory requirement scales with the number of time steps. 
By contrast, the proposed memory-efficient method's memory usage is independent of the number of time steps. Thus, even if the quality would deteriorate for very large numbers of time steps ($N\gg 10\,000$), the standard adjoint sensitivity computation would not even be applicable.

\subsection{Computation Time}

\begin{figure}[htbp]
    \centering
    \begin{subfigure}[t]{0.49\textwidth}
        \includegraphics[width=\textwidth]{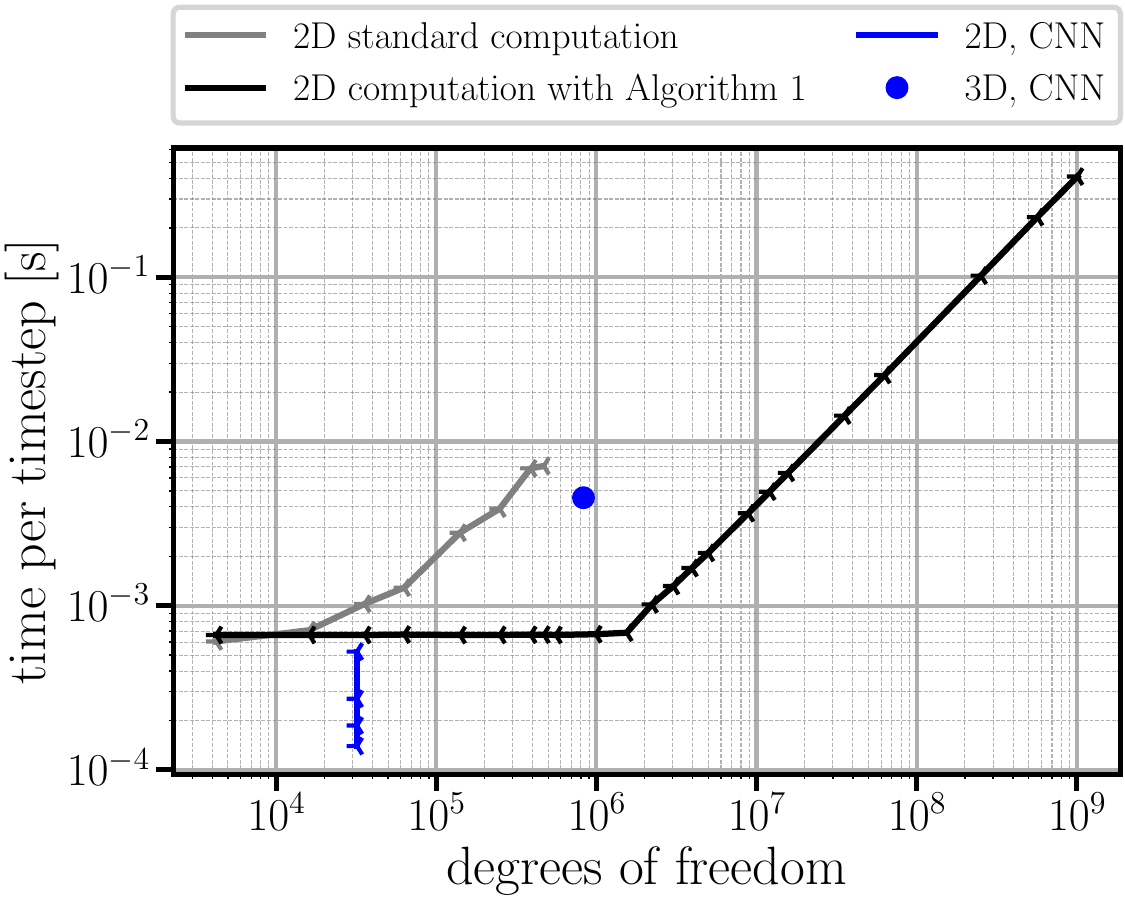}
    \end{subfigure}
    \caption{Computation time of the gradient computation (including one forward and one adjoint solve) versus number of degrees of freedom in one, two, and three dimensions. The timings rely on the CUDA$^{\text{\textregistered}}$ implementation and are compared to the timings from~\cite{herrmann_use_2023}, which relied on the CNN implementation. Here, the scalar wave equation from \Cref{eq:scalarwaveequation} and the sensitivity analysis for FWI from \Cref{ssec:fwisensitivity} are considered.}\label{fig:computationtime2}
\end{figure}

We now consider the computation time, as previously for the forward simulation in \Cref{fig:computationtime1}. A single sensitivity computation is composed of one forward and one adjoint solve. \Cref{fig:computationtime2} illustrates how the standard gradient computation (gray) described in \Cref{ssec:fwisensitivity} is limited in terms of memory requirements and thereby barely reaches $10^6$ degrees of freedom\footnote{The exact limit depends on the number of time steps, which in this study was 2\,000.}. 
This is alleviated by the memory-efficient adjoint sensitivity computation, that is summarized by \Cref{alg:memoryTrick}. 
Here, the gradient for up to $10^9$ degrees of freedom can be computed. 
The theoretical limit without implementational overheads is the memory footprint of four times the number of degrees of freedom, which with the $40$ GB of memory yields $40$ GB $/ 4 / 4$ B $=2.5\cdot 10^9$ parameters for single floating-point precision. 
It is four times, due to the three solution grids for the finite difference scheme and one grid for the sensitivity analysis, as laid out in \Cref{sec:MEM}. \\

Also, in terms of computation time, the memory-efficient scheme outperforms the standard computation scheme, which is most likely connected to more efficient memory access patterns linked to the lower memory usage. Furthermore, take note that the gradient computations are about one order of magnitude more expensive than the corresponding forward simulations in \Cref{fig:computationtime1}, despite consisting only of one forward computation and one adjoint computation. 
This is also linked to less efficient memory access patterns, which might be improved through more sophisticated implementations. \\

In \Cref{fig:computationtime2}, the memory-efficient scheme (black) is compared to the CNN implementation (blue). 
For low numbers of degrees of freedom, the CNN version is slightly better\footnote{Four computation times are provided in blue, as in~\cite{herrmann_use_2023}, four measurements are utilized in the FWI scheme. These four measurements can be simulated in parallel. This could essentially also be done for the CUDA$^{\text{\textregistered}}$ implementation, but would yield little benefit for larger problems. Thus, the slowest measurement of the CNN version should be considered when comparing it to the CUDA$^{\text{\textregistered}}$ implementation. The fastest timing is, however, the one provided in~\cite{herrmann_use_2023}.} than the memory-efficient scheme, which is probably linked to sophisticated implementations by PyTorch. 
For larger problems, however, at ${\sim}10^6$, the CNN scheme is one order of magnitude slower. 
This is also the limit in terms of the size of what is possible in PyTorch, due to the extensive memory overhead of PyTorch. Thus, also in the gradient computation, the CUDA$^{\text{\textregistered}}$ version (with the memory-efficient adjoint scheme) is superior. Nevertheless, the PyTorch CNN idea is beneficial for fast prototyping when working with smaller problems.

\section{Optimization Results}\label{sec:results}
To demonstrate the memory-efficient adjoint sensitivity in optimization scenarios, we consider an FWI example in three dimensions (\Cref{ssec:fwiresults}) and a TATO example in two dimensions (\Cref{ssec:topopt}).

\subsection{Full Waveform Inversion}\label{ssec:fwiresults}

\begin{figure}[htbp]
    \centering
    \includegraphics[width=0.4\textwidth]{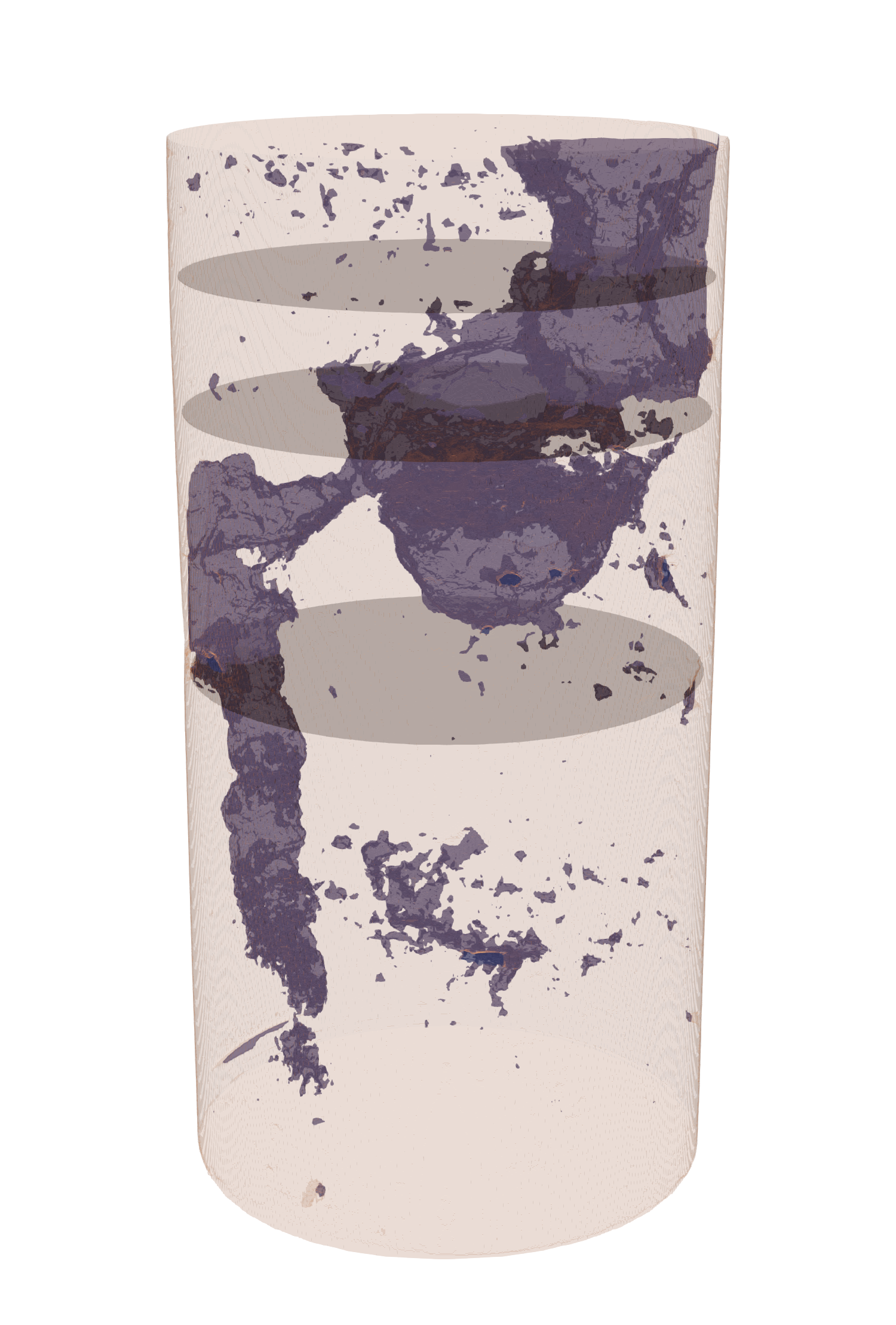}
    \caption{Cylindrical probe with internal voids, obtained with a CT-scan from~\cite{hug_three-field_2022}. The structure is divided into an eighth, a quarter, and a half substructure, as the circular slices indicate.}\label{fig:CTscan}
\end{figure}

We investigate the three-dimensional structure of a cylindrical probe~\cite{hug_three-field_2022}.
The task is to recover the internal voids seen in \Cref{fig:CTscan}. 
The geometry was obtained using computed tomography (CT). 
In the following, we aim to reconstruct the CT scan using FWI (which, in practice, would be a much cheaper alternative to generate a volumetric image of the sample). 
The full structure is characterized by the parameters in \Cref{tab:problemsetupFWI}. 
As the inversion with FWI is a numerically very challenging task, subparts of the structure are considered during the inversion. 
The upper eighth, quarter, half, and full structures are considered. 
The smaller substructures with $n_3=124, n_3=246$ and $n_3=490$ are simulated with correspondingly fewer time steps: $N=1\,300$, $N=2\,400$, and $N=3\,800$. 
To improve the direct comparison between the inversion results of the four structures, an additional void is added at the center, namely a hole in cubical shape composed of $16 \times 16\times 16$ voxels. \\

\begin{table}[htbp]
    \centering
    \caption{Problem parameters of the three-dimensional FWI problem based on the CT-scanned structure from \Cref{fig:CTscan} and measurement setup defined in \Cref{fig:BCCT1,fig:BCCT2}. This includes the domain size $\L_1\times L_2 \times L_3$, the grid points $n_1 \times n_2 \times n_3$, number of time steps $N$, time step size $\Delta t$, source parameters $f, n_c, \psi_0$, and learning rate $\alpha$ for Adam.}\label{tab:problemsetupFWI}
    \begin{tabular}{ccccccc}
    \multicolumn{6}{l}{\textbf{problem parameters}} \\
    \hline
    \hline
    $L_1$ & $L_2$ & $L_3$ & $f$ & $n_c$ & $\psi_0$ \\
    \hline
    0.05 m & 0.05 m & 0.0973 m & $2\cdot 10^6$ Hz & 2 & $10^{12}$ $\text{N}/\text{m}^2$\\
    \hline 
    \multicolumn{6}{l}{\textbf{discretization \& optimization parameters}} \\
    \hline 
    \hline
    $n_1$ & $n_2$ & $n_3$ & $N$ & $\Delta t$ & $\alpha$ \\
    \hline 
    503 & 503 & 976 & 7\,200 & $9\cdot 10^{-9}$ s & 0.2\\
    \hline
    \end{tabular}
\end{table}

\begin{figure}[htbp]
    \centering
    \begin{subfigure}[t]{0.32\textwidth}
        \includegraphics[width=\textwidth]{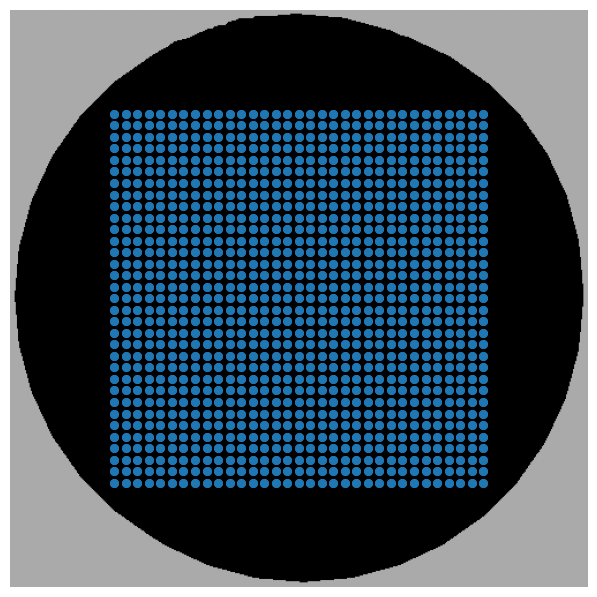}
        \caption{Sensor placement on top surface}\label{fig:BCCT1}
    \end{subfigure}
    \hspace{0.5cm}
    \begin{subfigure}[t]{0.32\textwidth}
        \includegraphics[width=\textwidth]{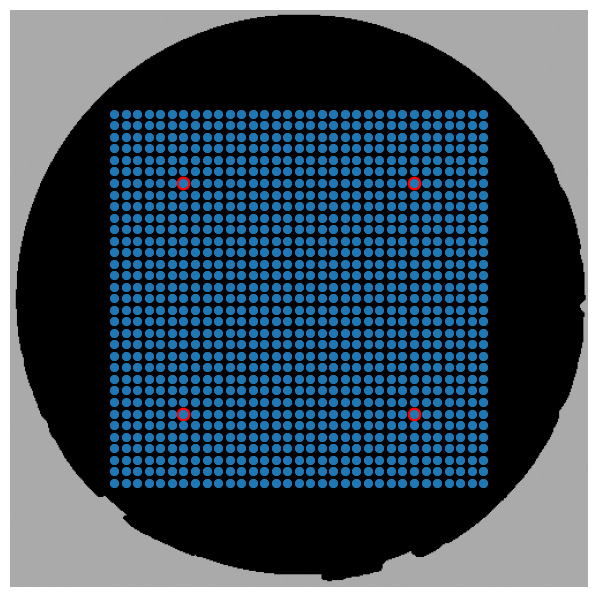}
        \caption{Source and sensor placement on bottom surface}\label{fig:BCCT2}
    \end{subfigure}
    \caption{Measurement setup. The two sensor $33\times 33$ grids on top and bottom surfaces are indicated in blue, while the $2\times 2$ sources (in red) are placed at the bottom.}\label{fig:BCCT}
\end{figure}

For the inversion, $33\times 33$ sensors are placed at the top and bottom surfaces of the cylinder, according to \Cref{fig:BCCT}. 
In total, four sources are applied at the bottom surface. 
The cylinder is simulated with the finite difference discretization described in \Cref{sec:forwardsolver}. 
The cylindrical shape is handled in an embedded sense~\cite{parvizian_finite_2007,duster__2017}, i.e., it is embedded in a larger rectangular cuboid, as illustrated in \Cref{fig:BCCT} for the top and bottom surfaces. 
The fictitious domain, indicated in gray, is handled by setting $\gamma=\epsilon$. Only the internal geometry shown in black is modified during the inversion procedure. \\

\begin{figure}
    \centering
    \begin{subfigure}[t]{0.29\textwidth}
        \includegraphics[width=\textwidth, trim={5cm 15cm 5cm 12cm}, clip]{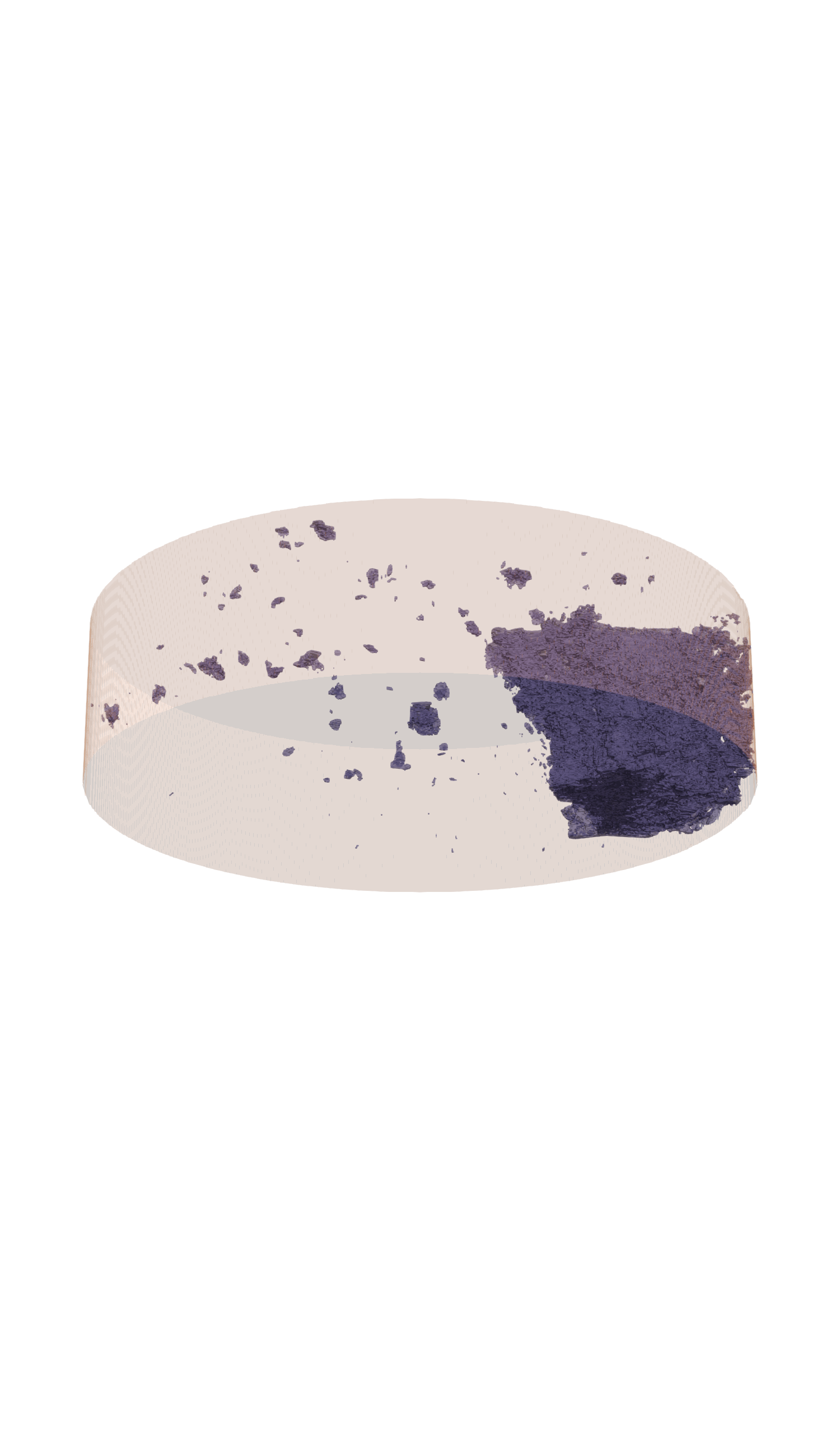}
        \caption{inversion ($1/8$)}\label{fig:FWIresults1}
    \end{subfigure}
    \hfill
    \begin{subfigure}[t]{0.29\textwidth}
        \includegraphics[width=\textwidth, trim={5cm 20cm 5cm 12cm}, clip]{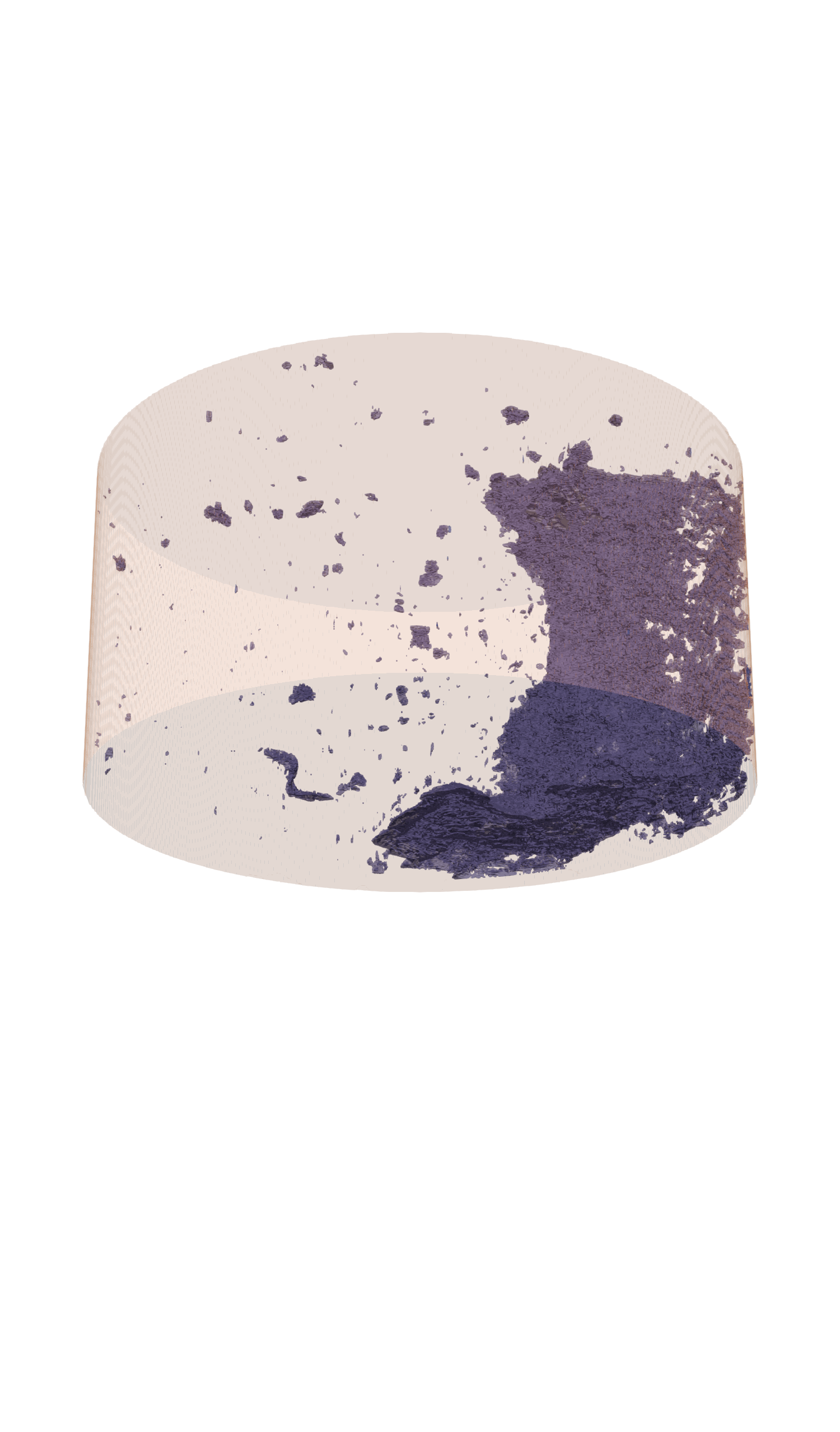}
        \caption{inversion ($1/4$)}\label{fig:FWIresults2}
    \end{subfigure}
    \hfill
    \begin{subfigure}[t]{0.29\textwidth}
        \includegraphics[width=\textwidth, trim={5cm 15cm 5cm 12cm}, clip]{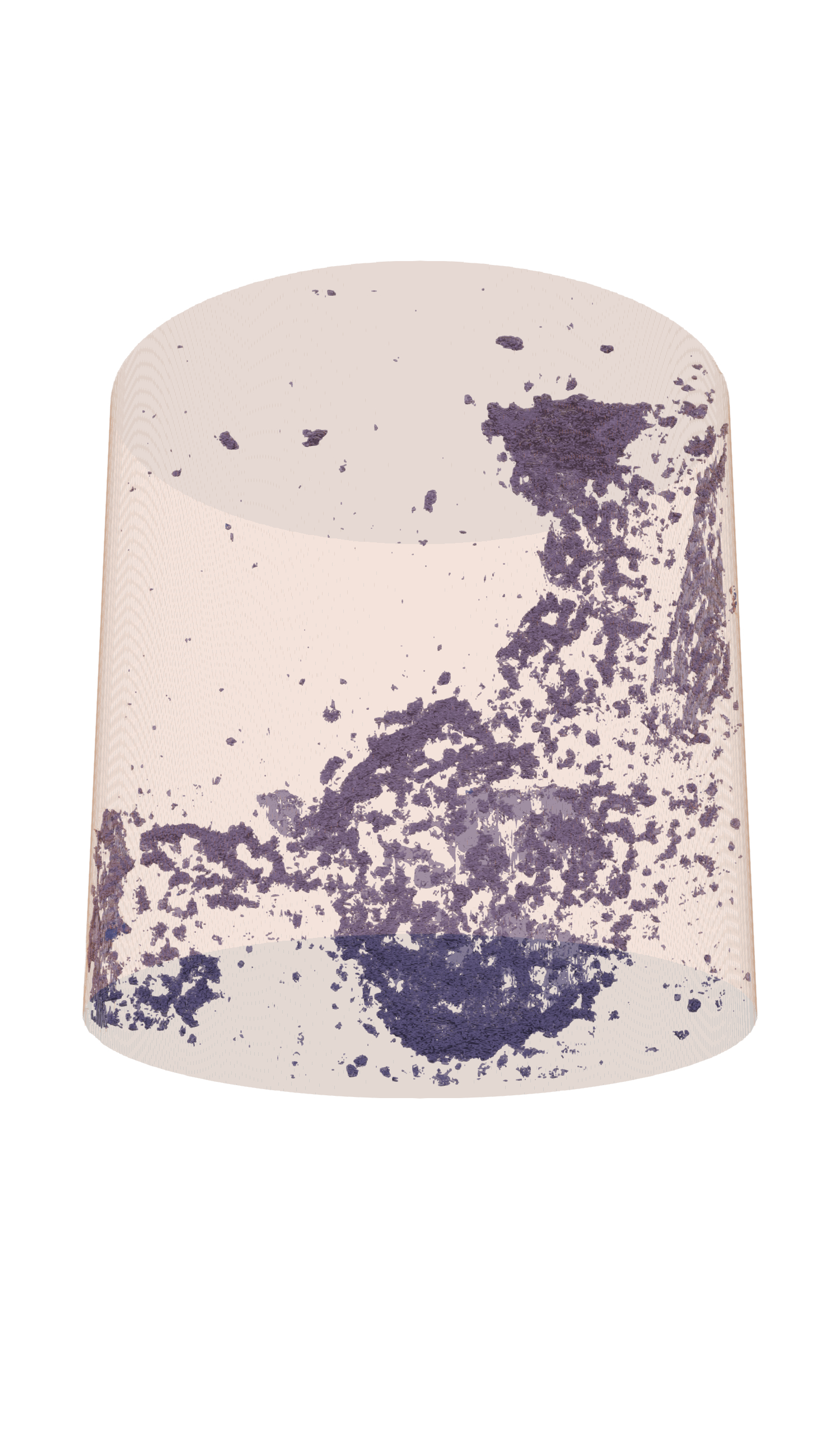}
        \caption{inversion ($1/2$)}\label{fig:FWIresults3}
    \end{subfigure}
    \\
    \begin{subfigure}[t]{0.29\textwidth}
        \includegraphics[width=\textwidth, trim={5cm 15cm 5cm 12cm}, clip]{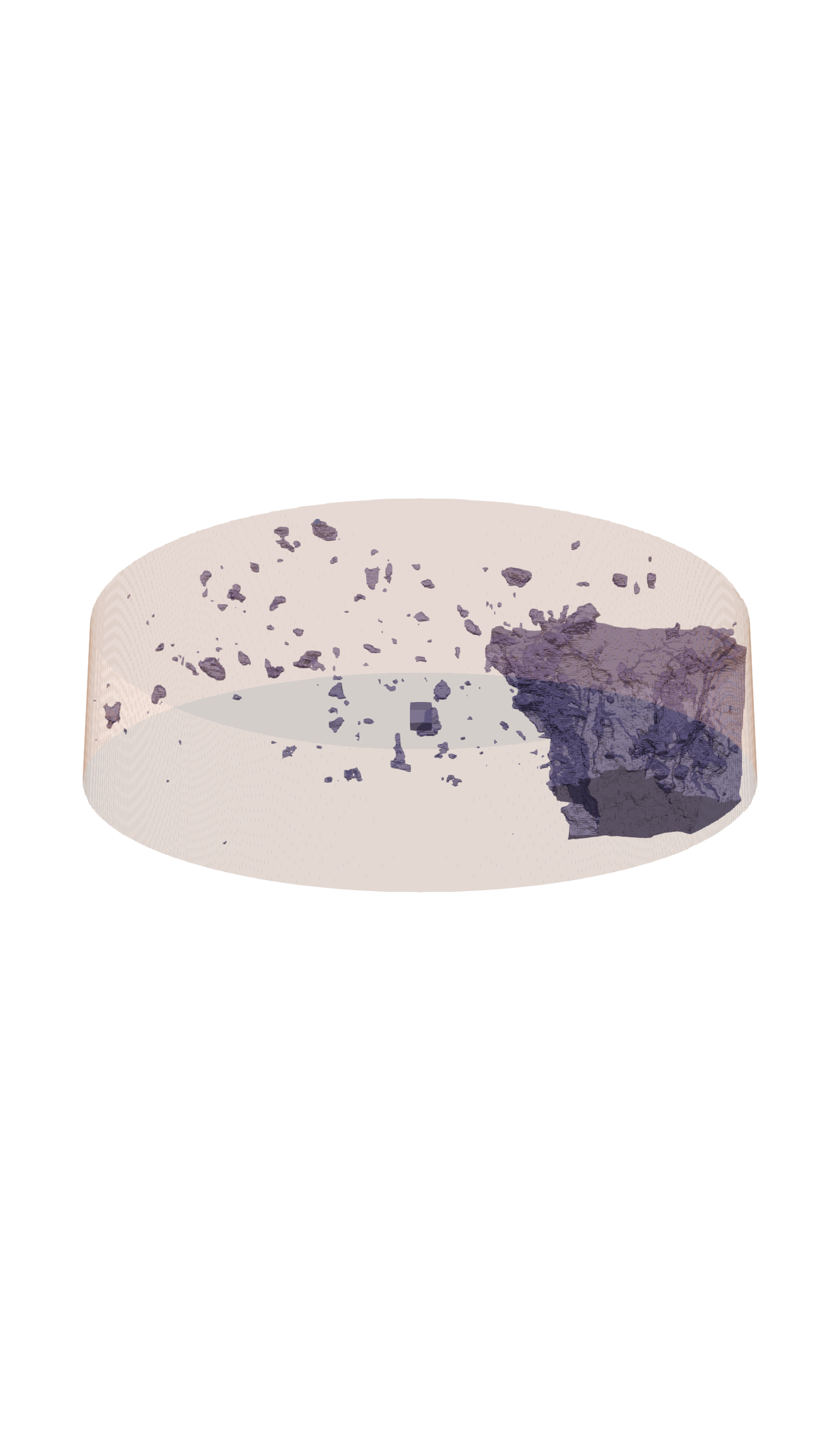}
        \caption{ground truth ($1/8$)}\label{fig:FWIresults4}
    \end{subfigure}
    \hfill
    \begin{subfigure}[t]{0.29\textwidth}
        \includegraphics[width=\textwidth, trim={5cm 20cm 5cm 12cm}, clip]{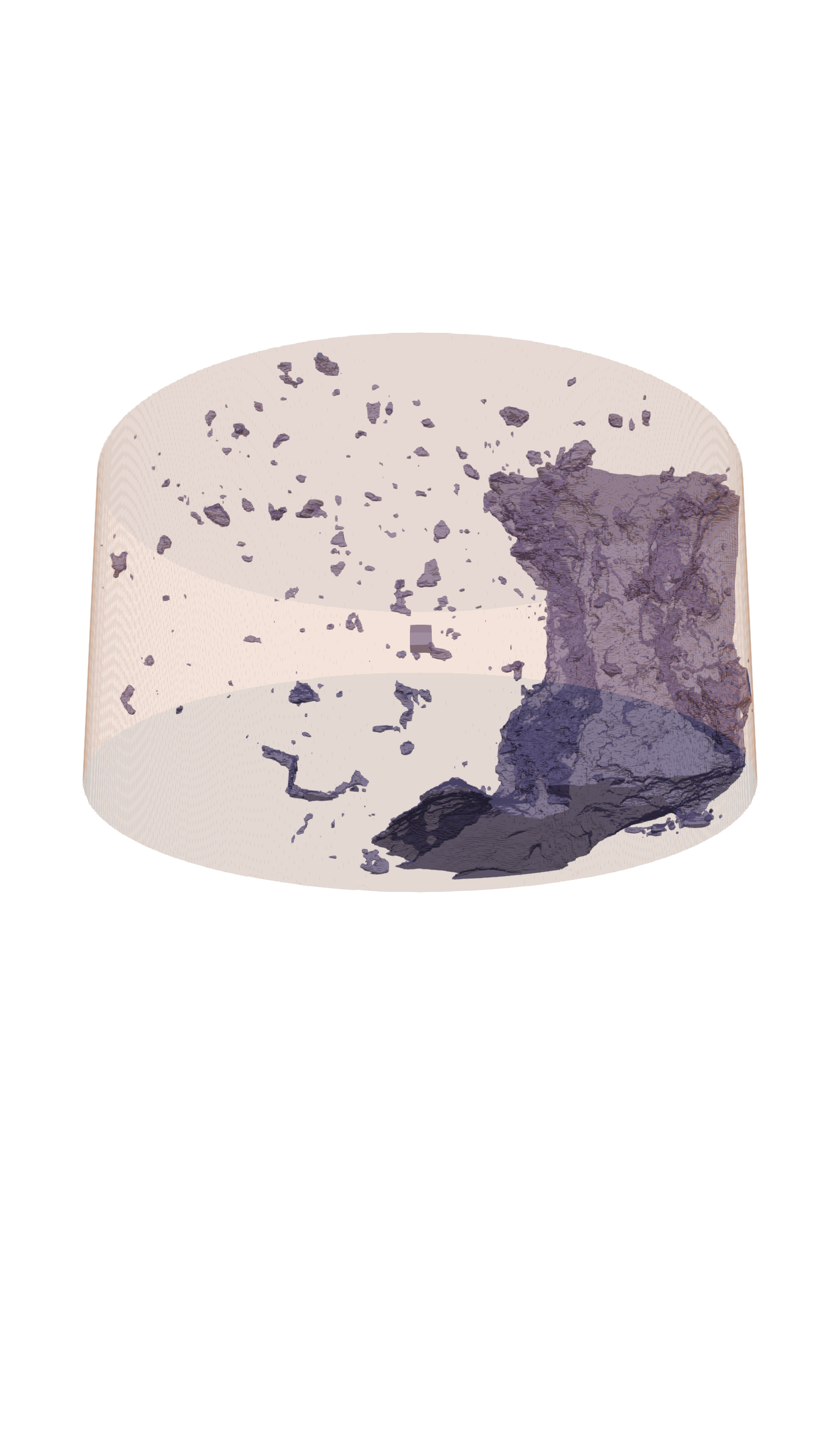}
        \caption{ground truth ($1/4$)}\label{fig:FWIresults5}
    \end{subfigure}
    \hfill
    \begin{subfigure}[t]{0.29\textwidth}
        \includegraphics[width=\textwidth, trim={5cm 15cm 5cm 12cm}, clip]{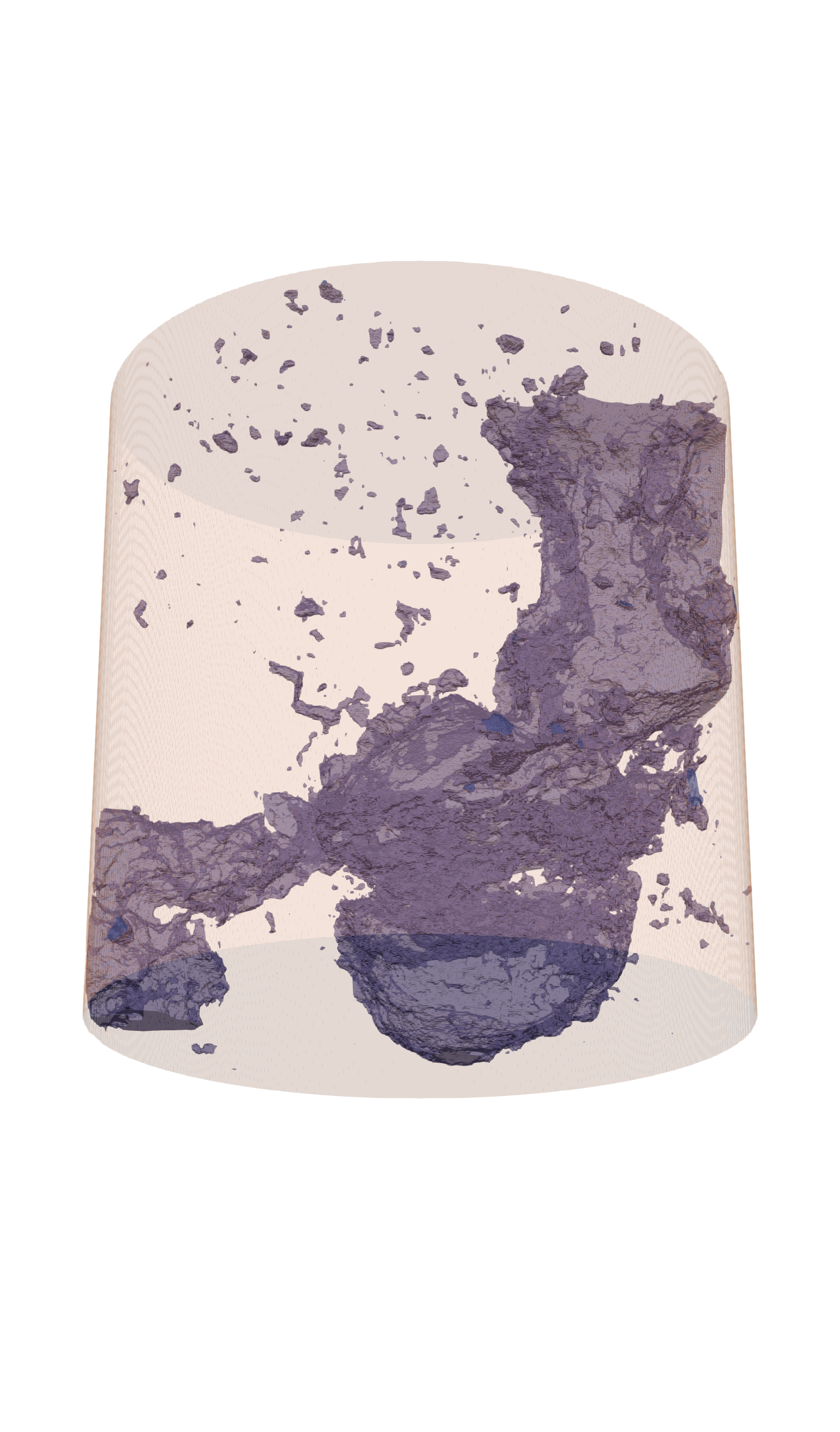}
        \caption{ground truth ($1/2$)}\label{fig:FWIresults6}
    \end{subfigure}
    \caption{Internal voids recovered with FWI (thresholded at $\gamma=0.1$) in comparison to the ground truth}\label{fig:FWIresults}
\end{figure}

\begin{figure}
    \centering
    \begin{subfigure}[t]{0.24\textwidth}
        \includegraphics[width=\textwidth]{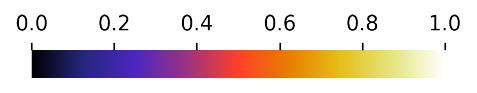}
    \end{subfigure}
    \hfill
    \begin{subfigure}[t]{0.72\textwidth}
    \end{subfigure}
    \\
    \begin{subfigure}[t]{0.24\textwidth}
        \includegraphics[width=\textwidth]{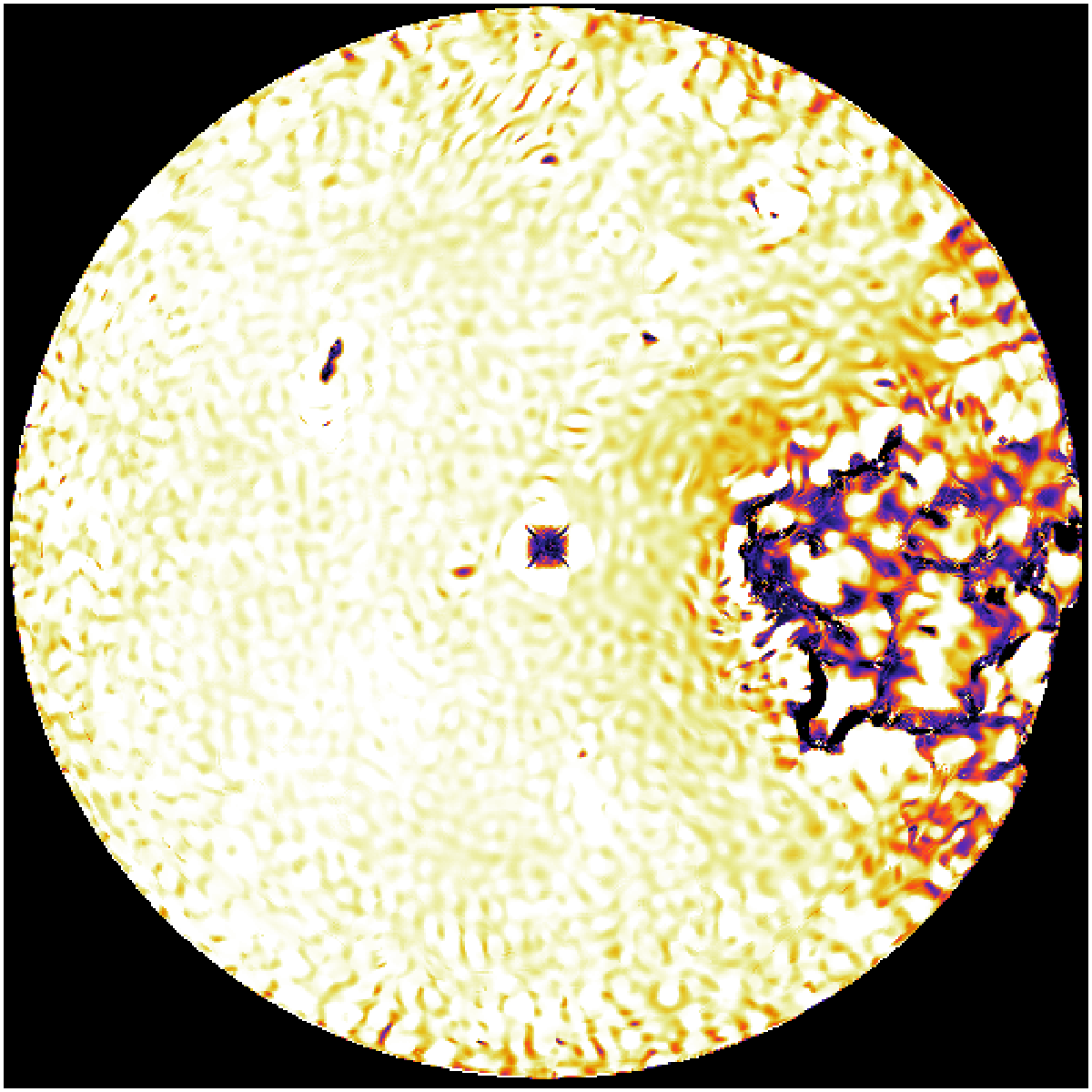}
        \caption{inversion ($1/8$)}\label{fig:slicesFWI3Da}
    \end{subfigure}
    \hfill
    \begin{subfigure}[t]{0.24\textwidth}
        \includegraphics[width=\textwidth]{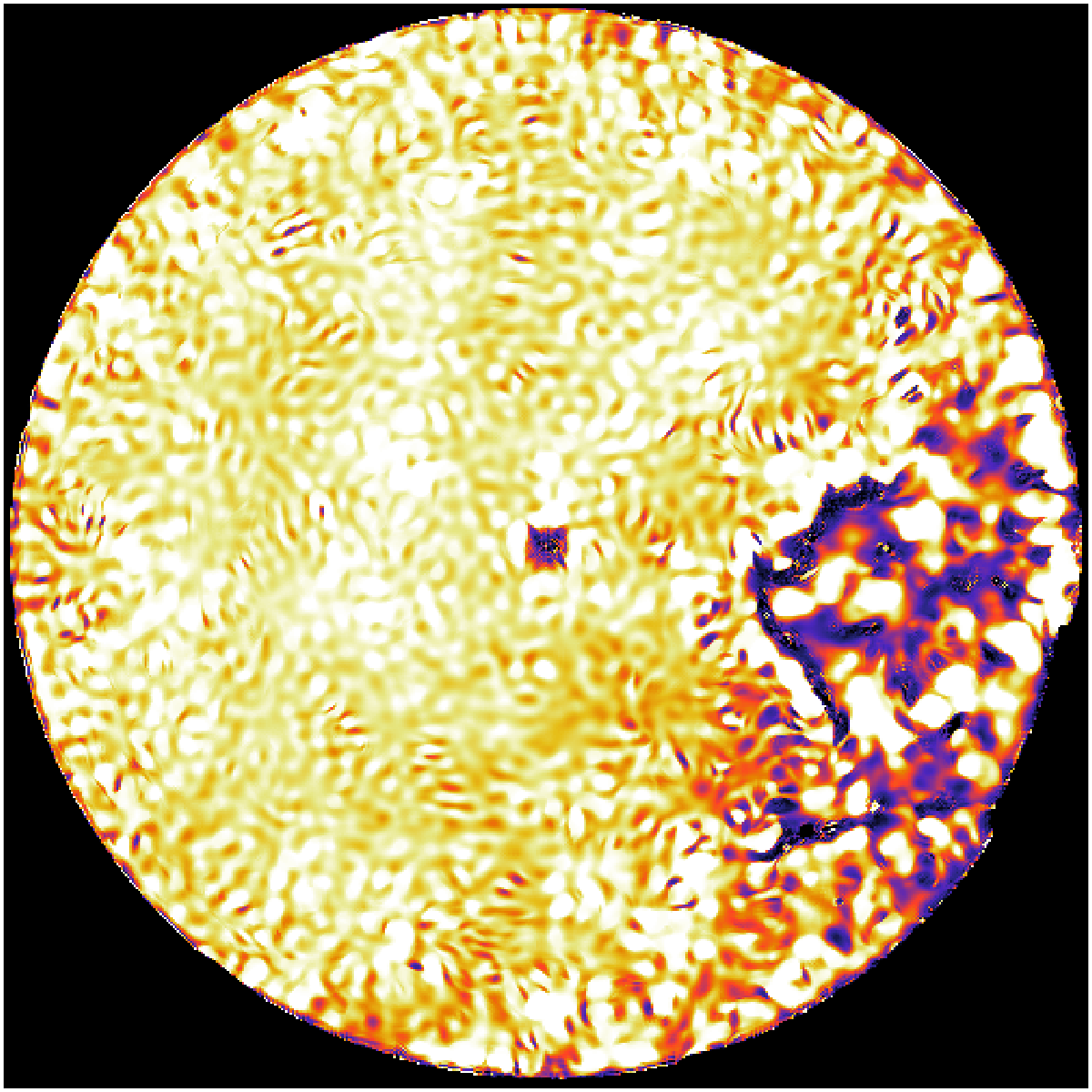}
        \caption{inversion ($1/4$)}\label{fig:slicesFWI3Db}
    \end{subfigure}
    \hfill
    \begin{subfigure}[t]{0.24\textwidth}
        \includegraphics[width=\textwidth]{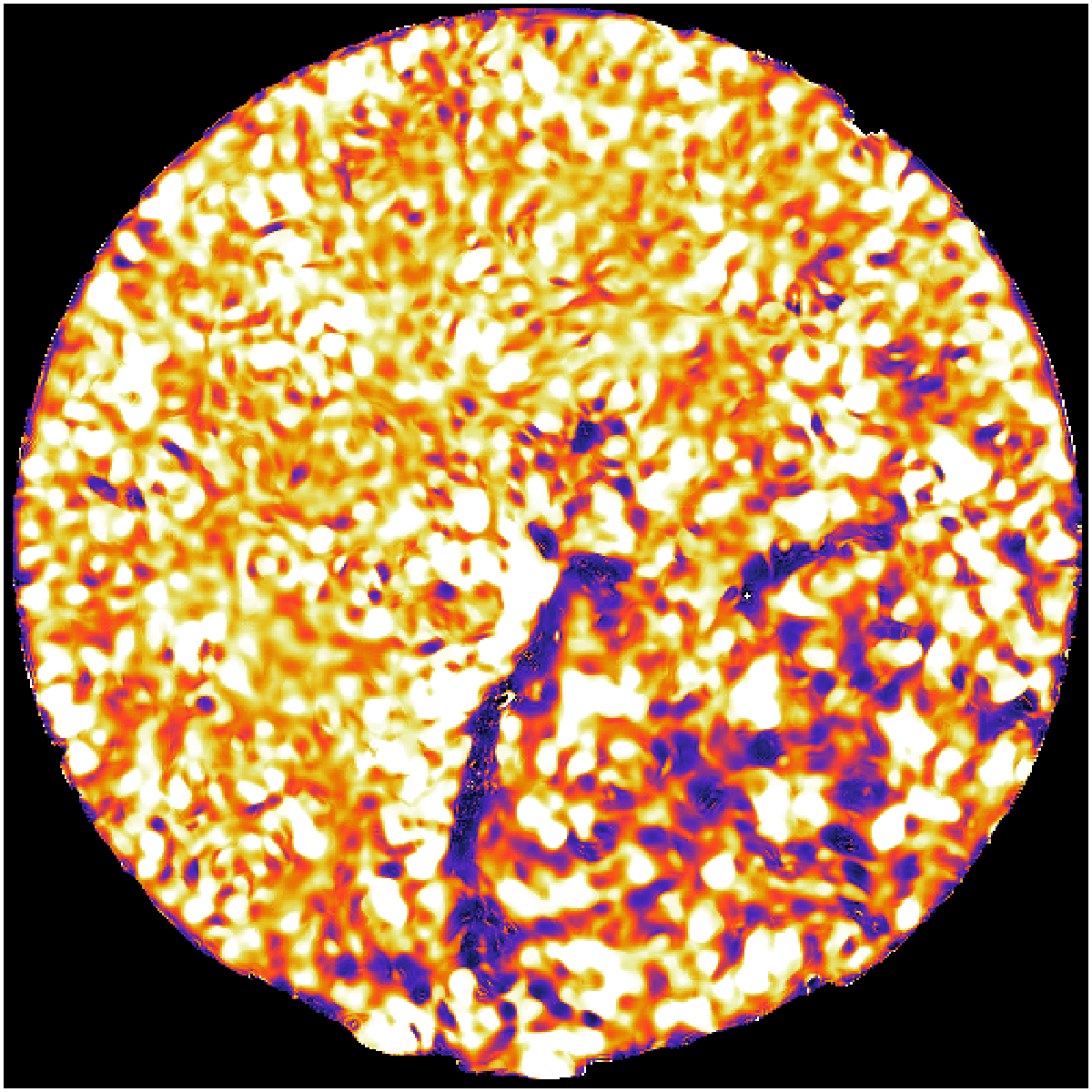}
        \caption{inversion ($1/2$)}\label{fig:slicesFWI3Dc}
    \end{subfigure}
    \hfill
    \begin{subfigure}[t]{0.24\textwidth}
        \includegraphics[width=\textwidth]{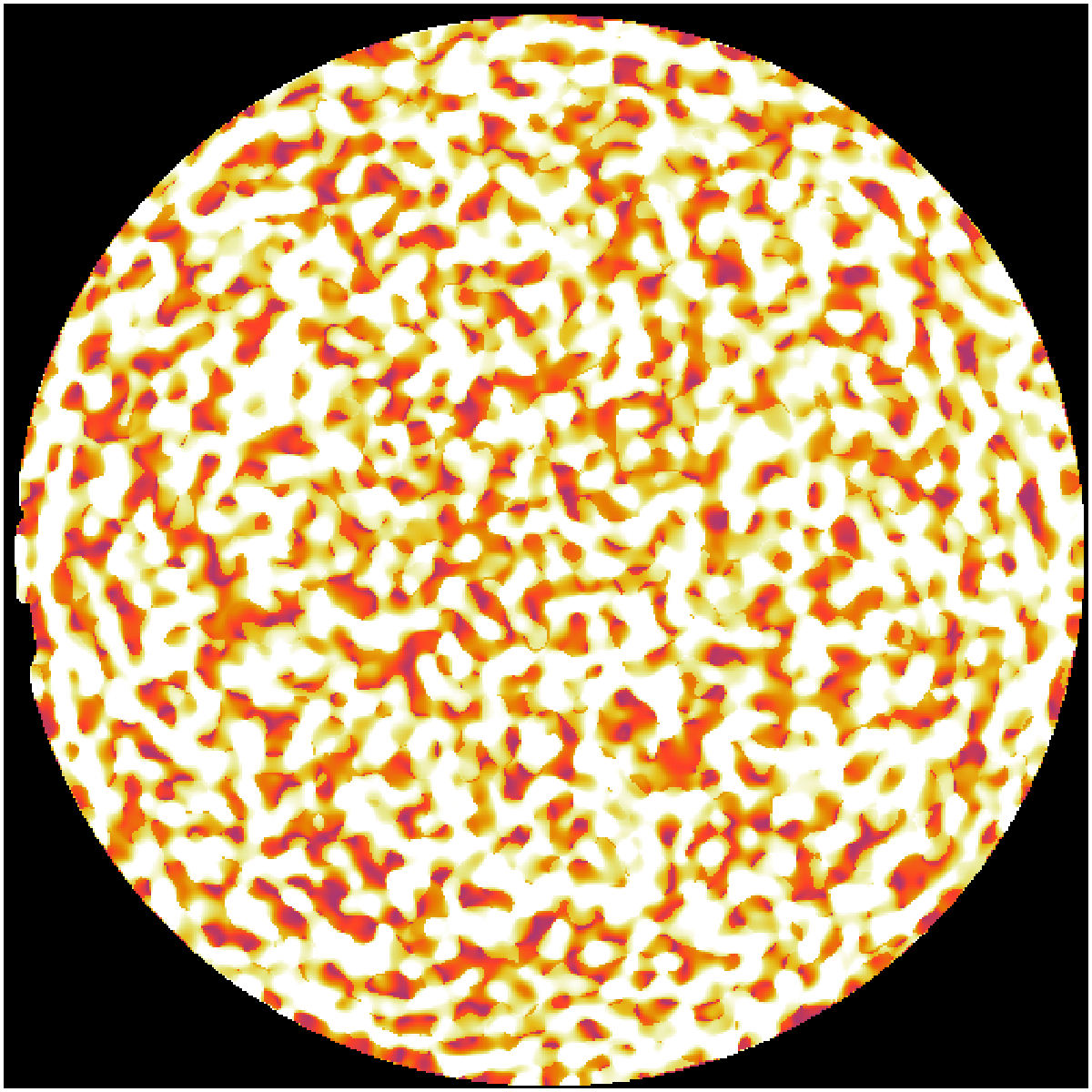}
        \caption{inversion ($1/1$)}\label{fig:slicesFWI3Dd}
    \end{subfigure}
    \\
    \begin{subfigure}[t]{0.24\textwidth}
        \includegraphics[width=\textwidth]{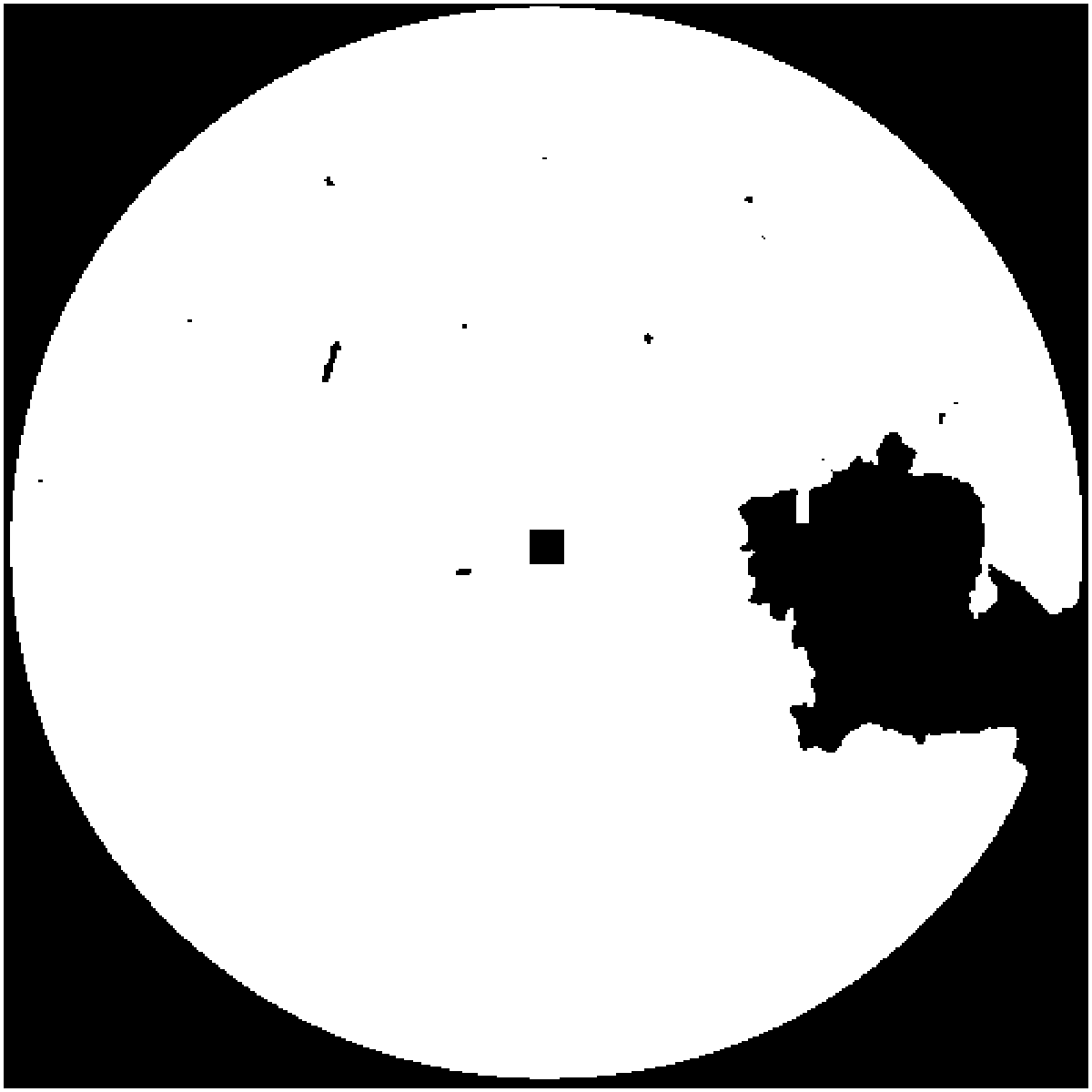}
        \caption{ground truth ($1/8$)}\label{fig:slicesFWI3De}
    \end{subfigure}
    \hfill
    \begin{subfigure}[t]{0.24\textwidth}
        \includegraphics[width=\textwidth]{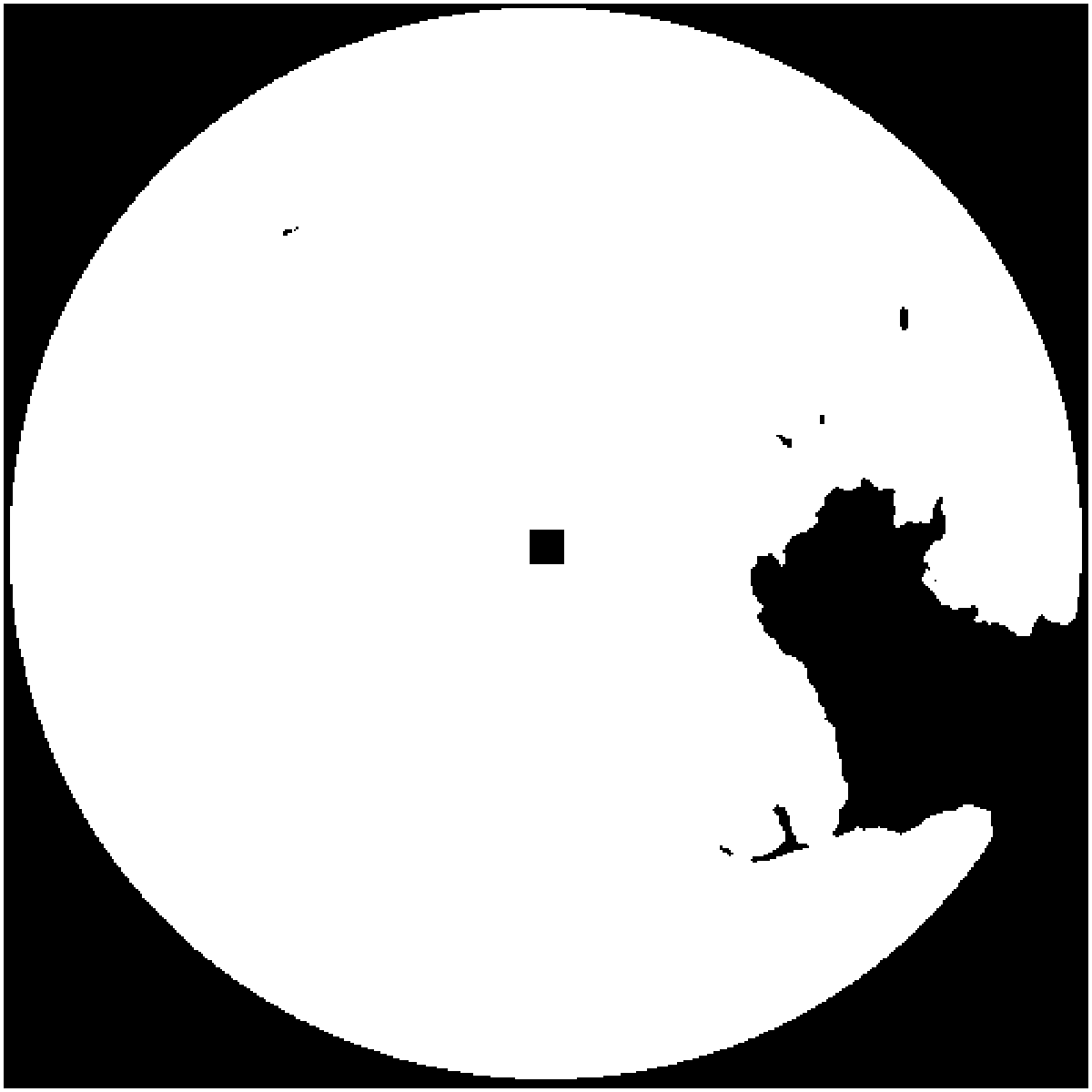}
        \caption{ground truth ($1/4$)}\label{fig:slicesFWI3Df}
    \end{subfigure}
    \hfill
    \begin{subfigure}[t]{0.24\textwidth}
        \includegraphics[width=\textwidth]{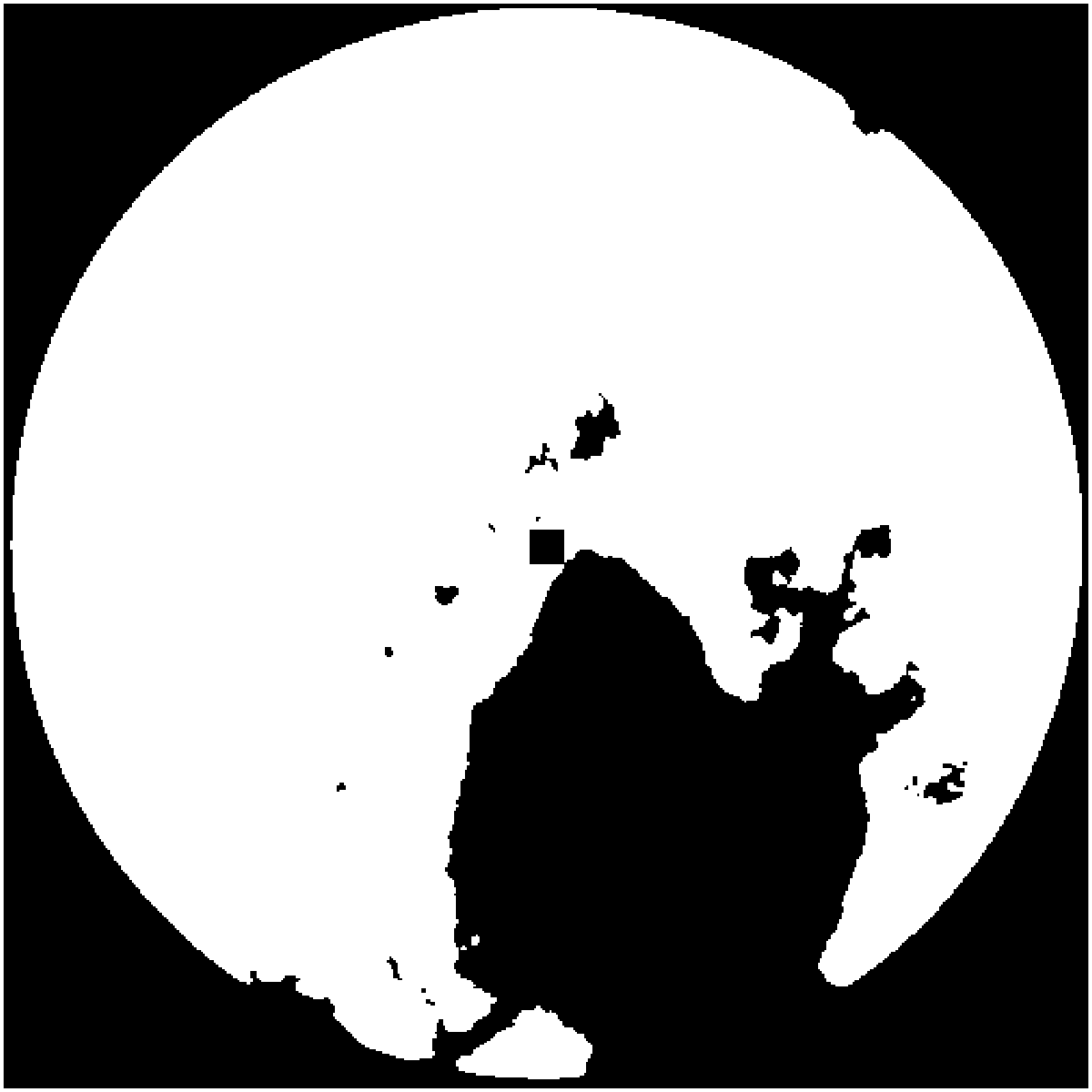}
        \caption{ground truth ($1/2$)}\label{fig:slicesFWI3Dg}
    \end{subfigure}
    \hfill
    \begin{subfigure}[t]{0.24\textwidth}
        \includegraphics[width=\textwidth]{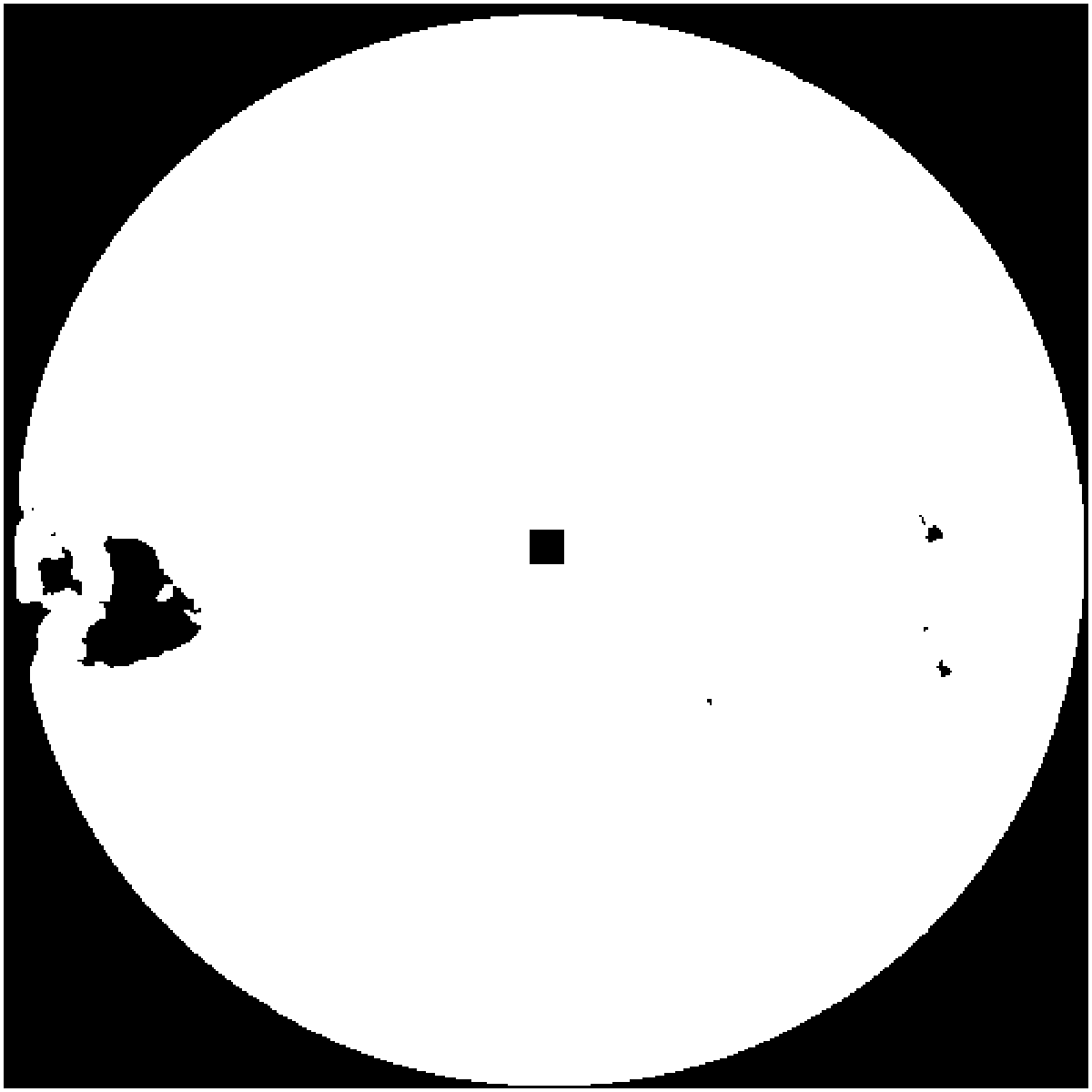}
        \caption{ground truth ($1/1$)}\label{fig:slicesFWI3Dh}
    \end{subfigure}
    \caption{Central slices comparing results obtained from inversion and the true internal voids for different substructures defined according to \Cref{fig:CTscan}. The inversion of (1/1) only ran for ten iterations without successful convergence.}\label{fig:slicesFWI3D}
\end{figure}

The resulting FWI results are provided in \Cref{fig:FWIresults} for the one eighth ($1/8$), one quarter ($1/4$), and one half ($1/2$) substructures. The computational time per iteration for each of the structures is provided in \Cref{tab:timings} together with the associated number of degrees of freedom. Note that all computations are impossible to carry out without employing \Cref{alg:memoryTrick} as even the smallest domain size necessitates 31\,373\,116 degrees of freedom. 
However, as seen in \Cref{fig:FWIresults3}, the inversion quality deteriorates for very large problem sizes. 
This is even clearer when viewing the central slices of each of the substructures (now including the full structure ($1/1$)) in \Cref{fig:slicesFWI3D}, where the centered cuboid aids in a direct qualitative comparison. 
As can be seen in \Cref{fig:slicesFWI3Dd}, we were unable to perform a successful inversion for the full structure, since the wave field information from waves traveling through the entire cylinder is insufficient. Yet, these are issues related to the well-posedness of FWI and not directly a result of the approximation used in \Cref{alg:memoryTrick}. Instead, the main advantage of \Cref{alg:memoryTrick} is that it enables to investigate FWI for such large structures on fast GPU implementations at all.

\FloatBarrier
\subsection{Transient Acoustic Topology Optimization}
\label{ssec:topopt}

\begin{figure}[htbp]
    \centering
    \begin{subfigure}[t]{0.49\textwidth}
        \includegraphics[width=\textwidth]{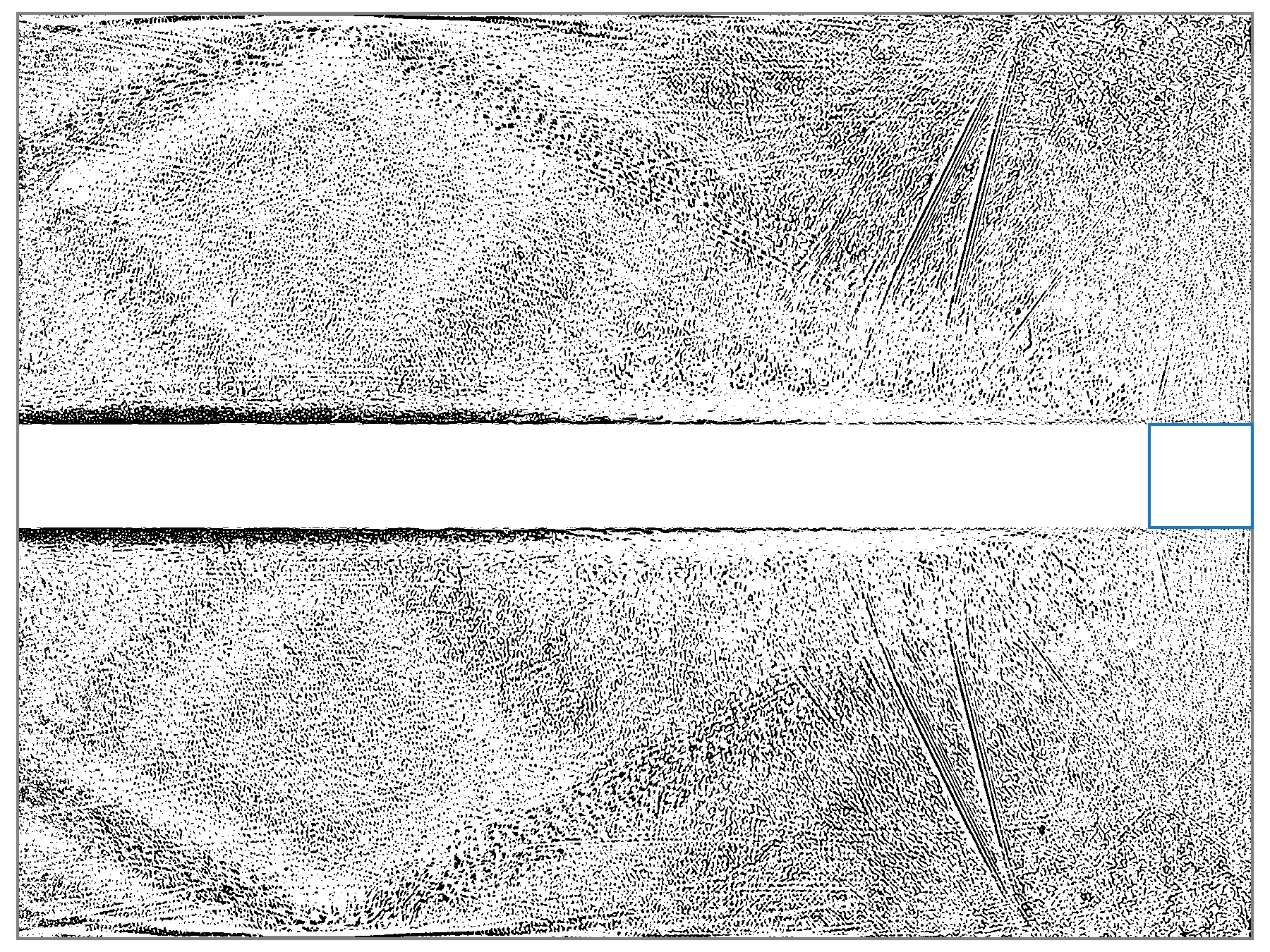}
        \caption{suppression}\label{fig:designs1}
    \end{subfigure}
    \hfill
    \begin{subfigure}[t]{0.49\textwidth}
        \includegraphics[width=\textwidth]{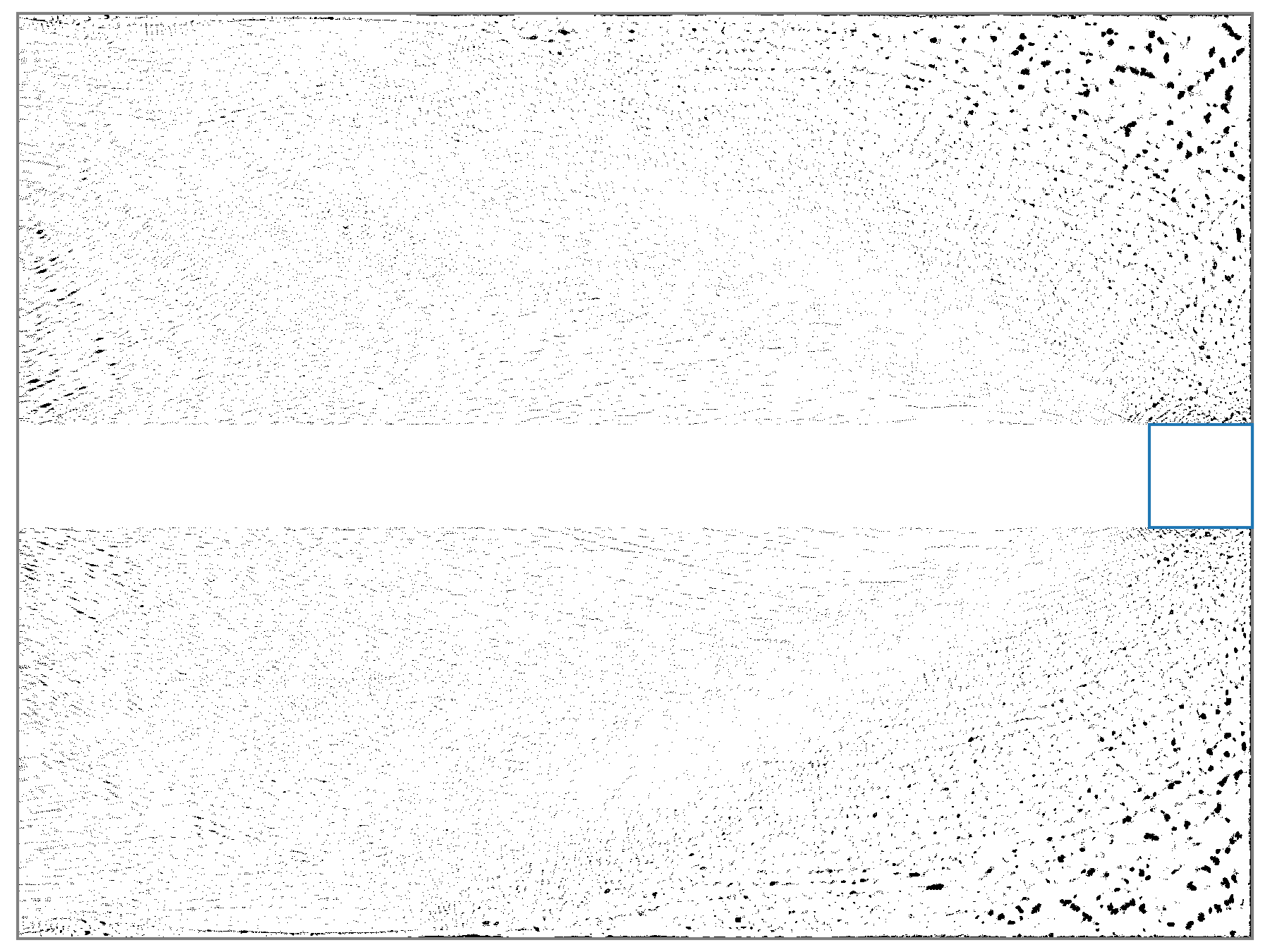}
        \caption{amplification}\label{fig:designs2}
    \end{subfigure}
    \caption{Optimized designs  (with filter radius 1.5) suppressing (left) or amplifying (right) the acoustic pressures inside the blue square ($\Omega_s$)}\label{fig:designs}
\end{figure}

We consider optimizing an acoustic black hole, following \Cref{fig:setupTopOpt}. 
The corresponding problem parameters are given in \Cref{tab:problemsetupTopOpt}. 
In addition to performing an optimization that suppresses the acoustic pressure in the blue domain $\Omega_s$ in \Cref{fig:setupTopOpt}, an optimization that amplifies the pressure is also performed. 
To this end, the signs of \Cref{eq:cost2,eq:adjointSource2} are simply inverted. 
While \Cref{fig:designs1} shows the design achieving noise suppression, the design of \Cref{fig:designs2} leads to amplification. \\

However, the obtained designs are clearly unmanufacturable. Larger filter radii (see Appendix~\ref{appendix:filter} for examples) amend the manufacturability issue to a certain extent, but also yield less desirable designs in terms of the objective function (\Cref{eq:cost2}). More sophisticated options are incorporating robustness into the optimization~\cite{sigmund_morphology-based_2007,sigmund_manufacturing_2009,christiansen_creating_2015} or extending it with structural integrity through a simultaneous compliance minimization~\cite{sigmund_99_2001,bendsoe_topology_2003}. Ensuring manufacturability in TATO is, however, out of the scope of the present work.\\

\begin{table}[htbp]
    \centering
    \caption{Problem parameters of the two-dimensional TATO problem defined in \Cref{fig:setupTopOpt}. This includes the domain size $L_1\times L_2$, the grid points $n_1 \times n_2$, number of time steps $N$, time step size $\Delta t$, source parameters $f, n_c, \psi_0$, and learning rate $\alpha$ for Adam.}\label{tab:problemsetupTopOpt}
    \begin{tabular}{cccccc}
    \multicolumn{5}{l}{\textbf{problem parameters}} \\
    \hline
    \hline
    $L_1$ & $L_2$ & $f$ & $n_c$ & $\psi_0$ \\
    \hline
    18 m & 9 m & $5\,000$ Hz & 2 & $10^{2} \text{ s}^{-2}$\\
    \hline 
    \multicolumn{5}{l}{\textbf{discretization \& optimization parameters}} \\
    \hline 
    \hline
    $n_1$ & $n_2$ & $N$ & $\Delta t$ & $\alpha$ \\
    \hline 
    5403 & 2703 & 17\,184 & ${\sim}4.58\cdot 10^{-6}$ s & 0.1\\
    \hline
    \end{tabular}
\end{table}



\begin{figure}[htbp]
    \centering
    \begin{subfigure}[t]{0.24\textwidth}
        \includegraphics[width=1.9\textwidth, angle=-90]{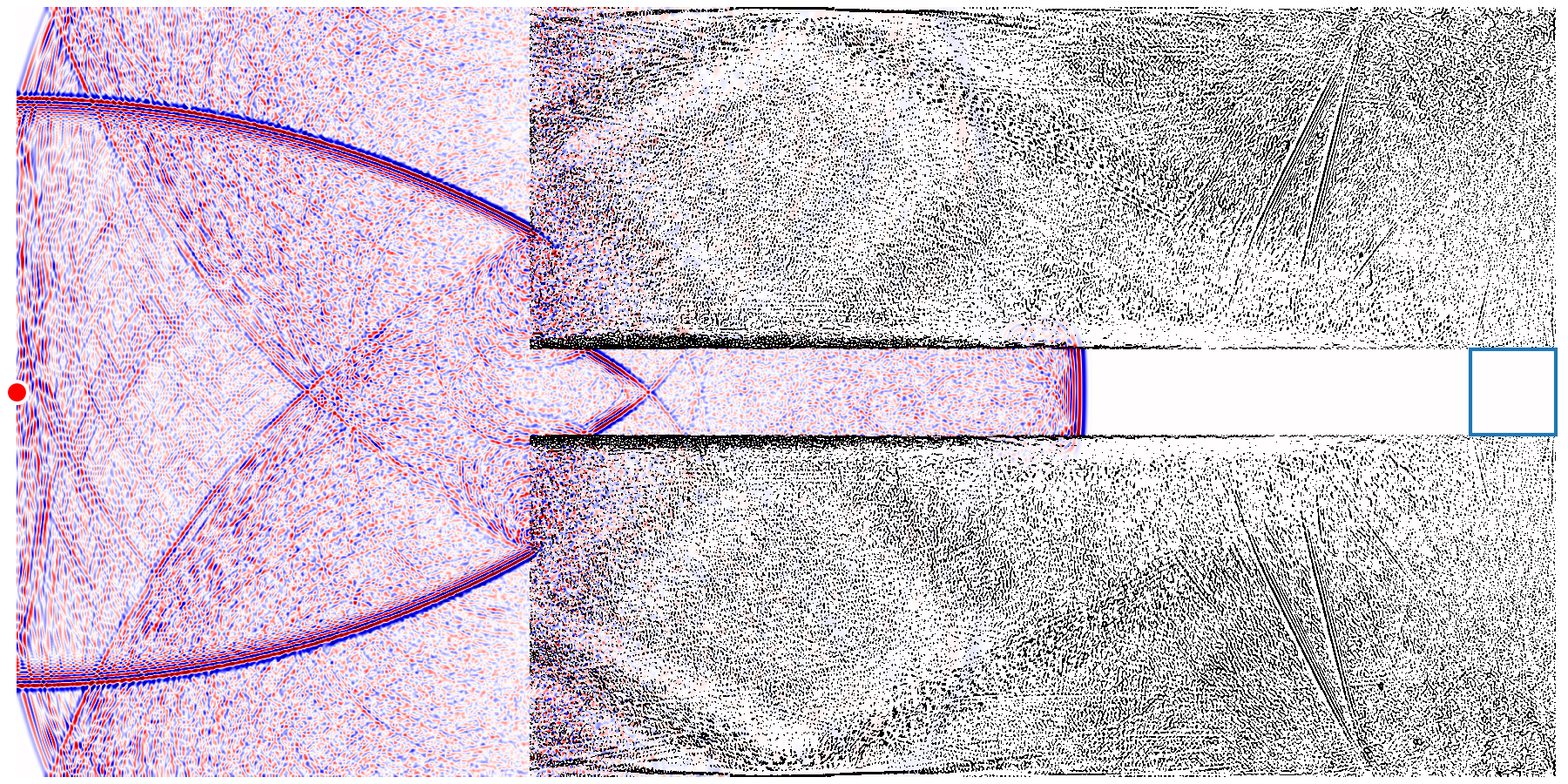}
        \caption{$t\approx3.66\cdot10^{-2}$ s}
    \end{subfigure}
    \hfill
    \begin{subfigure}[t]{0.24\textwidth}
        \includegraphics[width=1.9\textwidth, angle=-90]{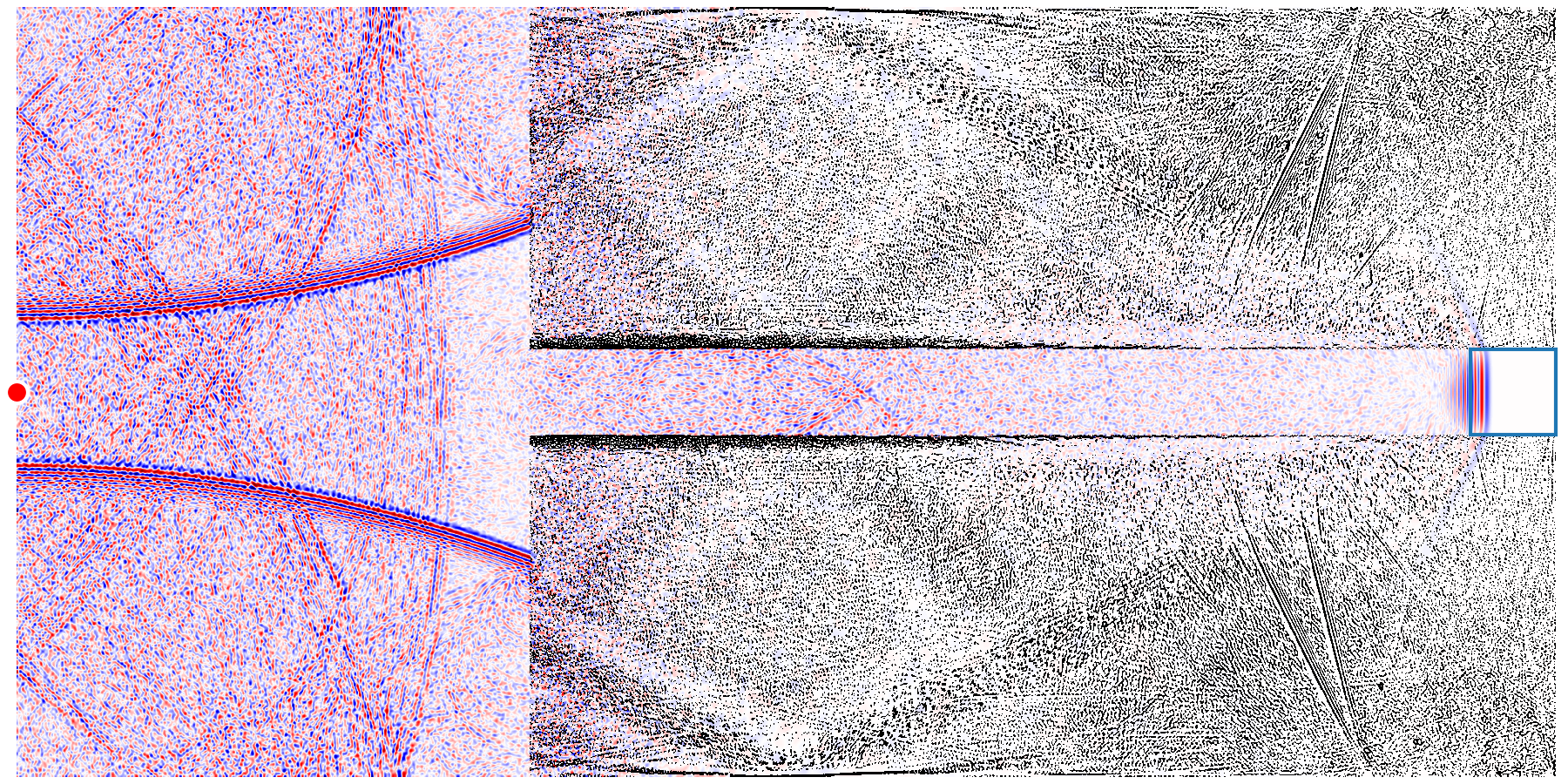}
        \caption{$t=\approx5.00\cdot10^{-2}$ s}
    \end{subfigure}
    \hfill
    \begin{subfigure}[t]{0.24\textwidth}
        \includegraphics[width=1.9\textwidth, angle=-90]{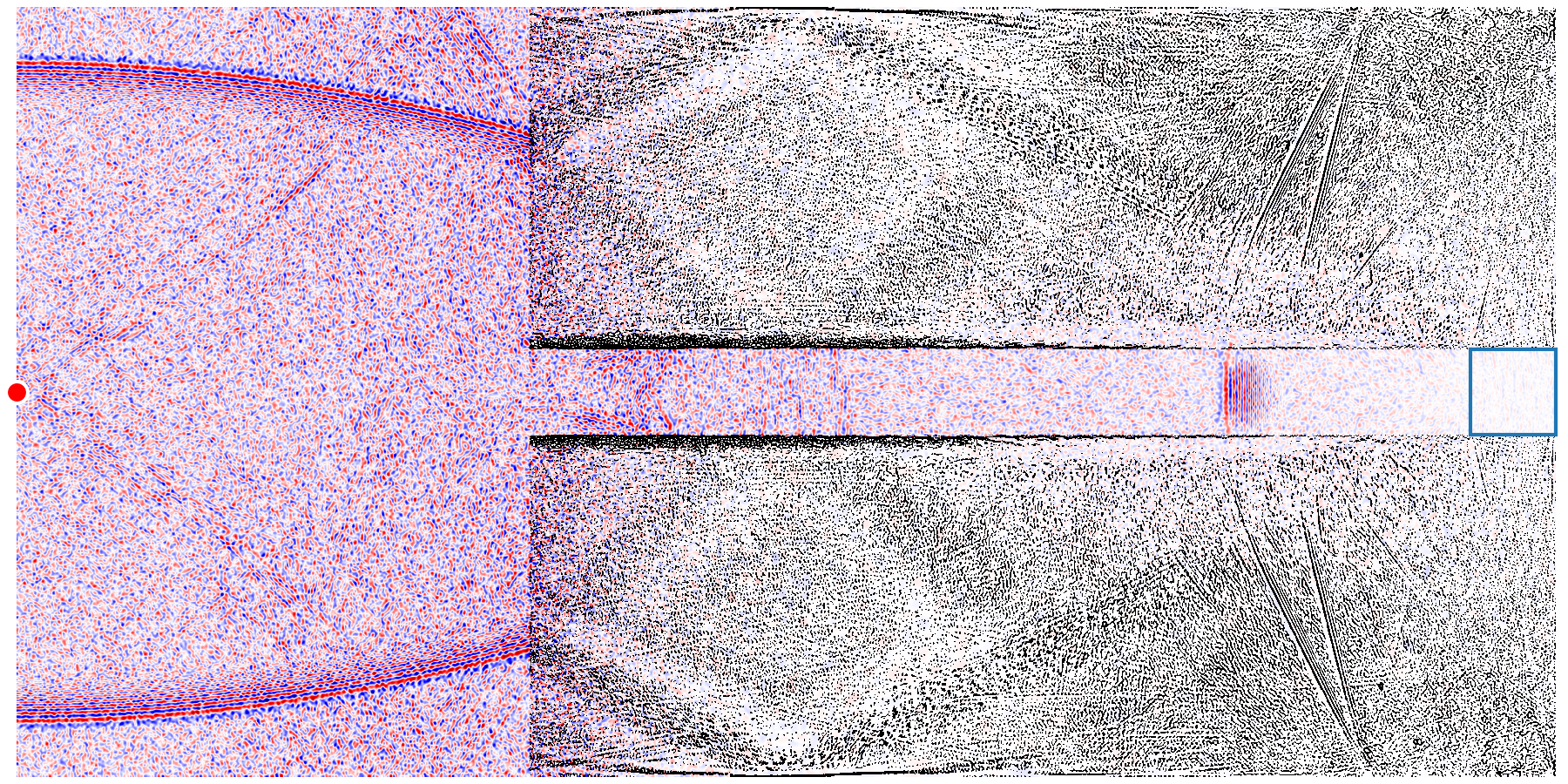}
        \caption{$t\approx6.41\cdot10^{-2}$ s}
    \end{subfigure}
    \hfill
    \begin{subfigure}[t]{0.24\textwidth}
        \includegraphics[width=1.9\textwidth, angle=-90]{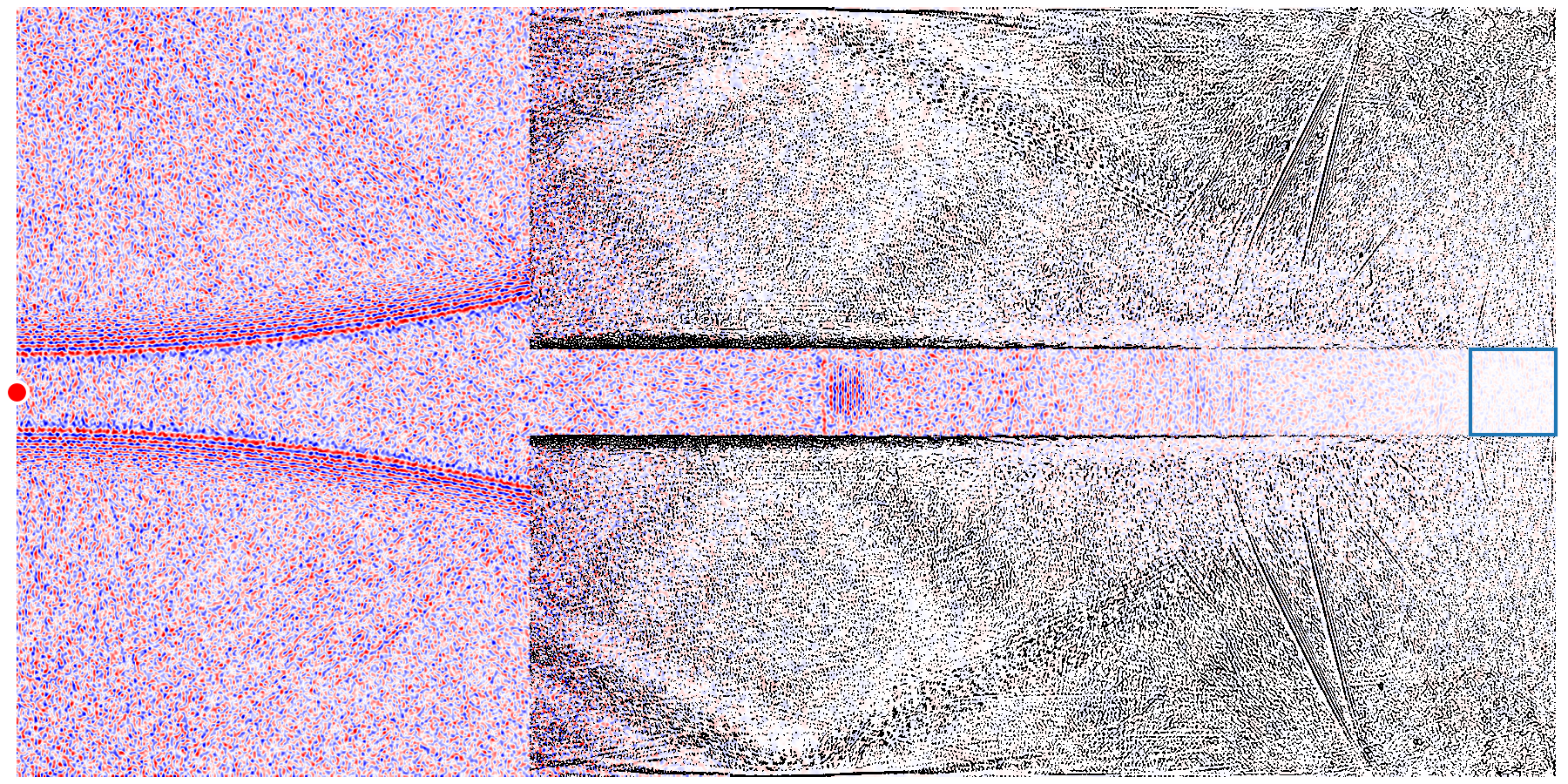}
        \caption{$t\approx7.78\cdot10^{-2}$ s}\label{fig:TopOptLarge1}
    \end{subfigure}
    \\
    \begin{subfigure}[t]{0.24\textwidth}
        \includegraphics[width=1.9\textwidth, angle=-90]{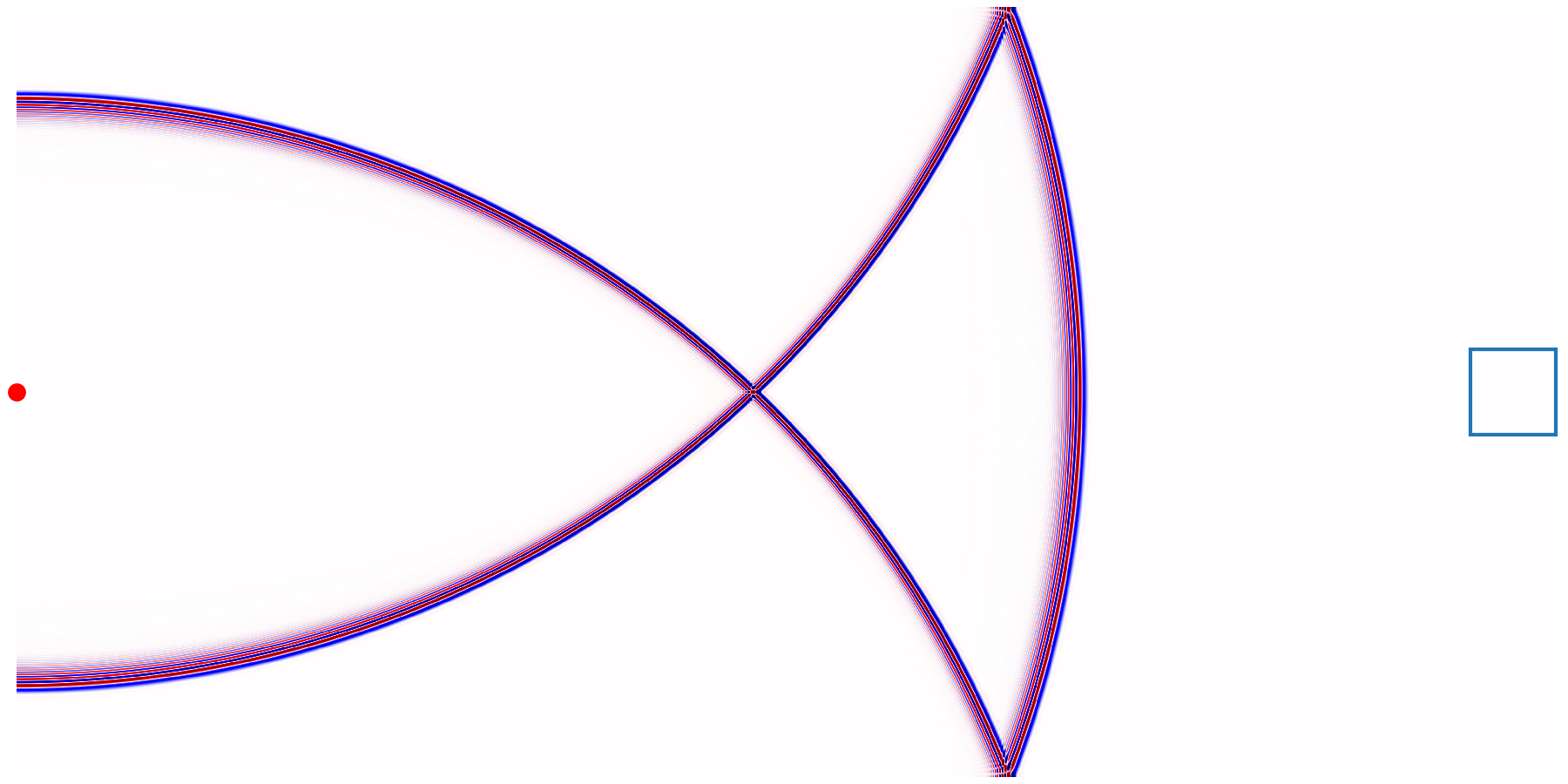}
        \caption{$t\approx3.66\cdot10^{-2}$ s}\label{fig:TopOptLarge22}
    \end{subfigure}
    \hfill
    \begin{subfigure}[t]{0.24\textwidth}
        \includegraphics[width=1.9\textwidth, angle=-90]{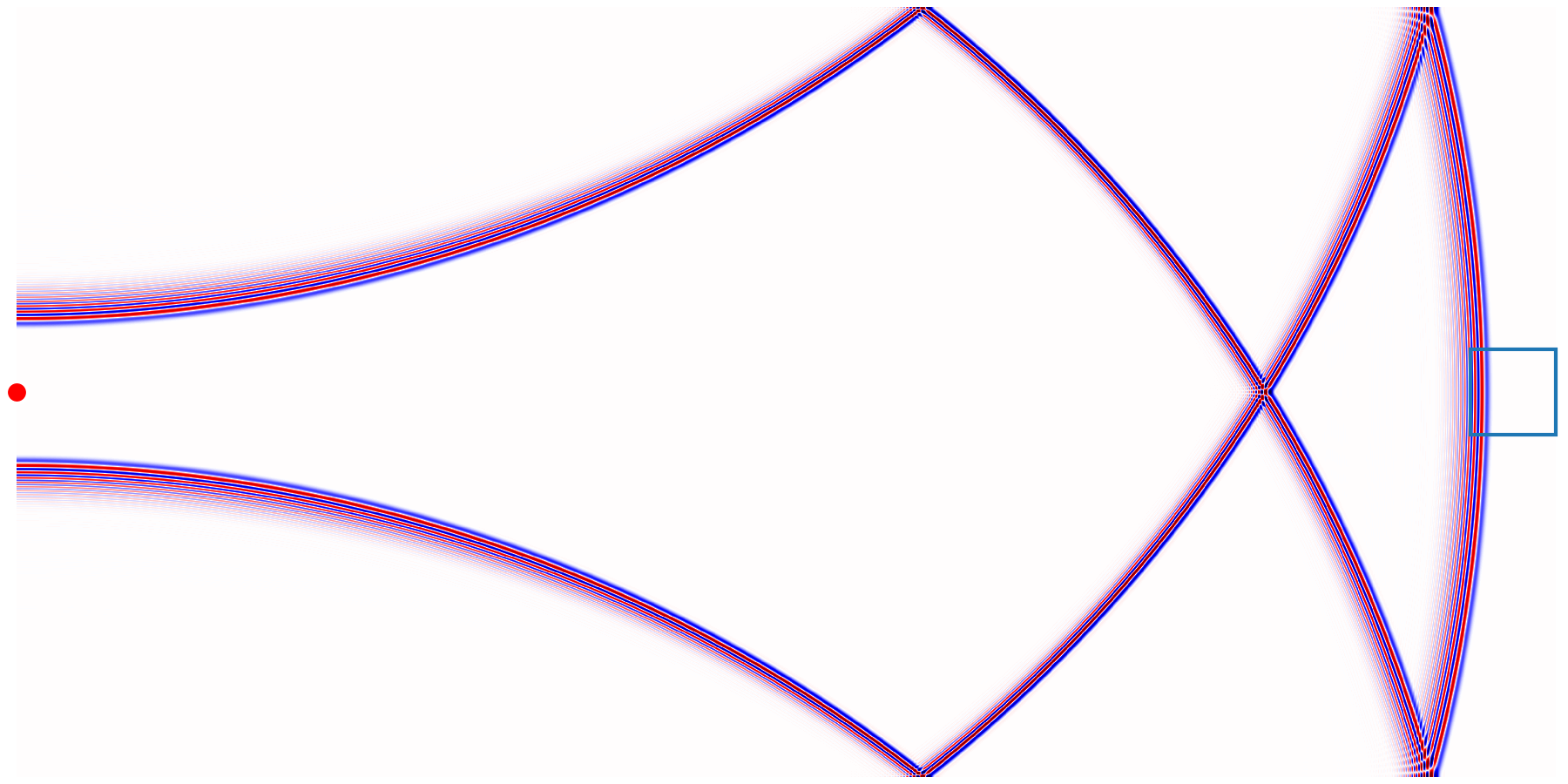}
        \caption{$t=\approx5.00\cdot10^{-2}$ s}\label{fig:TopOptLarge23}
    \end{subfigure}
    \hfill
    \begin{subfigure}[t]{0.24\textwidth}
        \includegraphics[width=1.9\textwidth, angle=-90]{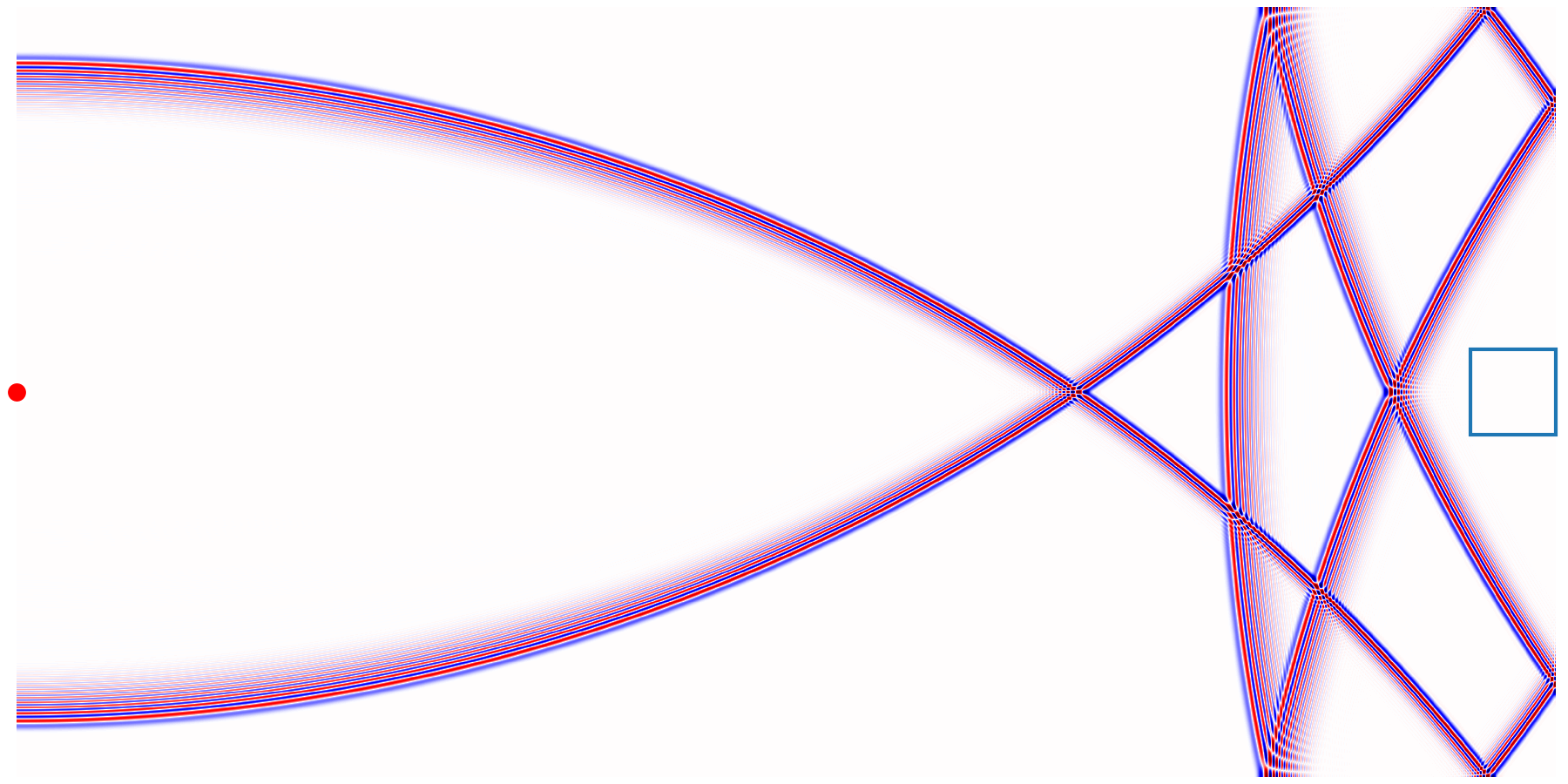}
        \caption{$t\approx6.41\cdot10^{-2}$ s}\label{fig:TopOptLarge24}
    \end{subfigure}
    \hfill
    \begin{subfigure}[t]{0.24\textwidth}
        \includegraphics[width=1.9\textwidth, angle=-90]{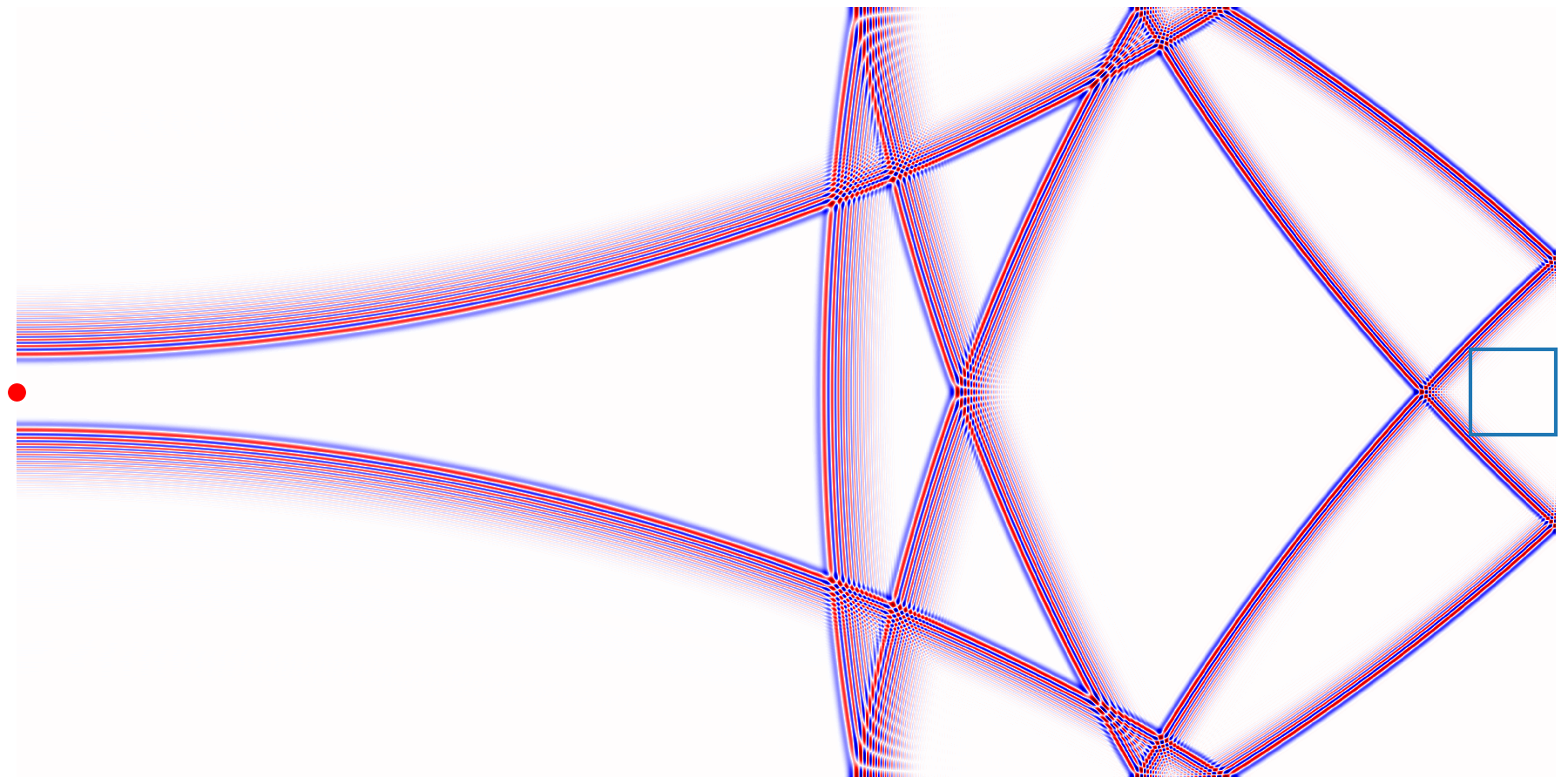}
        \caption{$t\approx7.78\cdot10^{-2}$ s}\label{fig:TopOptLarge2}
    \end{subfigure}
    \\
    \begin{subfigure}[t]{0.24\textwidth}
        \includegraphics[width=1.9\textwidth, angle=-90]{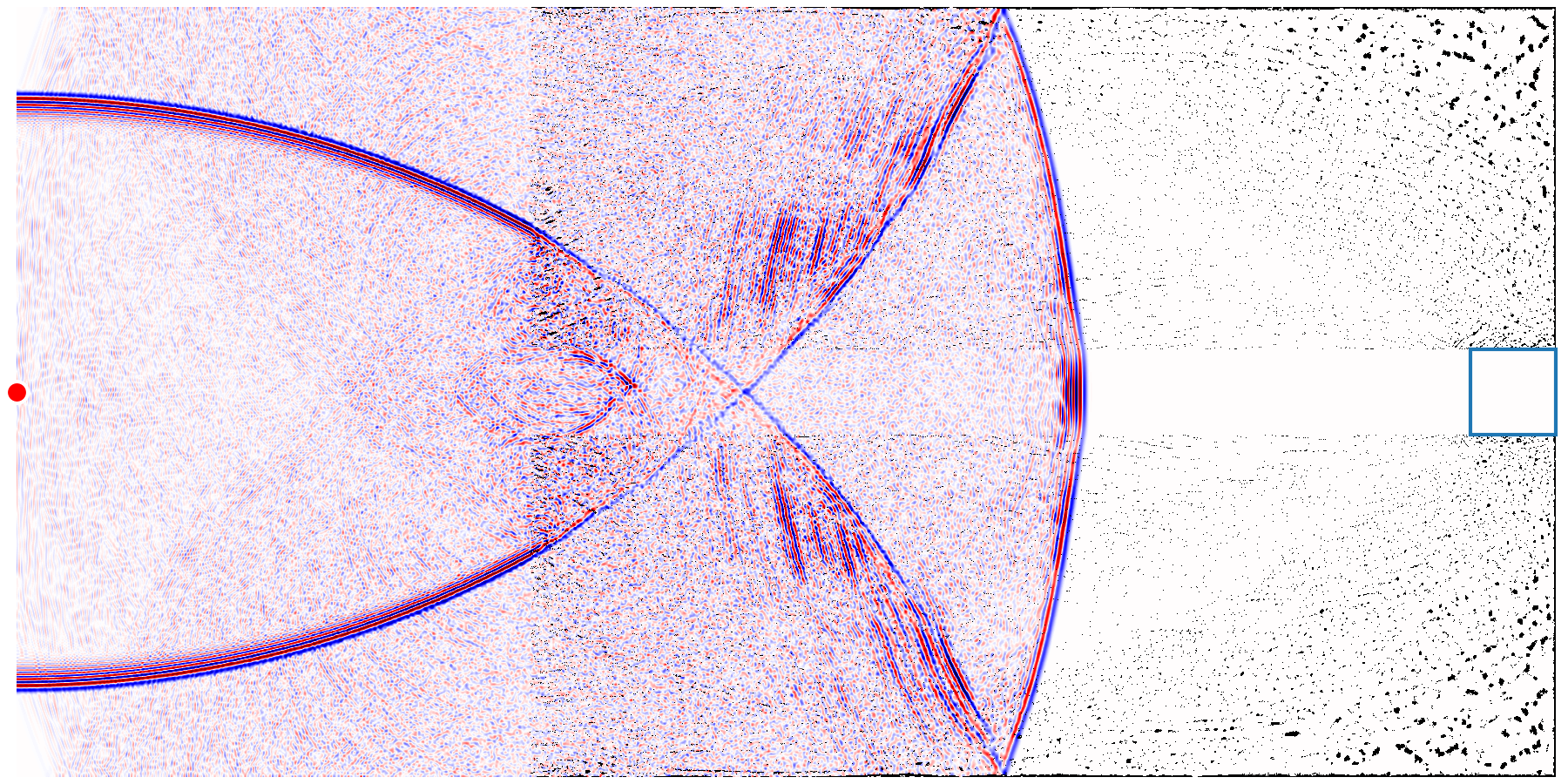}
        \caption{$t\approx3.66\cdot10^{-2}$ s}
    \end{subfigure}
    \hfill
    \begin{subfigure}[t]{0.24\textwidth}
        \includegraphics[width=1.9\textwidth, angle=-90]{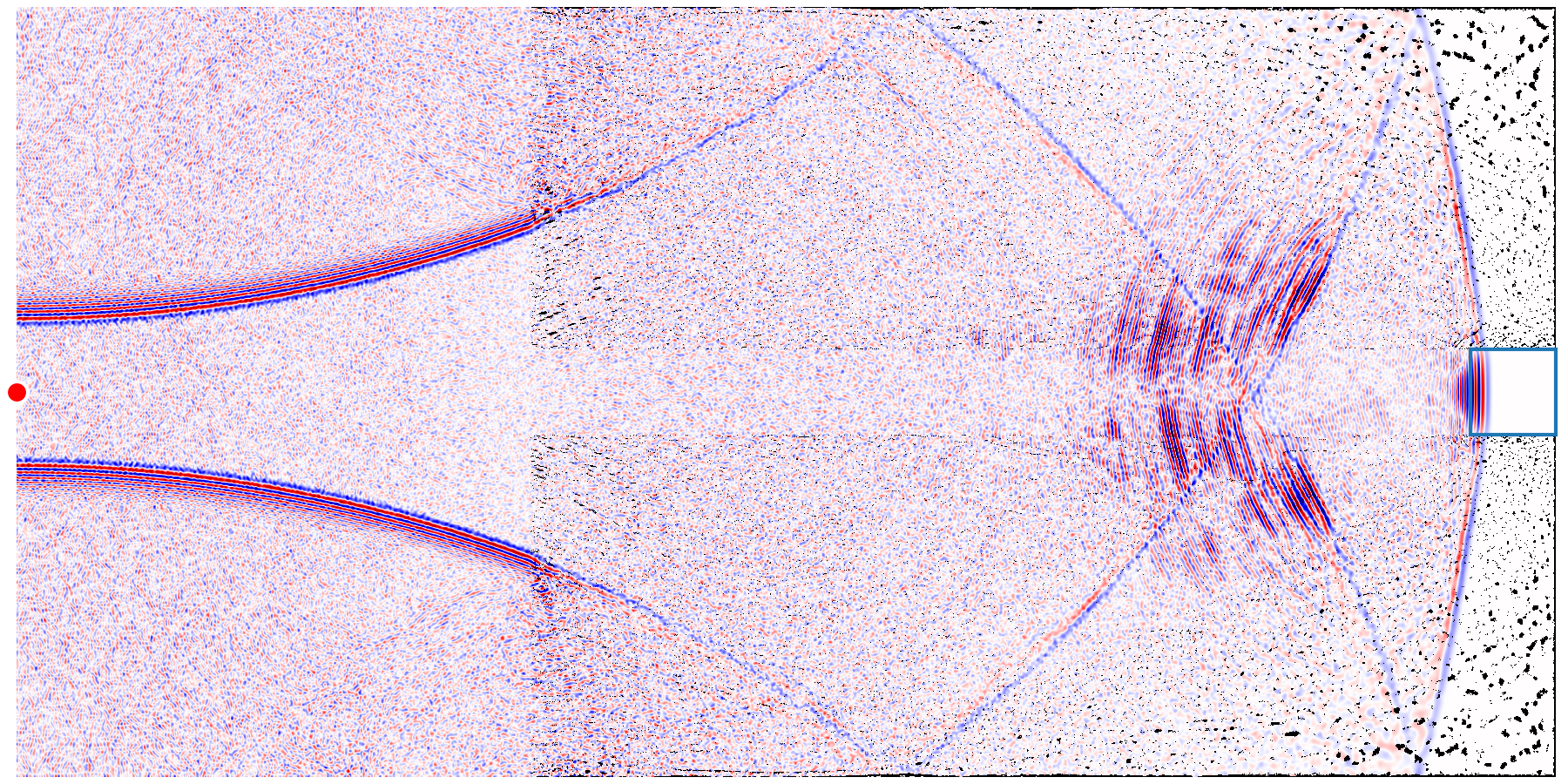}
        \caption{$t=\approx5.00\cdot10^{-2}$ s}
    \end{subfigure}
    \hfill
    \begin{subfigure}[t]{0.24\textwidth}
        \includegraphics[width=1.9\textwidth, angle=-90]{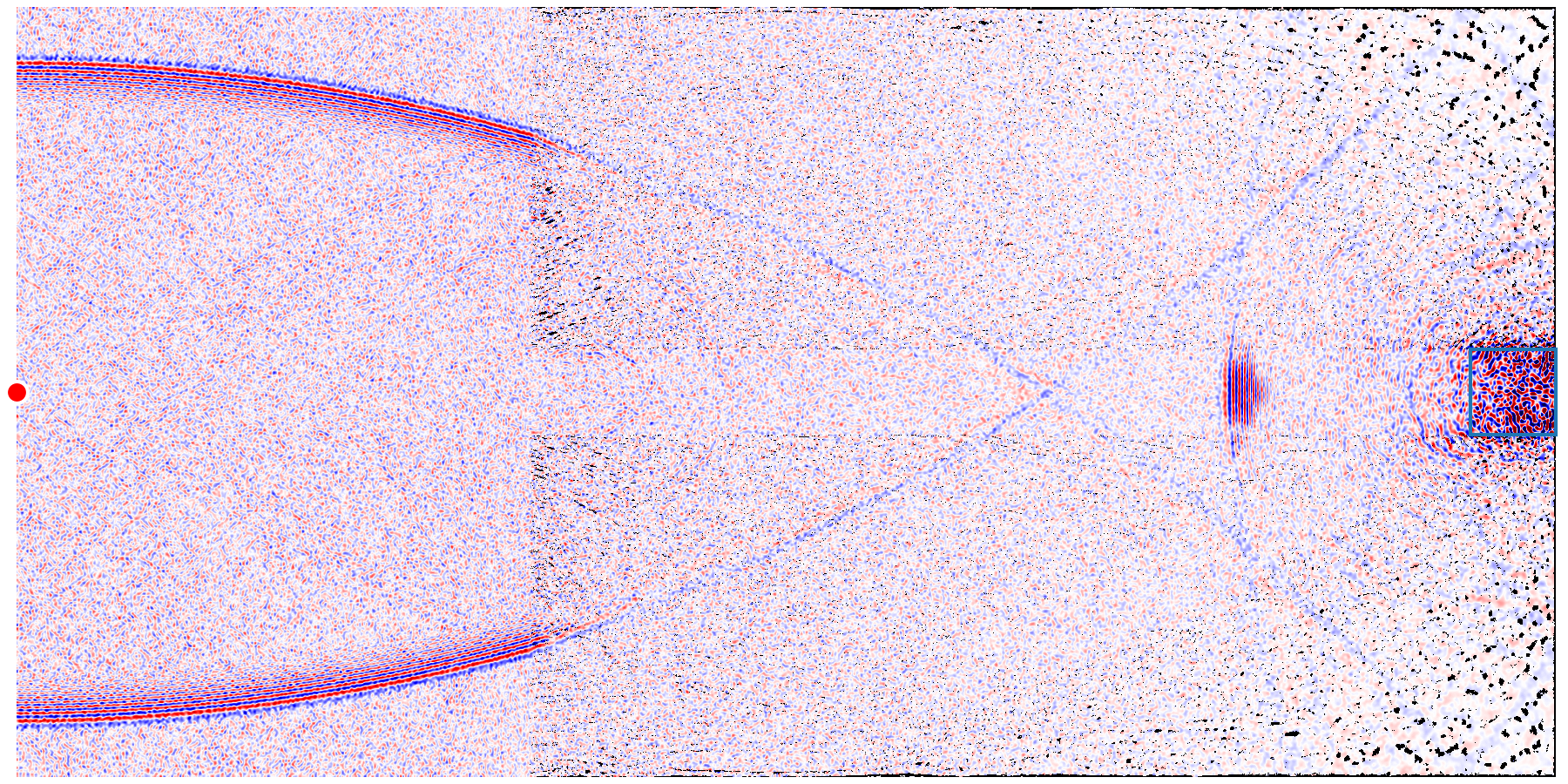}
        \caption{$t\approx6.41\cdot10^{-2}$ s}
    \end{subfigure}
    \hfill
    \begin{subfigure}[t]{0.24\textwidth}
        \includegraphics[width=1.9\textwidth, angle=-90]{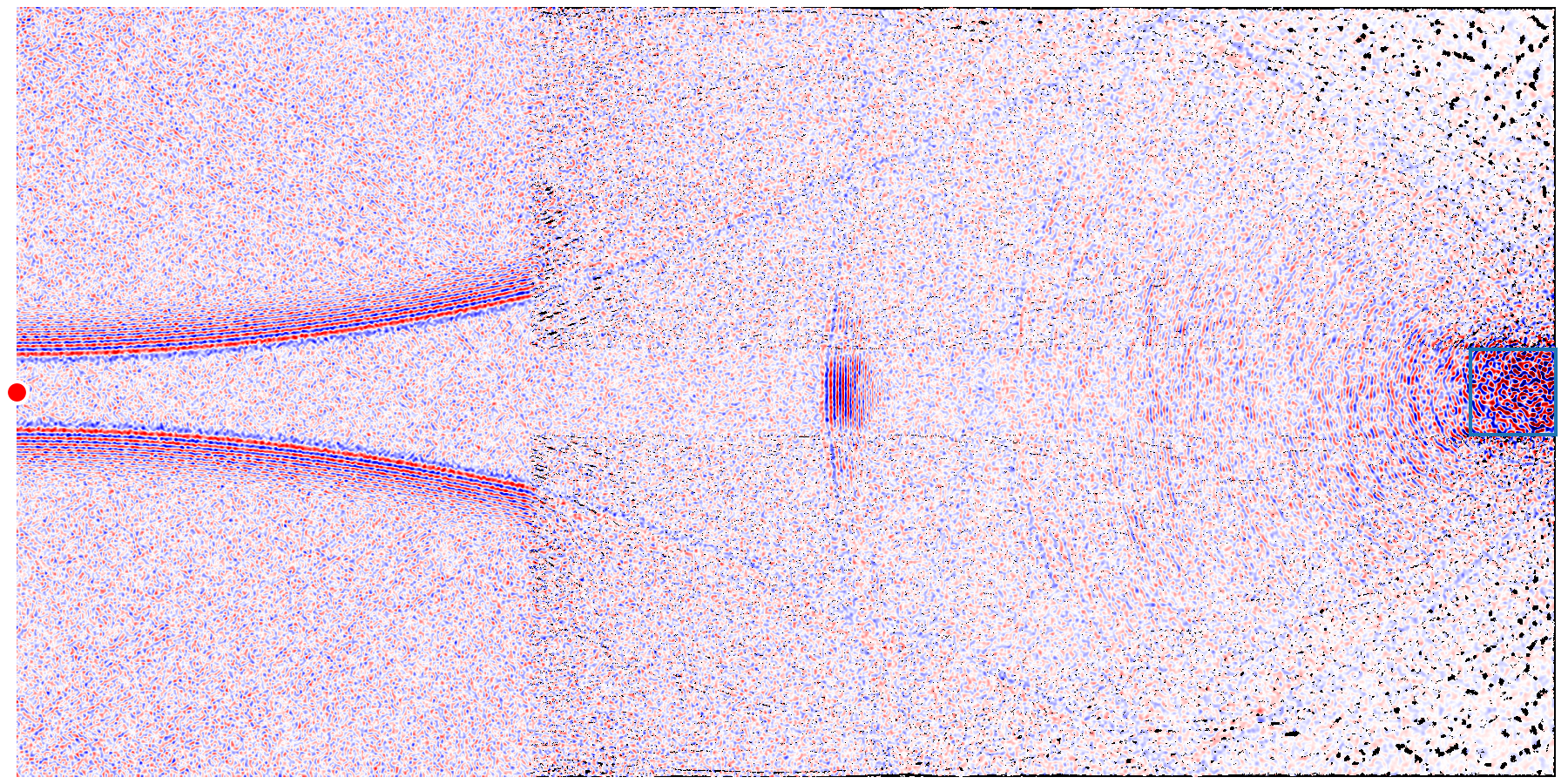}
        \caption{$t\approx7.78\cdot10^{-2}$ s}\label{fig:TopOptLarge3}
    \end{subfigure}
    \caption{Wave fields interacting with designs: suppression (top), no structure (center), amplification (bottom)}\label{fig:TopOptLarge}
\end{figure}

\begin{figure}[htbp]
    \centering
    \begin{subfigure}[t]{0.53\textwidth}
        \includegraphics[width=\textwidth]{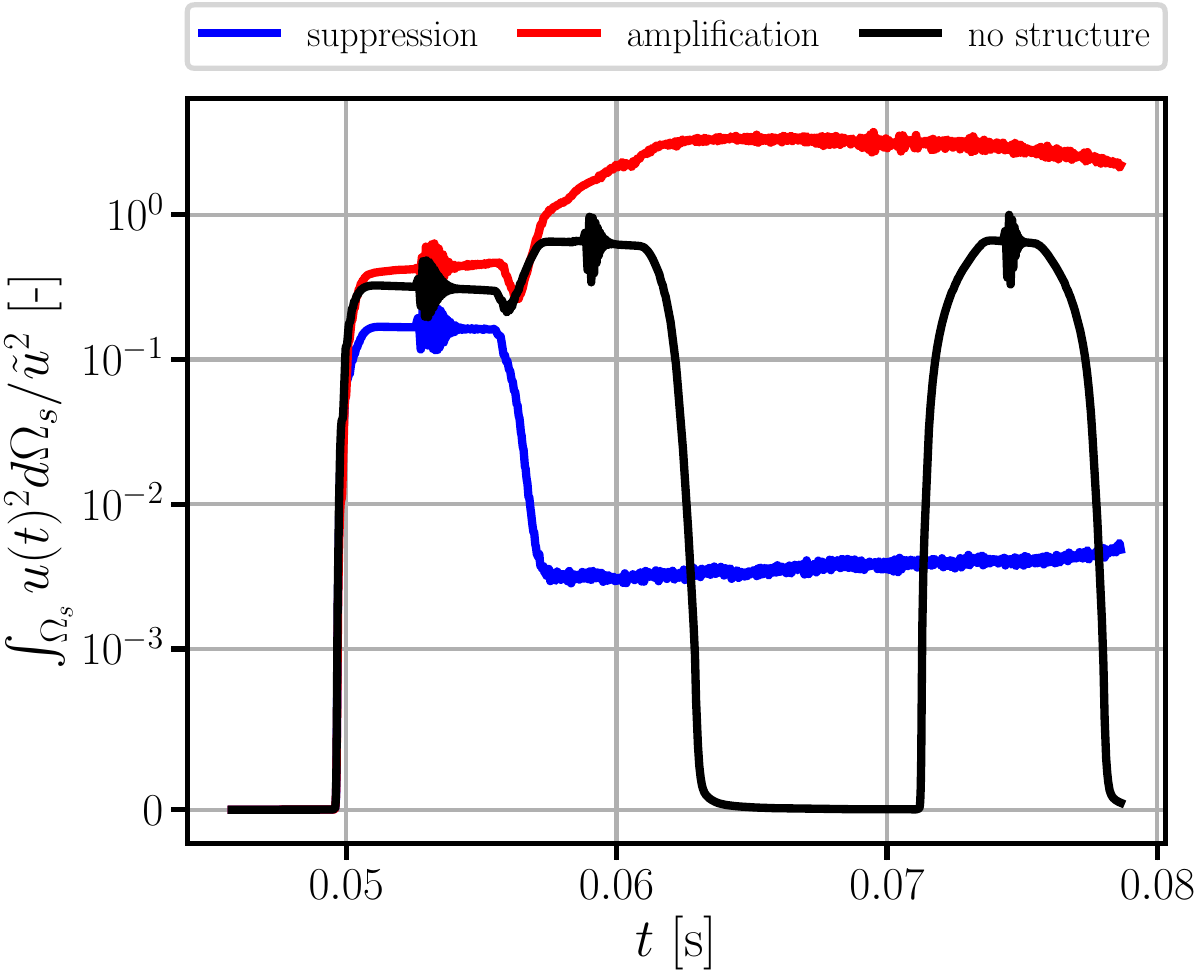}
    \end{subfigure}
    \caption{Transient response of integrated acoustic pressures in the domain of interest (blue domain in \Cref{fig:setupTopOpt,fig:designs}). The pressures are normalized by the maximum pressure $\tilde{u}$ reached by the transient response without a structure (\Cref{fig:TopOptLarge2,fig:TopOptLarge22,fig:TopOptLarge23,fig:TopOptLarge24}).}\label{fig:transientresponse}
\end{figure}

To illustrate how the optimized structures interact with the wave fields and achieve their corresponding goals, snapshots before and after the first wave fronts arrive at $\Omega_s$ are considered. \Cref{fig:TopOptLarge} visualizes these snapshots (at $t\approx3.66\cdot 10^{-2}$ s, $t\approx5.00\cdot 10^{-2}$ s, $t\approx6.41\cdot 10^{-2}$ s, and $t\approx7.78\cdot 10^{-2}$ s) for the suppressing structure (from \Cref{fig:designs1}), the amplifying structure (from \Cref{fig:designs1}), and no structure, i.e., the situation with only air. At $t>5\cdot 10^{-2}$ s, noise is clearly suppressed in \Cref{fig:TopOptLarge1} and amplified in \Cref{fig:TopOptLarge3} in comparison to \Cref{fig:TopOptLarge2}. This becomes even clearer by considering the transient response of the integrated acoustic pressures within $\Omega_s$, as shown by \Cref{fig:transientresponse}. \\

The corresponding computation time is again provided in \Cref{tab:timings} together with the degrees of freedom.

\begin{table}[htbp]
    \centering
    \caption{Computation times for all considered cases obtained on an NVIDIA$^{\text{\textregistered}}$ A100 40 GB GPU. The time is given per iteration (per time step) and, in the case of FWI per source, i.e., divided by four in all considered cases. The total time is obtained by multiplying the number of time steps $N$, number of sources, iteration time, and number of iterations.}\label{tab:timings}
    \setlength{\tabcolsep}{4pt}
    {\small
    \begin{tabular}{cccccccc}
    & Fig.~\ref{fig:FWI2D} & Fig.~\ref{fig:topopt2Dsmall} \&~\ref{fig:optimized2DTopOpt} & Fig.~\ref{fig:slicesFWI3Da} & Fig.~\ref{fig:slicesFWI3Db} & Fig.~\ref{fig:slicesFWI3Dc} & Fig.~\ref{fig:slicesFWI3Dd} & Fig.~\ref{fig:designs} \\
    \hline
    \hline
    dofs & 63\,001 & 131\,769 & 31\,373\,116 & 62\,240\,214 & 123\,974\,410 & 246\,936\,784 & 14\,604\,309 \\
    \hline
    $N$ & 3\,200 & 1\,800 & 1\,300 & 2\,400 & 3\,800 & 7\,200 & 17\,184 \\ 
    \hline
    sources & 4 & 1 & 4 & 4 & 4 & 4 & 1 \\
    \hline 
    \begin{tabular}{c}iter.\\time [s]\end{tabular}
    & ${\sim}1.8\cdot 10^{-4}$ & ${\sim}2.0\cdot 10^{-4}$ & ${\sim}1.5\cdot 10^{-2}$ & ${\sim}3.3\cdot 10^{-2}$ & ${\sim}6.6\cdot 10^{-2}$ & ${\sim}1.6\cdot 10^{-1}$ & ${\sim}1.5\cdot 10^{-3}$ \\
    \hline
    iters. & 50 & 300 & 200 & 200 & 100 & (10) & 300 \\ 
    \hline \hline
    \begin{tabular}{c}total\\time [s]\end{tabular} 
    & ${\sim}$115 s & ${\sim}$109 s & ${\sim}$4.3 h & ${\sim}$17.6 h & ${\sim}$27.9 h & (${\sim}$12.8 h) & ${\sim}$2.1 h \\ 
    \hline
    \end{tabular}
    }
\end{table}






\FloatBarrier
\section{Conclusion}
\label{sec:conclusion}

In this paper, we propose a memory-efficient approach to compute the adjoint gradient for dynamic optimization problems, such as full waveform inversion (FWI) and transient acoustic topology optimization (TATO). 
The approach exploits the linearity of the underlying problem by using a superposition of the adjoint wave field with the forward computation.
Based on this approach, we avoid the need to save the forward wave field for the gradient computation. 
The memory requirement is reduced from the number of degrees of freedom times the number of time steps to only the number of degrees of freedom. 
It thereby breaks the memory bound of classical adjoint approaches for transient problems which has been a pressing issue for efficient GPU-based implementations.\\

The sensitivity approximation comes with no additional computational effort in comparison to the standard adjoint sensitivity method. 
Since the approach relies on the time-reversibility of the forward simulation, it is limited to self-adjoint problems and time-reversible integration schemes.
The approximation quality depends on a hyperparameter $k$, but the quality of the inversion is only mildly dependent on its choice as the range of the admissible values of $k$ spans multiple orders of magnitude. 
Moreover, $k$ can easily be estimated prior to the optimization, by reducing $k$ in the first optimization as much as possible until the computed sensitivity starts changing. 
A verification can also be performed by comparing a single precision computation with a double precision computation.\\ 

Relying on an efficient GPU-based finite difference solver (available at~\cite{herrmann_memory_2025}), we have successfully applied the approximation to the inverse problem of FWI (up to 123\,974\,410 degrees of freedom in three dimensions) and the optimization problem of TATO (up to 14\,604\,309 degrees of freedom in two dimensions). Sensitivities were computed for problems up to ${\sim}10^9$ parameters. The theoretical limit, however, lies at the memory capacity divided by the memory footprint of four times the number of degrees of freedom, i.e., $40$ GB $/ 4 / 4$ B $=2.5\cdot 10^9$ parameters for an NVIDIA$^{\text{\textregistered}}$ A100 GPU with 40 GB. This can even be pushed further with multi-GPU implementations\footnote{These implementations are made straightforward through packages such as JAX~\cite{bradbury_jax_2018} in Python or dedicated languages such as Julia~\cite{bezanson_julia_2017}. A prominent simulation-based multi-GPU package in Julia is \texttt{ParallelStencil.jl}~\cite{omlin_high-performance_2024,omlin_distributed_2024} (\href{https://github.com/omlins/ParallelStencil.jl}{https://github.com/omlins/ParallelStencil.jl}) on which the impressive wave equation solver \texttt{SeismicWaves.jl}~\cite{aloisi_seimicwavesjl_2024} (\href{https://github.com/GinvLab/SeismicWaves.jl}{https://github.com/GinvLab/SeismicWaves.jl}) is built.} or potentially larger memory capacities of future GPU architectures\footnote{Driven by the ever longer context-sizes of large language models~\cite{vaswani_attention_2017,sevilla_compute_2022}, GPU memory sizes are expected to increase drastically, as seen by the current NVIDIA$^{\text{\textregistered}}$ H200 (141 GB) and NVIDIA$^{\text{\textregistered}}$ Blackwell Ultra (288 GB), but also the announced NVIDIA$^{\text{\textregistered}}$ Rubin Ultra (1 TB). This would correspondingly expand the possible degrees of freedom from $1.5\cdot 10^9$ to ${\sim}8.8\cdot 10^9$ (141 GB), $1.8 \cdot 10^{10}$ and ${\sim} 6.3 \cdot 10^{10}$ (1 TB).}.\\

Beyond the direct applicability in standard dynamic optimization, the memory-efficient sensitivity approximation is useful in the so-called neural reparametrizations --- also known as neural topology optimization (see~\cite{xu_neural_2019,berg_neural_2021} for inverse problems in general,~\cite{zhu_integrating_2022,herrmann_use_2023,jiang_full_2024,singh_accelerating_2025,liu_deep_2025,norder_pentagonal_2025} for FWI, ~\cite{hoyer_neural_2019,deng_topology_2020,chen_new_2021,chandrasekhar_tounn_2021,chandrasekhar_multi-material_2021,chandrasekhar_approximate_2022,mallon_neural_2024,sanu_neural_2024,kuszczak_meta-neural_2025} for topology optimization\footnote{Note that the memory-efficient sensitivity computation is only helpful for transient optimization problems.} in general, and~\cite{herrmann_neural_2024} for acoustic topology optimization), where neural networks are employed to reparametrize design variables. Neural networks are preferably implemented on GPUs, where the memory capacity is a limitation. Thus, GPU-based forward solvers and memory-efficient sensitivity computations are of practical relevance to these techniques.
Finally, we expect that our approach can be combined with random boundaries proposed in \cite{Clapp2009, Shen2015} to mimic absorbing boundary conditions.

\section*{Acknowledgements}
The authors gratefully acknowledge the funding provided by the Deutsche Forschungsgemeinschaft under Project no. 438252876, Grant KO 4570/1-2 which supports Leon Herrmann and RA 624/29-2, which supports Tim Bürchner. Furthermore, we thank the Georg Nemetschek Institut (GNI) for the support provided in the joint research project DeepMonitor.

\section*{Declarations}
\textbf{Conflict of interest} No potential conflict of interest was reported by
the authors.\\

\section*{Replication of Results}
We provide a PyTorch and CUDA$^{\text{\textregistered}}$ implementation of the finite difference solver and the memory-efficient adjoint sensitivity method in~\cite{herrmann_memory_2025}. Together with the CNN implementation from~\cite{herrmann_use_2024}, all results can be replicated.

\renewcommand\thesection{\Alph{section}}
\setcounter{section}{0}




\section{Finite Difference Stencils as Convolutional Kernels}\label{appendix:conv}


The finite difference stencils from \Cref{eq:FDstencil1,eq:FDstencil2} can be rewritten in terms of convolutions. Consider the wave equation discretization from \Cref{eq:FDstencil1}, rewritten as
\begin{equation}
    \begin{split}
        \boldsymbol{u}^{n+1}=&-\boldsymbol{u}^{n-1}+2\boldsymbol{u}^{n}\\
        &+\frac{2}{\boldsymbol{\gamma}}\Bigg( \frac{1}{\frac{1}{\boldsymbol{\gamma}}*\boldsymbol{K}_{\gamma_{0}}}(\boldsymbol{u}^{n}*\boldsymbol{K}_{u_{0}})-\frac{1}{\frac{1}{\boldsymbol{\gamma}}*\boldsymbol{K}_{\gamma_{1}}}(\boldsymbol{u}^{n}*\boldsymbol{K}_{u_{1}})\\
        &+\frac{\Delta t^2}{2 \rho_0}\boldsymbol{f}^{n}\Bigg),
    \end{split}\label{eq:convolutionalFDstencil}
\end{equation}
where the corresponding kernels are
\begin{align*}
\boldsymbol{K}_{\gamma_{0}}&=\begin{pmatrix}
0 & 1 & 1
\end{pmatrix},\\
\boldsymbol{K}_{u_{0}}&=\begin{pmatrix}
0 & -\left(\frac{c_0\Delta t}{\Delta x}\right)^2 & \left(\frac{c_0\Delta t}{\Delta x}\right)^2 
\end{pmatrix}, \\
\boldsymbol{K}_{\gamma_{1}}&=\begin{pmatrix}
1 & 1 & 0 
\end{pmatrix},\\
\boldsymbol{K}_{u_{1}}&=\begin{pmatrix}
-\left(\frac{c_0\Delta t}{\Delta x}\right)^2 & \left(\frac{c_0\Delta t}{\Delta x}\right)^2 & 0 
\end{pmatrix}.
\end{align*}
These kernels act on the solution, force, and indicator vectors
\begin{equation}
    \begin{split}
        \boldsymbol{u}^n&=\begin{pmatrix} 
        u_{i-1}^n, u_i^n, u_{i+1}^n
        \end{pmatrix}, \\
        \boldsymbol{f}^n&=\begin{pmatrix}
        f_{i-1}^n, f_i^n, f_{i+1}^n
        \end{pmatrix}, \\
        \boldsymbol{\gamma}&=\begin{pmatrix}
        \gamma_{i-1}, \gamma_i, \gamma_{i+1}
        \end{pmatrix}.
    \end{split}
\end{equation}
Through the formulation through convolutions in \Cref{eq:convolutionalFDstencil}, the forward solver can be implemented as a recurrent CNN, as outlined in detail in~\cite{herrmann_use_2023} with a corresponding code available at~\cite{herrmann_use_2024}.

\section{Designs with Filter Radius 12}\label{appendix:filter}
\begin{figure}[htbp]
    \centering
    \begin{subfigure}[t]{0.49\textwidth}
        \includegraphics[width=\textwidth]{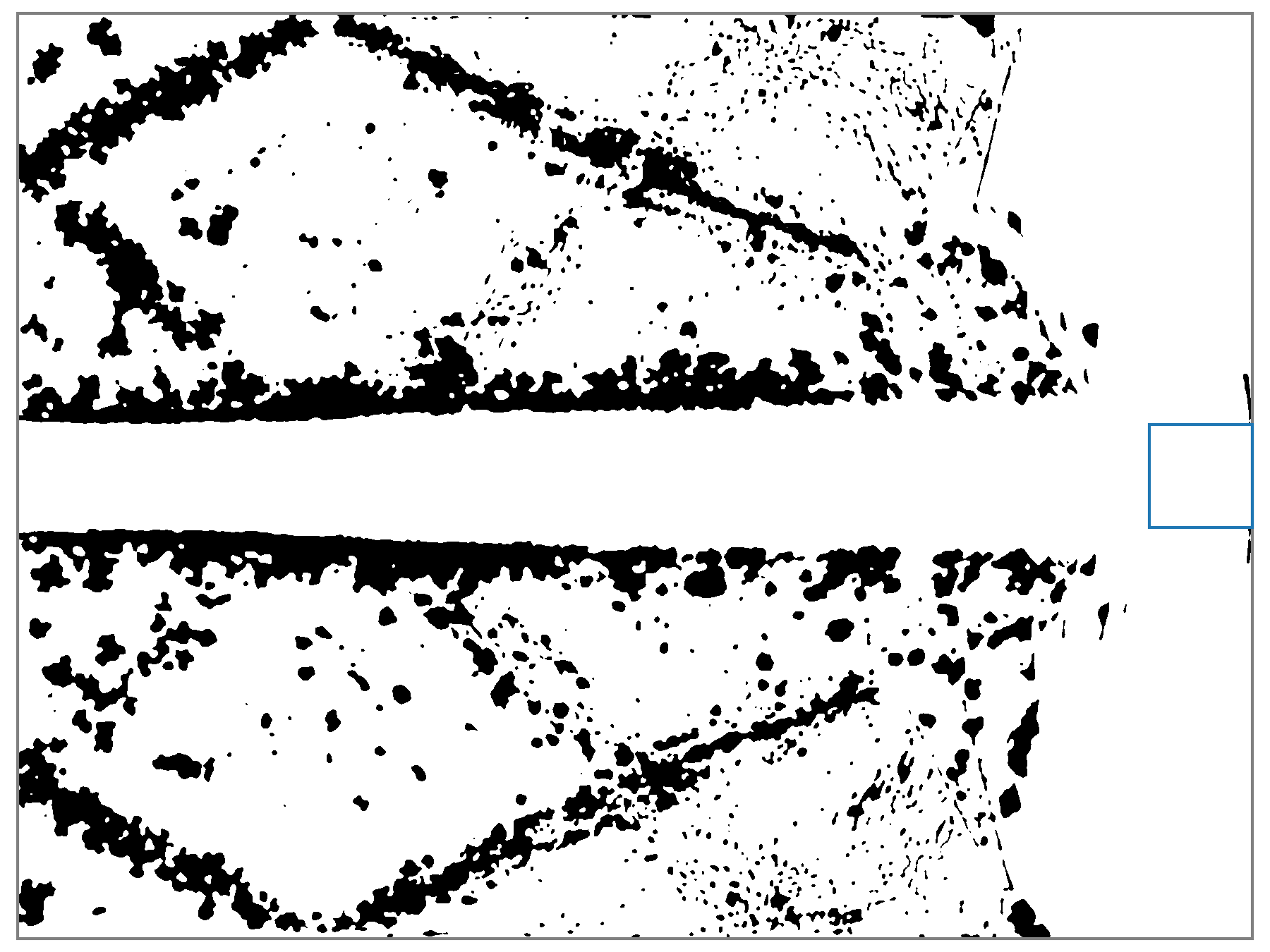}
        \caption{suppression}\label{fig:designs3}
    \end{subfigure}
    \hfill
    \begin{subfigure}[t]{0.49\textwidth}
        \includegraphics[width=\textwidth]{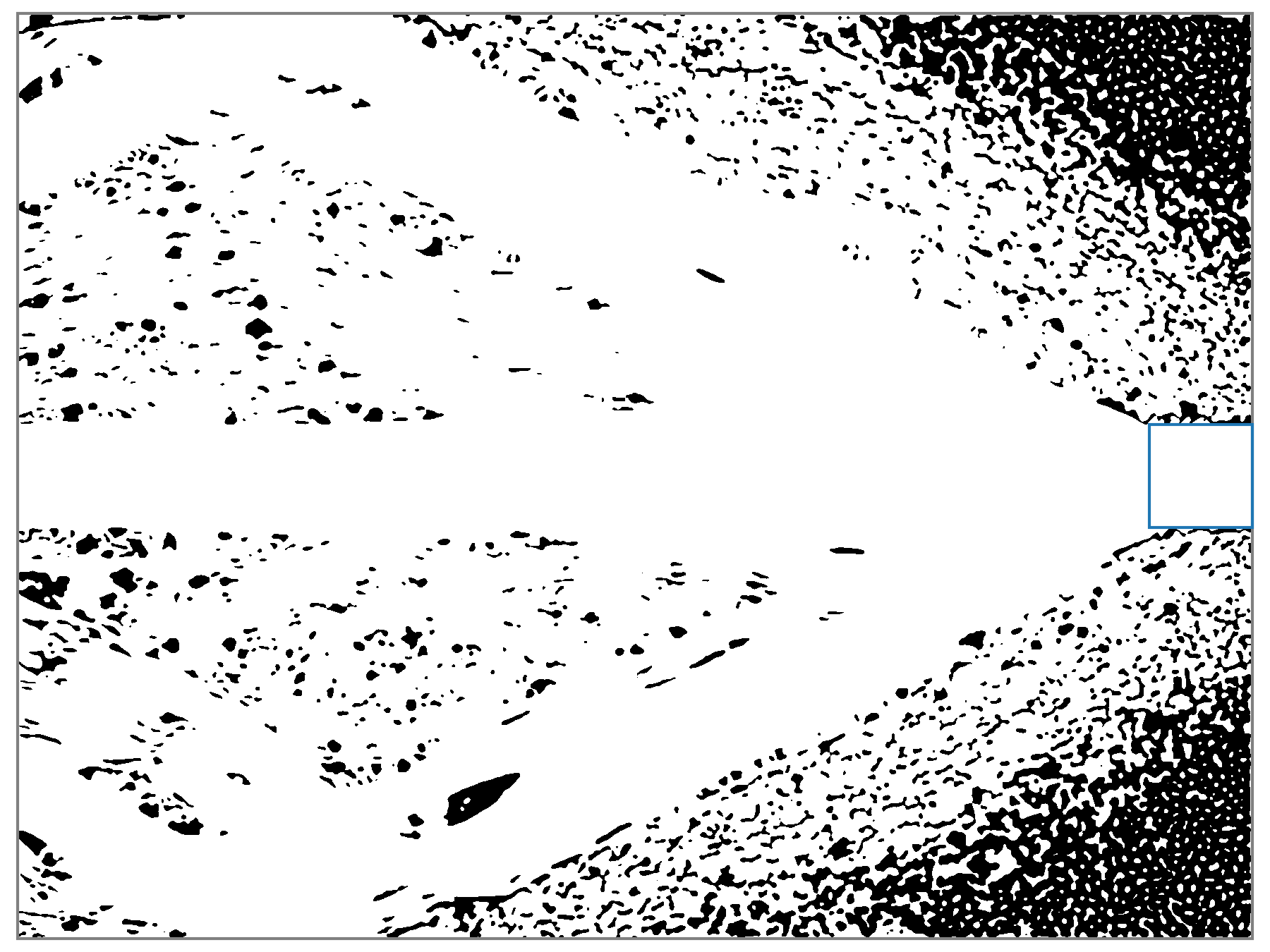}
        \caption{amplification}\label{fig:designs4}
    \end{subfigure}
    \caption{Optimized designs  (with filter radius 12) suppressing (left) or amplifying (right) the acoustic pressures inside the blue square ($\Omega_s$)}\label{fig:designs12}
\end{figure}

\begin{figure}[htbp]
    \centering
    \begin{subfigure}[t]{0.53\textwidth}
        \includegraphics[width=\textwidth]{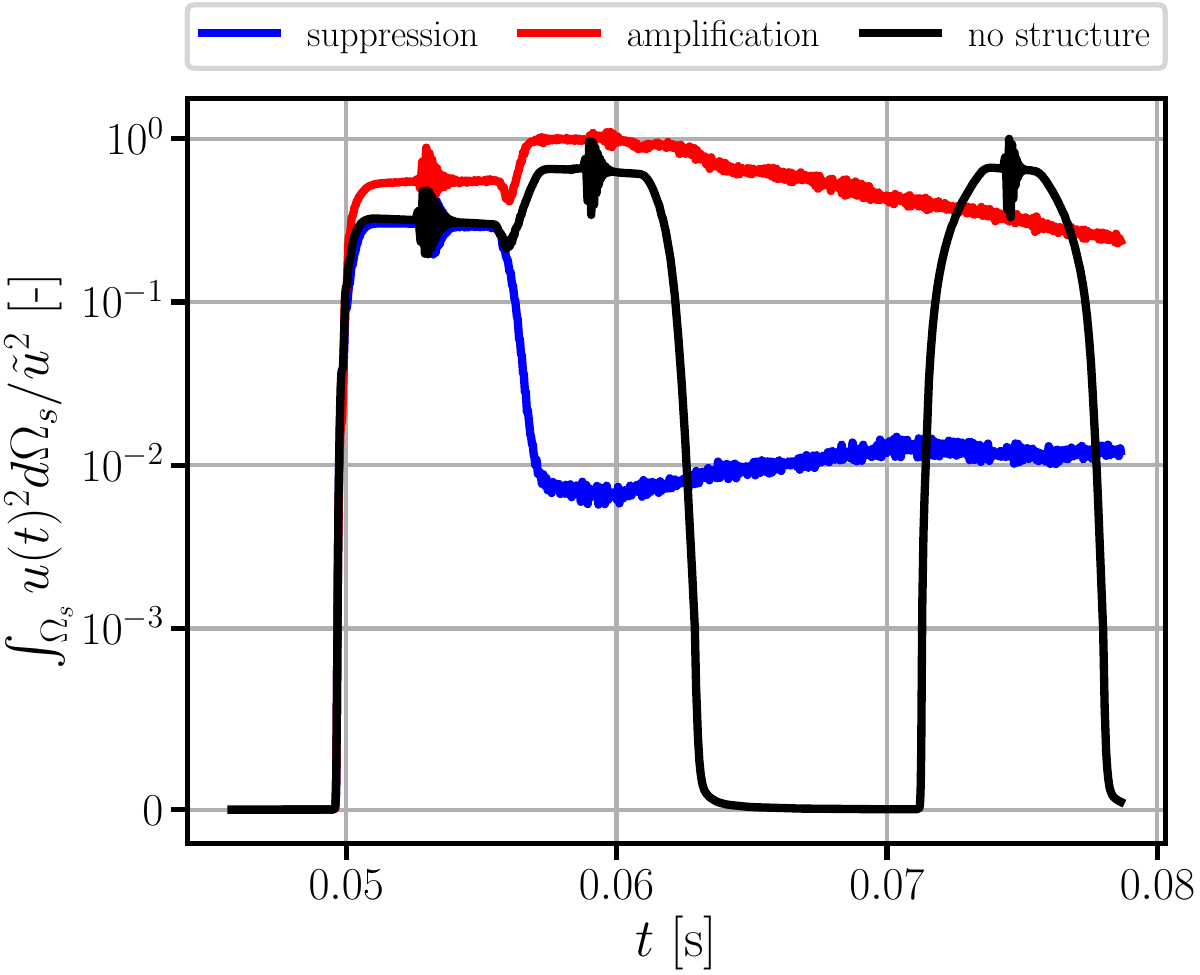}
    \end{subfigure}
    \caption{Transient response of integrated acoustic pressures in the domain of interest (blue domain in \Cref{fig:setupTopOpt,fig:designs12}). The pressures are normalized by the maximum pressure $\tilde{u}$ reached by the transient response without a structure (\Cref{fig:TopOptLarge2,fig:TopOptLarge22,fig:TopOptLarge23,fig:TopOptLarge24}). In contrast to \Cref{fig:transientresponse12}, the designs are filtered with a radius of 12.}\label{fig:transientresponse12}
\end{figure}

Performing the same optimization used for the designs in \Cref{fig:designs}, but with a filter radius of 12 instead of 1.5, yields the structures in \Cref{fig:designs12}\footnote{The optimization in the amplification relies on $\alpha=0.2$ instead of $\alpha=0.1$. Overall, the optimization in TATO is challenging and needs to be investigated further with different optimization tools (at smaller scales, i.e., with fewer degrees of freedom).}. Corresponding transient responses --- less pronounced than the ones in \Cref{fig:transientresponse} --- are given by \Cref{fig:transientresponse12}.

\newpage

\bibliographystyle{unsrtnat}

\setlength{\bibsep}{3pt}
\setlength{\bibhang}{0.75cm}{\fontsize{9}{9}\selectfont\bibliography{2025_gpuacceleratedinversion}}

\begin{thebibliography}{105}
\providecommand{\natexlab}[1]{#1}
\providecommand{\url}[1]{\texttt{#1}}
\expandafter\ifx\csname urlstyle\endcsname\relax
  \providecommand{\doi}[1]{doi: #1}\else
  \providecommand{\doi}{doi: \begingroup \urlstyle{rm}\Url}\fi

\bibitem[Michéa and Komatitsch(2010)]{michea_accelerating_2010}
David Michéa and Dimitri Komatitsch.
\newblock Accelerating a three-dimensional finite-difference wave propagation
  code using {GPU} graphics cards: {Accelerating} a wave propagation code using
  {GPUs}.
\newblock \emph{Geophysical Journal International}, pages no--no, May 2010.
\newblock ISSN 0956540X, 1365246X.
\newblock \doi{10.1111/j.1365-246X.2010.04616.x}.
\newblock URL
  \url{https://academic.oup.com/gji/article-lookup/doi/10.1111/j.1365-246X.2010.04616.x}.

\bibitem[Hamilton and Webb(2013)]{hamilton_room_2013}
Brian Hamilton and Craig~J. Webb.
\newblock Room acoustics modelling using {GPU}-accelerated finite difference
  and finite volume methods on a face-centered cubic grid.
\newblock June 2013.
\newblock URL \url{https://hgpu.org/?p=9655}.

\bibitem[Ye et~al.(2022)Ye, Zhang, Wan, Yan, and Sun]{ye_accelerating_2022}
Chuang-Chao Ye, Peng-Jun-Yi Zhang, Zhen-Hua Wan, Rui Yan, and De-Jun Sun.
\newblock Accelerating {CFD} simulation with high order finite difference
  method on curvilinear coordinates for modern {GPU} clusters.
\newblock \emph{Advances in Aerodynamics}, 4\penalty0 (1):\penalty0 7, February
  2022.
\newblock ISSN 2524-6992.
\newblock \doi{10.1186/s42774-021-00098-3}.
\newblock URL \url{https://doi.org/10.1186/s42774-021-00098-3}.

\bibitem[Schweiger(2011)]{schweiger_gpu-accelerated_2011}
Martin Schweiger.
\newblock {GPU}-{Accelerated} {Finite} {Element} {Method} for {Modelling}
  {Light} {Transport} in {Diffuse} {Optical} {Tomography}.
\newblock \emph{International Journal of Biomedical Imaging}, 2011:\penalty0
  403892, 2011.
\newblock ISSN 1687-4196.
\newblock \doi{10.1155/2011/403892}.

\bibitem[Träff et~al.(2023)Träff, Rydahl, Karlsson, Sigmund, and
  Aage]{traff_simple_2023}
Erik~A. Träff, Anton Rydahl, Sven Karlsson, Ole Sigmund, and Niels Aage.
\newblock Simple and efficient {GPU} accelerated topology optimisation: {Codes}
  and applications.
\newblock \emph{Computer Methods in Applied Mechanics and Engineering},
  410:\penalty0 116043, May 2023.
\newblock ISSN 0045-7825.
\newblock \doi{10.1016/j.cma.2023.116043}.
\newblock URL
  \url{https://www.sciencedirect.com/science/article/pii/S0045782523001676}.

\bibitem[Afanasiev et~al.(2019)Afanasiev, Boehm, van Driel, Krischer,
  Rietmann, May, Knepley, and Fichtner]{afanasiev_modular_2019}
Michael Afanasiev, Christian Boehm, Martin van Driel, Lion Krischer, Max
  Rietmann, Dave~A May, Matthew~G Knepley, and Andreas Fichtner.
\newblock Modular and flexible spectral-element waveform modelling in two and
  three dimensions.
\newblock \emph{Geophysical Journal International}, 216\penalty0 (3):\penalty0
  1675--1692, March 2019.
\newblock ISSN 0956-540X, 1365-246X.
\newblock \doi{10.1093/gji/ggy469}.
\newblock URL \url{https://academic.oup.com/gji/article/216/3/1675/5174970}.

\bibitem[Xu et~al.(2016)Xu, Fu, Gan, Yang, Xue, Xu, Zhao, Wang, Chen, and
  Yang]{xu_generalized_2016}
Jingheng Xu, Haohuan Fu, Lin Gan, Chao Yang, Wei Xue, Shizhen Xu, Wenlai Zhao,
  Xinliang Wang, Bingwei Chen, and Guangwen Yang.
\newblock Generalized {GPU} {Acceleration} for {Applications} {Employing}
  {Finite}-{Volume} {Methods}.
\newblock In \emph{2016 16th {IEEE}/{ACM} {International} {Symposium} on
  {Cluster}, {Cloud} and {Grid} {Computing} ({CCGrid})}, pages 126--135, May
  2016.
\newblock \doi{10.1109/CCGrid.2016.30}.
\newblock URL \url{https://ieeexplore.ieee.org/document/7515679}.

\bibitem[Goodfellow et~al.(2016)Goodfellow, Bengio, and
  Courville]{goodfellow_deep_2016}
Ian Goodfellow, Yoshua Bengio, and Aaron Courville.
\newblock \emph{Deep {Learning}}.
\newblock MIT Press, 2016.
\newblock ISBN 0-262-03561-8.
\newblock URL \url{http://www.deeplearningbook.org}.

\bibitem[Bishop and Bishop(2024)]{bishop_deep_2024}
Christopher~M. Bishop and Hugh Bishop.
\newblock \emph{Deep {Learning}: {Foundations} and {Concepts}}.
\newblock Springer International Publishing, Cham, 2024.
\newblock ISBN 978-3-031-45467-7 978-3-031-45468-4.
\newblock \doi{10.1007/978-3-031-45468-4}.
\newblock URL \url{https://link.springer.com/10.1007/978-3-031-45468-4}.

\bibitem[Abadi et~al.(2016)Abadi, Agarwal, Barham, Brevdo, Chen, Citro,
  Corrado, Davis, Dean, Devin, Ghemawat, Goodfellow, Harp, Irving, Isard, Jia,
  Jozefowicz, Kaiser, Kudlur, Levenberg, Mane, Monga, Moore, Murray, Olah,
  Schuster, Shlens, Steiner, Sutskever, Talwar, Tucker, Vanhoucke, Vasudevan,
  Viegas, Vinyals, Warden, Wattenberg, Wicke, Yu, and
  Zheng]{abadi_tensorflow_2016}
Martín Abadi, Ashish Agarwal, Paul Barham, Eugene Brevdo, Zhifeng Chen, Craig
  Citro, Greg~S. Corrado, Andy Davis, Jeffrey Dean, Matthieu Devin, Sanjay
  Ghemawat, Ian Goodfellow, Andrew Harp, Geoffrey Irving, Michael Isard,
  Yangqing Jia, Rafal Jozefowicz, Lukasz Kaiser, Manjunath Kudlur, Josh
  Levenberg, Dan Mane, Rajat Monga, Sherry Moore, Derek Murray, Chris Olah,
  Mike Schuster, Jonathon Shlens, Benoit Steiner, Ilya Sutskever, Kunal Talwar,
  Paul Tucker, Vincent Vanhoucke, Vijay Vasudevan, Fernanda Viegas, Oriol
  Vinyals, Pete Warden, Martin Wattenberg, Martin Wicke, Yuan Yu, and Xiaoqiang
  Zheng.
\newblock {TensorFlow}: {Large}-{Scale} {Machine} {Learning} on {Heterogeneous}
  {Distributed} {Systems}, 2016.
\newblock URL \url{https://arxiv.org/abs/1603.04467}.

\bibitem[Bradbury et~al.(2018)Bradbury, Frostig, Hawkins, Johnson, Leary,
  Maclaurin, Necula, Paszke, VanderPlas, Wanderman-Milne, and
  Zhang]{bradbury_jax_2018}
James Bradbury, Roy Frostig, Peter Hawkins, Matthew~James Johnson, Chris Leary,
  Dougal Maclaurin, George Necula, Adam Paszke, Jake VanderPlas, Skye
  Wanderman-Milne, and Qiao Zhang.
\newblock {JAX}: composable transformations of {Python}+{NumPy} programs, 2018.
\newblock URL \url{http://github.com/google/jax}.

\bibitem[Paszke et~al.(2019)Paszke, Gross, Massa, Lerer, Bradbury, Chanan,
  Killeen, Lin, Gimelshein, Antiga, Desmaison, Köpf, Yang, DeVito, Raison,
  Tejani, Chilamkurthy, Steiner, Fang, Bai, and Chintala]{paszke_pytorch_2019}
Adam Paszke, Sam Gross, Francisco Massa, Adam Lerer, James Bradbury, Gregory
  Chanan, Trevor Killeen, Zeming Lin, Natalia Gimelshein, Luca Antiga, Alban
  Desmaison, Andreas Köpf, Edward Yang, Zach DeVito, Martin Raison, Alykhan
  Tejani, Sasank Chilamkurthy, Benoit Steiner, Lu~Fang, Junjie Bai, and Soumith
  Chintala.
\newblock {PyTorch}: {An} {Imperative} {Style}, {High}-{Performance} {Deep}
  {Learning} {Library}, December 2019.
\newblock URL \url{http://arxiv.org/abs/1912.01703}.

\bibitem[Kollmannsberger et~al.(2021)Kollmannsberger, D'Angella, Jokeit, and
  Herrmann]{kollmannsberger_deep_2021}
Stefan Kollmannsberger, Davide D'Angella, Moritz Jokeit, and Leon Herrmann.
\newblock \emph{Deep {Learning} in {Computational} {Mechanics}: {An}
  {Introductory} {Course}}, volume 977 of \emph{Studies in {Computational}
  {Intelligence}}.
\newblock Springer International Publishing, Cham, 2021.
\newblock ISBN 978-3-030-76586-6 978-3-030-76587-3.
\newblock \doi{10.1007/978-3-030-76587-3}.
\newblock URL \url{https://link.springer.com/10.1007/978-3-030-76587-3}.

\bibitem[Herrmann and Kollmannsberger(2024)]{herrmann_deep_2024}
Leon Herrmann and Stefan Kollmannsberger.
\newblock Deep learning in computational mechanics: a review.
\newblock \emph{Computational Mechanics}, January 2024.
\newblock ISSN 0178-7675, 1432-0924.
\newblock \doi{10.1007/s00466-023-02434-4}.
\newblock URL \url{https://link.springer.com/10.1007/s00466-023-02434-4}.

\bibitem[Takeuchi and Kosugi(1994)]{takeuchi_neural_1994}
Jun Takeuchi and Yukio Kosugi.
\newblock Neural network representation of finite element method.
\newblock \emph{Neural Networks}, 7\penalty0 (2):\penalty0 389--395, January
  1994.
\newblock ISSN 0893-6080.
\newblock \doi{10.1016/0893-6080(94)90031-0}.
\newblock URL
  \url{https://www.sciencedirect.com/science/article/pii/0893608094900310}.

\bibitem[Ramuhalli et~al.(2005)Ramuhalli, Udpa, and
  Udpa]{ramuhalli_finite-element_2005}
P.~Ramuhalli, L.~Udpa, and S.S. Udpa.
\newblock Finite-element neural networks for solving differential equations.
\newblock \emph{IEEE Transactions on Neural Networks}, 16\penalty0
  (6):\penalty0 1381--1392, November 2005.
\newblock ISSN 1941-0093.
\newblock \doi{10.1109/TNN.2005.857945}.
\newblock URL \url{https://ieeexplore.ieee.org/document/1528518}.

\bibitem[Yao et~al.(2019)Yao, Ren, and Liu]{yao_fea-net_2019}
Houpu Yao, Yi~Ren, and Yongming Liu.
\newblock {FEA}-{Net}: {A} {Deep} {Convolutional} {Neural} {Network} {With}
  {PhysicsPrior} {For} {Efficient} {Data} {Driven} {PDE} {Learning}.
\newblock In \emph{{AIAA} {Scitech} 2019 {Forum}}, San Diego, California,
  January 2019. American Institute of Aeronautics and Astronautics.
\newblock ISBN 978-1-62410-578-4.
\newblock \doi{10.2514/6.2019-0680}.
\newblock URL \url{https://arc.aiaa.org/doi/10.2514/6.2019-0680}.

\bibitem[Yao et~al.(2020)Yao, Gao, and Liu]{yao_fea-net_2020}
Houpu Yao, Yi~Gao, and Yongming Liu.
\newblock {FEA}-{Net}: {A} physics-guided data-driven model for efficient
  mechanical response prediction.
\newblock \emph{Computer Methods in Applied Mechanics and Engineering},
  363:\penalty0 112892, May 2020.
\newblock ISSN 0045-7825.
\newblock \doi{10.1016/j.cma.2020.112892}.
\newblock URL
  \url{https://www.sciencedirect.com/science/article/pii/S0045782520300748}.

\bibitem[Park et~al.(2023)Park, Lu, Saha, Xue, Guo, Mojumder, Apley, Wagner,
  and Liu]{park_convolution_2023}
Chanwook Park, Ye~Lu, Sourav Saha, Tianju Xue, Jiachen Guo, Satyajit Mojumder,
  Daniel~W. Apley, Gregory~J. Wagner, and Wing~Kam Liu.
\newblock Convolution hierarchical deep-learning neural network ({C}-{HiDeNN})
  with graphics processing unit ({GPU}) acceleration.
\newblock \emph{Computational Mechanics}, 72\penalty0 (2):\penalty0 383--409,
  August 2023.
\newblock ISSN 1432-0924.
\newblock \doi{10.1007/s00466-023-02329-4}.
\newblock URL \url{https://doi.org/10.1007/s00466-023-02329-4}.

\bibitem[Mishra and Hall(2005)]{mishra_nfdtd_2005}
R.K. Mishra and P.S. Hall.
\newblock {NFDTD} concept.
\newblock \emph{IEEE Transactions on Neural Networks}, 16\penalty0
  (2):\penalty0 484--490, March 2005.
\newblock ISSN 1941-0093.
\newblock \doi{10.1109/TNN.2004.841799}.
\newblock URL \url{https://ieeexplore.ieee.org/document/1402508}.

\bibitem[Richardson(2018)]{richardson_seismic_2018}
Alan Richardson.
\newblock Seismic {Full}-{Waveform} {Inversion} {Using} {Deep} {Learning}
  {Tools} and {Techniques}, January 2018.
\newblock URL \url{http://arxiv.org/abs/1801.07232}.
\newblock arXiv:1801.07232.

\bibitem[Sun et~al.(2020)Sun, Niu, Innanen, Li, and
  Trad]{sun_theory-guided_2020}
Jian Sun, Zhan Niu, Kristopher~A. Innanen, Junxiao Li, and Daniel~O. Trad.
\newblock A theory-guided deep-learning formulation and optimization of seismic
  waveform inversion.
\newblock \emph{GEOPHYSICS}, 85\penalty0 (2):\penalty0 R87--R99, March 2020.
\newblock ISSN 0016-8033, 1942-2156.
\newblock \doi{10.1190/geo2019-0138.1}.
\newblock URL \url{https://library.seg.org/doi/10.1190/geo2019-0138.1}.

\bibitem[Herrmann et~al.(2023)Herrmann, Bürchner, Dietrich, and
  Kollmannsberger]{herrmann_use_2023}
Leon Herrmann, Tim Bürchner, Felix Dietrich, and Stefan Kollmannsberger.
\newblock On the use of neural networks for full waveform inversion.
\newblock \emph{Computer Methods in Applied Mechanics and Engineering},
  415:\penalty0 116278, October 2023.
\newblock ISSN 00457825.
\newblock \doi{10.1016/j.cma.2023.116278}.
\newblock URL
  \url{https://linkinghub.elsevier.com/retrieve/pii/S0045782523004024}.

\bibitem[Plessix(2006)]{plessix_review_2006}
R.-E. Plessix.
\newblock A review of the adjoint-state method for computing the gradient of a
  functional with geophysical applications.
\newblock \emph{Geophysical Journal International}, 167\penalty0 (2):\penalty0
  495--503, November 2006.
\newblock ISSN 0956540X, 1365246X.
\newblock \doi{10.1111/j.1365-246X.2006.02978.x}.
\newblock URL
  \url{https://academic.oup.com/gji/article-lookup/doi/10.1111/j.1365-246X.2006.02978.x}.

\bibitem[Fichtner et~al.(2006{\natexlab{a}})Fichtner, Bunge, and
  Igel]{Fichtner2006a}
A.~Fichtner, H.-P. Bunge, and H.~Igel.
\newblock The adjoint method in seismology: I. theory.
\newblock \emph{Physics of the Earth and Planetary Interiors}, 157\penalty0
  (1):\penalty0 86--104, 2006{\natexlab{a}}.
\newblock ISSN 0031-9201.
\newblock \doi{https://doi.org/10.1016/j.pepi.2006.03.016}.
\newblock URL
  \url{https://www.sciencedirect.com/science/article/pii/S0031920106001051}.

\bibitem[Fichtner et~al.(2006{\natexlab{b}})Fichtner, Bunge, and
  Igel]{Fichtner2006b}
A.~Fichtner, H.-P. Bunge, and H.~Igel.
\newblock The adjoint method in seismology—: Ii. applications: traveltimes
  and sensitivity functionals.
\newblock \emph{Physics of the Earth and Planetary Interiors}, 157\penalty0
  (1):\penalty0 105--123, 2006{\natexlab{b}}.
\newblock ISSN 0031-9201.
\newblock \doi{https://doi.org/10.1016/j.pepi.2006.03.018}.
\newblock URL
  \url{https://www.sciencedirect.com/science/article/pii/S0031920106001038}.

\bibitem[Givoli(2021)]{givoli_tutorial_2021}
Dan Givoli.
\newblock A tutorial on the adjoint method for inverse problems.
\newblock \emph{Computer Methods in Applied Mechanics and Engineering},
  380:\penalty0 113810, July 2021.
\newblock ISSN 0045-7825.
\newblock \doi{10.1016/j.cma.2021.113810}.
\newblock URL
  \url{https://www.sciencedirect.com/science/article/pii/S0045782521001468}.

\bibitem[Griewank(1992)]{griewank_achieving_1992}
Andreas Griewank.
\newblock Achieving logarithmic growth of temporal and spatial complexity in
  reverse automatic differentiation.
\newblock \emph{Optimization Methods and Software}, 1\penalty0 (1):\penalty0
  35--54, January 1992.
\newblock ISSN 1055-6788, 1029-4937.
\newblock \doi{10.1080/10556789208805505}.
\newblock URL
  \url{https://www.tandfonline.com/doi/full/10.1080/10556789208805505}.

\bibitem[Griewank and Walther(2000)]{griewank_algorithm_2000}
Andreas Griewank and Andrea Walther.
\newblock Algorithm 799: revolve: an implementation of checkpointing for the
  reverse or adjoint mode of computational differentiation.
\newblock \emph{ACM Transactions on Mathematical Software}, 26\penalty0
  (1):\penalty0 19--45, March 2000.
\newblock ISSN 0098-3500, 1557-7295.
\newblock \doi{10.1145/347837.347846}.
\newblock URL \url{https://dl.acm.org/doi/10.1145/347837.347846}.

\bibitem[Symes(2007)]{symes_reverse_2007}
William~W. Symes.
\newblock Reverse time migration with optimal checkpointing.
\newblock \emph{GEOPHYSICS}, 72\penalty0 (5):\penalty0 SM213--SM221, September
  2007.
\newblock ISSN 0016-8033, 1942-2156.
\newblock \doi{10.1190/1.2742686}.
\newblock URL \url{https://library.seg.org/doi/10.1190/1.2742686}.

\bibitem[Anderson et~al.(2012)Anderson, Tan, and
  Wang]{anderson_time-reversal_2012}
John~E. Anderson, Lijian Tan, and Don Wang.
\newblock Time-reversal checkpointing methods for {RTM} and {FWI}.
\newblock \emph{GEOPHYSICS}, 77\penalty0 (4):\penalty0 S93--S103, July 2012.
\newblock ISSN 0016-8033, 1942-2156.
\newblock \doi{10.1190/geo2011-0114.1}.
\newblock URL \url{https://library.seg.org/doi/10.1190/geo2011-0114.1}.

\bibitem[Boehm et~al.(2016)Boehm, Hanzich, de~la Puente, and
  Fichtner]{Boehm2016}
Christian Boehm, Mauricio Hanzich, Josep de~la Puente, and Andreas Fichtner.
\newblock Wavefield compression for adjoint methods in full-waveform inversion.
\newblock \emph{GEOPHYSICS}, 81\penalty0 (6):\penalty0 R385--R397, 2016.
\newblock \doi{10.1190/geo2015-0653.1}.
\newblock URL \url{https://doi.org/10.1190/geo2015-0653.1}.

\bibitem[Silva et~al.(2019)Silva, Zhang, Kumar, and Herrmann]{Silva2019}
Curt~Da Silva, Yiming Zhang, Rajiv Kumar, and Felix~J. Herrmann.
\newblock Applications of low-rank compressed seismic data to full-waveform
  inversion and extended image volumes.
\newblock \emph{GEOPHYSICS}, 84\penalty0 (3):\penalty0 R371--R383, 2019.
\newblock \doi{10.1190/geo2018-0116.1}.
\newblock URL \url{https://doi.org/10.1190/geo2018-0116.1}.

\bibitem[Omlin and Räss(2024)]{omlin_high-performance_2024}
Samuel Omlin and Ludovic Räss.
\newblock High-performance {xPU} {Stencil} {Computations} in {Julia}.
\newblock \emph{JuliaCon Proceedings}, 6\penalty0 (64):\penalty0 138, October
  2024.
\newblock ISSN 2642-4029.
\newblock \doi{10.21105/jcon.00138}.
\newblock URL
  \url{https://proceedings.juliacon.org/papers/10.21105/jcon.00138}.

\bibitem[Omlin et~al.(2024)Omlin, Räss, and Utkin]{omlin_distributed_2024}
Samuel Omlin, Ludovic Räss, and Ivan Utkin.
\newblock Distributed {Parallelization} of {xPU} {Stencil} {Computations} in
  {Julia}.
\newblock \emph{JuliaCon Proceedings}, 6\penalty0 (65):\penalty0 137, November
  2024.
\newblock ISSN 2642-4029.
\newblock \doi{10.21105/jcon.00137}.
\newblock URL
  \url{https://proceedings.juliacon.org/papers/10.21105/jcon.00137}.

\bibitem[Aloisi et~al.(2024)Aloisi, Zunino, and
  Fichtner]{aloisi_seimicwavesjl_2024}
Giacomo Aloisi, Andrea Zunino, and Andreas Fichtner.
\newblock {SeimicWaves}.jl: an efficient yet user-friendly {Julia} package for
  {Full}-{Waveform} {Inversion} on multi-{xPUs}.
\newblock Technical Report EGU24-10916, Copernicus Meetings, March 2024.
\newblock URL
  \url{https://meetingorganizer.copernicus.org/EGU24/EGU24-10916.html}.

\bibitem[Clapp(2009)]{Clapp2009}
Robert~G. Clapp.
\newblock \emph{Reverse time migration with random boundaries}, pages
  2809--2813.
\newblock 2009.
\newblock \doi{10.1190/1.3255432}.
\newblock URL \url{https://library.seg.org/doi/abs/10.1190/1.3255432}.

\bibitem[Shen and Clapp(2015)]{Shen2015}
Xukai Shen and Robert Clapp.
\newblock Random boundary condition for memory-efficient waveform inversion
  gradient computation.
\newblock \emph{GEOPHYSICS}, 80:\penalty0 R351--R359, 11 2015.
\newblock \doi{10.1190/geo2014-0542.1}.

\bibitem[Tarantola(1984)]{tarantola_inversion_1984}
Albert Tarantola.
\newblock Inversion of seismic reflection data in the acoustic approximation.
\newblock \emph{GEOPHYSICS}, 49\penalty0 (8):\penalty0 1259--1266, August 1984.
\newblock ISSN 0016-8033, 1942-2156.
\newblock \doi{10.1190/1.1441754}.
\newblock URL \url{https://library.seg.org/doi/10.1190/1.1441754}.

\bibitem[Fichtner(2011)]{Fichtner2011}
Andreas Fichtner.
\newblock \emph{Full Seismic Waveform Modelling and Inversion}.
\newblock Advances in Geophysical and Environmental Mechanics and Mathematics.
  Springer Berlin Heidelberg, 2011.
\newblock ISBN 978-3-642-15806-3.
\newblock \doi{10.1007/978-3-642-15807-0}.

\bibitem[Wadbro and Berggren(2006)]{wadbro_topology_2006}
Eddie Wadbro and Martin Berggren.
\newblock Topology optimization of an acoustic horn.
\newblock \emph{Computer Methods in Applied Mechanics and Engineering},
  196\penalty0 (1-3):\penalty0 420--436, December 2006.
\newblock ISSN 00457825.
\newblock \doi{10.1016/j.cma.2006.05.005}.
\newblock URL
  \url{https://linkinghub.elsevier.com/retrieve/pii/S0045782506001745}.

\bibitem[Lee and Kim(2009)]{lee_rigid_2009}
Jin~Woo Lee and Yoon~Young Kim.
\newblock Rigid body modeling issue in acoustical topology optimization.
\newblock \emph{Computer Methods in Applied Mechanics and Engineering},
  198\penalty0 (9-12):\penalty0 1017--1030, February 2009.
\newblock ISSN 00457825.
\newblock \doi{10.1016/j.cma.2008.11.008}.
\newblock URL
  \url{https://linkinghub.elsevier.com/retrieve/pii/S004578250800412X}.

\bibitem[Du and Olhoff(2007)]{du_minimization_2007}
Jianbin Du and Niels Olhoff.
\newblock Minimization of sound radiation from vibrating bi-material structures
  using topology optimization.
\newblock \emph{Structural and Multidisciplinary Optimization}, 33\penalty0
  (4-5):\penalty0 305--321, February 2007.
\newblock ISSN 1615-147X, 1615-1488.
\newblock \doi{10.1007/s00158-006-0088-9}.
\newblock URL \url{http://link.springer.com/10.1007/s00158-006-0088-9}.

\bibitem[Yoon et~al.(2007)Yoon, Jensen, and Sigmund]{yoon_topology_2007}
Gil~Ho Yoon, Jakob~Søndergaard Jensen, and Ole Sigmund.
\newblock Topology optimization of acoustic–structure interaction problems
  using a mixed finite element formulation.
\newblock \emph{International Journal for Numerical Methods in Engineering},
  70\penalty0 (9):\penalty0 1049--1075, May 2007.
\newblock ISSN 0029-5981, 1097-0207.
\newblock \doi{10.1002/nme.1900}.
\newblock URL \url{https://onlinelibrary.wiley.com/doi/10.1002/nme.1900}.

\bibitem[Dühring et~al.(2008)Dühring, Jensen, and
  Sigmund]{duhring_acoustic_2008}
Maria~B. Dühring, Jakob~S. Jensen, and Ole Sigmund.
\newblock Acoustic design by topology optimization.
\newblock \emph{Journal of Sound and Vibration}, 317\penalty0 (3-5):\penalty0
  557--575, November 2008.
\newblock ISSN 0022460X.
\newblock \doi{10.1016/j.jsv.2008.03.042}.
\newblock URL
  \url{https://linkinghub.elsevier.com/retrieve/pii/S0022460X08002812}.

\bibitem[Kook et~al.(2012)Kook, Koo, Hyun, Jensen, and
  Wang]{kook_acoustical_2012}
Junghwan Kook, Kunmo Koo, Jaeyub Hyun, Jakob~S. Jensen, and Semyung Wang.
\newblock Acoustical topology optimization for {Zwicker}’s loudness model –
  {Application} to noise barriers.
\newblock \emph{Computer Methods in Applied Mechanics and Engineering},
  237-240:\penalty0 130--151, September 2012.
\newblock ISSN 00457825.
\newblock \doi{10.1016/j.cma.2012.05.004}.
\newblock URL
  \url{https://linkinghub.elsevier.com/retrieve/pii/S0045782512001521}.

\bibitem[Jensen and Sigmund(2011)]{jensen_topology_2011}
J.S. Jensen and O.~Sigmund.
\newblock Topology optimization for nano‐photonics.
\newblock \emph{Laser \& Photonics Reviews}, 5\penalty0 (2):\penalty0 308--321,
  March 2011.
\newblock ISSN 1863-8880, 1863-8899.
\newblock \doi{10.1002/lpor.201000014}.
\newblock URL \url{https://onlinelibrary.wiley.com/doi/10.1002/lpor.201000014}.

\bibitem[Christiansen and
  Sigmund(2021{\natexlab{a}})]{christiansen_inverse_2021}
Rasmus~E. Christiansen and Ole Sigmund.
\newblock Inverse design in photonics by topology optimization: tutorial.
\newblock \emph{Journal of the Optical Society of America B}, 38\penalty0
  (2):\penalty0 496, February 2021{\natexlab{a}}.
\newblock ISSN 0740-3224, 1520-8540.
\newblock \doi{10.1364/JOSAB.406048}.
\newblock URL \url{https://opg.optica.org/abstract.cfm?URI=josab-38-2-496}.

\bibitem[Christiansen and
  Sigmund(2021{\natexlab{b}})]{christiansen_compact_2021}
Rasmus~E. Christiansen and Ole Sigmund.
\newblock Compact 200 line {MATLAB} code for inverse design in photonics by
  topology optimization: tutorial.
\newblock \emph{Journal of the Optical Society of America B}, 38\penalty0
  (2):\penalty0 510, February 2021{\natexlab{b}}.
\newblock ISSN 0740-3224, 1520-8540.
\newblock \doi{10.1364/JOSAB.405955}.
\newblock URL \url{https://opg.optica.org/abstract.cfm?URI=josab-38-2-510}.

\bibitem[Nickolls et~al.(2008)Nickolls, Buck, Garland, and
  Skadron]{nickolls_scalable_2008}
John Nickolls, Ian Buck, Michael Garland, and Kevin Skadron.
\newblock Scalable parallel programming with {CUDA}.
\newblock In \emph{{ACM} {SIGGRAPH} 2008 classes}, pages 1--14, Los Angeles
  California, August 2008. ACM.
\newblock ISBN 978-1-4503-7845-1.
\newblock \doi{10.1145/1401132.1401152}.
\newblock URL \url{https://dl.acm.org/doi/10.1145/1401132.1401152}.

\bibitem[Herrmann et~al.(2025)Herrmann, Bürchner, Kudela, and
  Kollmannsberger]{herrmann_memory_2025}
Leon Herrmann, Tim Bürchner, László Kudela, and Stefan Kollmannsberger.
\newblock A {Memory} {Efficient} {Adjoint} {Method} to {Enable} {Billion}
  {Parameter} {Optimization} on a {Single} {GPU} in {Dynamic} {Problems}
  [{Software}], September 2025.
\newblock URL \url{https://zenodo.org/doi/10.5281/zenodo.17157434}.

\bibitem[Kingma and Ba(2017)]{kingma_adam_2017}
Diederik~P. Kingma and Jimmy Ba.
\newblock Adam: {A} {Method} for {Stochastic} {Optimization}.
\newblock \emph{arXiv:1412.6980 [cs]}, January 2017.
\newblock URL \url{http://arxiv.org/abs/1412.6980}.

\bibitem[Liu and Nocedal(1989)]{liu_limited_1989}
Dong~C. Liu and Jorge Nocedal.
\newblock On the limited memory {BFGS} method for large scale optimization.
\newblock \emph{Mathematical Programming}, 45\penalty0 (1):\penalty0 503--528,
  August 1989.
\newblock ISSN 1436-4646.
\newblock \doi{10.1007/BF01589116}.
\newblock URL \url{https://doi.org/10.1007/BF01589116}.

\bibitem[Svanberg(1987)]{svanberg_method_1987}
Krister Svanberg.
\newblock The method of moving asymptotes—a new method for structural
  optimization.
\newblock \emph{International Journal for Numerical Methods in Engineering},
  24\penalty0 (2):\penalty0 359--373, February 1987.
\newblock ISSN 0029-5981, 1097-0207.
\newblock \doi{10.1002/nme.1620240207}.
\newblock URL \url{https://onlinelibrary.wiley.com/doi/10.1002/nme.1620240207}.

\bibitem[Virtanen et~al.(2020)Virtanen, Gommers, Oliphant, Haberland, Reddy,
  Cournapeau, Burovski, Peterson, Weckesser, Bright, Van Der~Walt, Brett,
  Wilson, Millman, Mayorov, Nelson, Jones, Kern, Larson, Carey, Polat, Feng,
  Moore, VanderPlas, Laxalde, Perktold, Cimrman, Henriksen, Quintero, Harris,
  Archibald, Ribeiro, Pedregosa, Van~Mulbregt, {SciPy 1.0 Contributors},
  Vijaykumar, Bardelli, Rothberg, Hilboll, Kloeckner, Scopatz, Lee, Rokem,
  Woods, Fulton, Masson, Häggström, Fitzgerald, Nicholson, Hagen, Pasechnik,
  Olivetti, Martin, Wieser, Silva, Lenders, Wilhelm, Young, Price, Ingold,
  Allen, Lee, Audren, Probst, Dietrich, Silterra, Webber, Slavič, Nothman,
  Buchner, Kulick, Schönberger, De~Miranda~Cardoso, Reimer, Harrington,
  Rodríguez, Nunez-Iglesias, Kuczynski, Tritz, Thoma, Newville, Kümmerer,
  Bolingbroke, Tartre, Pak, Smith, Nowaczyk, Shebanov, Pavlyk, Brodtkorb, Lee,
  McGibbon, Feldbauer, Lewis, Tygier, Sievert, Vigna, Peterson, More, Pudlik,
  Oshima, Pingel, Robitaille, Spura, Jones, Cera, Leslie, Zito, Krauss,
  Upadhyay, Halchenko, and Vázquez-Baeza]{virtanen_scipy_2020}
Pauli Virtanen, Ralf Gommers, Travis~E. Oliphant, Matt Haberland, Tyler Reddy,
  David Cournapeau, Evgeni Burovski, Pearu Peterson, Warren Weckesser, Jonathan
  Bright, Stéfan~J. Van Der~Walt, Matthew Brett, Joshua Wilson, K.~Jarrod
  Millman, Nikolay Mayorov, Andrew R.~J. Nelson, Eric Jones, Robert Kern, Eric
  Larson, C~J Carey, İlhan Polat, Yu~Feng, Eric~W. Moore, Jake VanderPlas,
  Denis Laxalde, Josef Perktold, Robert Cimrman, Ian Henriksen, E.~A. Quintero,
  Charles~R. Harris, Anne~M. Archibald, Antônio~H. Ribeiro, Fabian Pedregosa,
  Paul Van~Mulbregt, {SciPy 1.0 Contributors}, Aditya Vijaykumar,
  Alessandro~Pietro Bardelli, Alex Rothberg, Andreas Hilboll, Andreas
  Kloeckner, Anthony Scopatz, Antony Lee, Ariel Rokem, C.~Nathan Woods, Chad
  Fulton, Charles Masson, Christian Häggström, Clark Fitzgerald, David~A.
  Nicholson, David~R. Hagen, Dmitrii~V. Pasechnik, Emanuele Olivetti, Eric
  Martin, Eric Wieser, Fabrice Silva, Felix Lenders, Florian Wilhelm, G.~Young,
  Gavin~A. Price, Gert-Ludwig Ingold, Gregory~E. Allen, Gregory~R. Lee, Hervé
  Audren, Irvin Probst, Jörg~P. Dietrich, Jacob Silterra, James~T Webber,
  Janko Slavič, Joel Nothman, Johannes Buchner, Johannes Kulick, Johannes~L.
  Schönberger, José~Vinícius De~Miranda~Cardoso, Joscha Reimer, Joseph
  Harrington, Juan Luis~Cano Rodríguez, Juan Nunez-Iglesias, Justin Kuczynski,
  Kevin Tritz, Martin Thoma, Matthew Newville, Matthias Kümmerer, Maximilian
  Bolingbroke, Michael Tartre, Mikhail Pak, Nathaniel~J. Smith, Nikolai
  Nowaczyk, Nikolay Shebanov, Oleksandr Pavlyk, Per~A. Brodtkorb, Perry Lee,
  Robert~T. McGibbon, Roman Feldbauer, Sam Lewis, Sam Tygier, Scott Sievert,
  Sebastiano Vigna, Stefan Peterson, Surhud More, Tadeusz Pudlik, Takuya
  Oshima, Thomas~J. Pingel, Thomas~P. Robitaille, Thomas Spura, Thouis~R.
  Jones, Tim Cera, Tim Leslie, Tiziano Zito, Tom Krauss, Utkarsh Upadhyay,
  Yaroslav~O. Halchenko, and Yoshiki Vázquez-Baeza.
\newblock {SciPy} 1.0: fundamental algorithms for scientific computing in
  {Python}.
\newblock \emph{Nature Methods}, 17\penalty0 (3):\penalty0 261--272, March
  2020.
\newblock ISSN 1548-7091, 1548-7105.
\newblock \doi{10.1038/s41592-019-0686-2}.
\newblock URL \url{https://www.nature.com/articles/s41592-019-0686-2}.

\bibitem[Bürchner et~al.(2023{\natexlab{a}})Bürchner, Kopp, Kollmannsberger,
  and Rank]{burchner_immersed_2023}
Tim Bürchner, Philipp Kopp, Stefan Kollmannsberger, and Ernst Rank.
\newblock Immersed boundary parametrizations for full waveform inversion.
\newblock \emph{Computer Methods in Applied Mechanics and Engineering},
  406:\penalty0 115893, March 2023{\natexlab{a}}.
\newblock ISSN 0045-7825.
\newblock \doi{10.1016/j.cma.2023.115893}.
\newblock URL
  \url{https://www.sciencedirect.com/science/article/pii/S0045782523000166}.

\bibitem[Fichtner et~al.(2006{\natexlab{c}})Fichtner, Bunge, and
  Igel]{fichtner_adjoint_2006-1}
A.~Fichtner, H.~P. Bunge, and H.~Igel.
\newblock The adjoint method in seismology: {I}. {Theory}.
\newblock \emph{Physics of the Earth and Planetary Interiors}, 157\penalty0
  (1):\penalty0 86--104, August 2006{\natexlab{c}}.
\newblock ISSN 0031-9201.
\newblock \doi{10.1016/j.pepi.2006.03.016}.
\newblock URL
  \url{https://www.sciencedirect.com/science/article/pii/S0031920106001051}.

\bibitem[Fichtner et~al.(2006{\natexlab{d}})Fichtner, Bunge, and
  Igel]{fichtner_adjoint_2006}
A.~Fichtner, H.~P. Bunge, and H.~Igel.
\newblock The adjoint method in seismology—: {II}. {Applications}:
  traveltimes and sensitivity functionals.
\newblock \emph{Physics of the Earth and Planetary Interiors}, 157\penalty0
  (1):\penalty0 105--123, August 2006{\natexlab{d}}.
\newblock ISSN 0031-9201.
\newblock \doi{10.1016/j.pepi.2006.03.018}.
\newblock URL
  \url{https://www.sciencedirect.com/science/article/pii/S0031920106001038}.

\bibitem[Bürchner et~al.(2023{\natexlab{b}})Bürchner, Kopp, Kollmannsberger,
  and Rank]{burchner_isogeometric_2023}
Tim Bürchner, Philipp Kopp, Stefan Kollmannsberger, and Ernst Rank.
\newblock Isogeometric multi-resolution full waveform inversion based on the
  finite cell method.
\newblock \emph{Computer Methods in Applied Mechanics and Engineering},
  417:\penalty0 116286, December 2023{\natexlab{b}}.
\newblock ISSN 0045-7825.
\newblock \doi{10.1016/j.cma.2023.116286}.
\newblock URL
  \url{https://www.sciencedirect.com/science/article/pii/S0045782523004103}.

\bibitem[Arridge et~al.(2019)Arridge, Maass, Öktem, and
  Schönlieb]{arridge_solving_2019}
Simon Arridge, Peter Maass, Ozan Öktem, and Carola-Bibiane Schönlieb.
\newblock Solving inverse problems using data-driven models.
\newblock \emph{Acta Numerica}, 28:\penalty0 1--174, May 2019.
\newblock ISSN 0962-4929, 1474-0508.
\newblock \doi{10.1017/S0962492919000059}.
\newblock URL
  \url{https://www.cambridge.org/core/journals/acta-numerica/article/solving-inverse-problems-using-datadriven-models/CE5B3725869AEAF46E04874115B0AB15}.

\bibitem[Guest et~al.(2004)Guest, Prévost, and
  Belytschko]{guest_achieving_2004}
J.~K. Guest, J.~H. Prévost, and T.~Belytschko.
\newblock Achieving minimum length scale in topology optimization using nodal
  design variables and projection functions.
\newblock \emph{International Journal for Numerical Methods in Engineering},
  61\penalty0 (2):\penalty0 238--254, September 2004.
\newblock ISSN 0029-5981, 1097-0207.
\newblock \doi{10.1002/nme.1064}.
\newblock URL \url{https://onlinelibrary.wiley.com/doi/10.1002/nme.1064}.

\bibitem[Wang et~al.(2011)Wang, Lazarov, and Sigmund]{wang_projection_2011}
Fengwen Wang, Boyan~Stefanov Lazarov, and Ole Sigmund.
\newblock On projection methods, convergence and robust formulations in
  topology optimization.
\newblock \emph{Structural and Multidisciplinary Optimization}, 43\penalty0
  (6):\penalty0 767--784, June 2011.
\newblock ISSN 1615-1488.
\newblock \doi{10.1007/s00158-010-0602-y}.
\newblock URL \url{https://doi.org/10.1007/s00158-010-0602-y}.

\bibitem[Li and Khandelwal(2015)]{li_volume_2015}
Lei Li and Kapil Khandelwal.
\newblock Volume preserving projection filters and continuation methods in
  topology optimization.
\newblock \emph{Engineering Structures}, 85:\penalty0 144--161, February 2015.
\newblock ISSN 0141-0296.
\newblock \doi{10.1016/j.engstruct.2014.10.052}.
\newblock URL
  \url{https://www.sciencedirect.com/science/article/pii/S0141029614006579}.

\bibitem[Bendsøe(1989)]{bendsoe_optimal_1989}
M.~P. Bendsøe.
\newblock Optimal shape design as a material distribution problem.
\newblock \emph{Structural optimization}, 1\penalty0 (4):\penalty0 193--202,
  December 1989.
\newblock ISSN 1615-1488.
\newblock \doi{10.1007/BF01650949}.
\newblock URL \url{https://doi.org/10.1007/BF01650949}.

\bibitem[Bendsøe and Sigmund(1999)]{bendsoe_material_1999}
M.~P. Bendsøe and O.~Sigmund.
\newblock Material interpolation schemes in topology optimization.
\newblock \emph{Archive of Applied Mechanics}, 69\penalty0 (9):\penalty0
  635--654, November 1999.
\newblock ISSN 1432-0681.
\newblock \doi{10.1007/s004190050248}.
\newblock URL \url{https://doi.org/10.1007/s004190050248}.

\bibitem[Bendsøe and Sigmund(2003)]{bendsoe_topology_2003}
Martin~P. Bendsøe and O.~Sigmund.
\newblock \emph{Topology optimization: theory, methods, and applications}.
\newblock Springer, Berlin ; New York, 2003.
\newblock ISBN 978-3-540-42992-0.

\bibitem[Hyun and Kim(2021)]{hyun_transient_2021}
Jaeyub Hyun and H.~Alicia Kim.
\newblock Transient level-set topology optimization of a planar acoustic lens
  working with short-duration pulse.
\newblock \emph{The Journal of the Acoustical Society of America}, 149\penalty0
  (5):\penalty0 3010--3026, May 2021.
\newblock ISSN 0001-4966, 1520-8524.
\newblock \doi{10.1121/10.0004819}.
\newblock URL
  \url{https://pubs.aip.org/jasa/article/149/5/3010/607500/Transient-level-set-topology-optimization-of-a}.

\bibitem[Dilgen and Aage(2024)]{dilgen_topology_2024}
Cetin~B. Dilgen and Niels Aage.
\newblock Topology optimization of transient vibroacoustic problems for
  broadband filter design using cut elements.
\newblock \emph{Finite Elements in Analysis and Design}, 234:\penalty0 104123,
  July 2024.
\newblock ISSN 0168-874X.
\newblock \doi{10.1016/j.finel.2024.104123}.
\newblock URL
  \url{https://www.sciencedirect.com/science/article/pii/S0168874X24000179}.

\bibitem[Pelat et~al.(2020)Pelat, Gautier, Conlon, and
  Semperlotti]{pelat_acoustic_2020}
Adrien Pelat, François Gautier, Stephen~C. Conlon, and Fabio Semperlotti.
\newblock The acoustic black hole: {A} review of theory and applications.
\newblock \emph{Journal of Sound and Vibration}, 476:\penalty0 115316, June
  2020.
\newblock ISSN 0022-460X.
\newblock \doi{10.1016/j.jsv.2020.115316}.
\newblock URL
  \url{https://www.sciencedirect.com/science/article/pii/S0022460X20301474}.

\bibitem[Mousavi et~al.(2024)Mousavi, Berggren, Hägg, and
  Wadbro]{mousavi_topology_2024}
Abbas Mousavi, Martin Berggren, Linus Hägg, and Eddie Wadbro.
\newblock Topology optimization of a waveguide acoustic black hole for enhanced
  wave focusing.
\newblock \emph{The Journal of the Acoustical Society of America}, 155\penalty0
  (1):\penalty0 742--756, January 2024.
\newblock ISSN 1520-8524.
\newblock \doi{10.1121/10.0024470}.

\bibitem[Herrmann et~al.(2024{\natexlab{a}})Herrmann, Sigmund, Li, Vogl, and
  Kollmannsberger]{herrmann_neural_2024}
Leon Herrmann, Ole Sigmund, Viola~Muning Li, Christian Vogl, and Stefan
  Kollmannsberger.
\newblock On neural networks for generating better local optima in topology
  optimization.
\newblock \emph{Structural and Multidisciplinary Optimization}, 67\penalty0
  (11):\penalty0 192, November 2024{\natexlab{a}}.
\newblock ISSN 1615-1488.
\newblock \doi{10.1007/s00158-024-03908-6}.
\newblock URL \url{https://doi.org/10.1007/s00158-024-03908-6}.

\bibitem[Sigmund and Petersson(1998)]{sigmund_numerical_1998}
O.~Sigmund and J.~Petersson.
\newblock Numerical instabilities in topology optimization: {A} survey on
  procedures dealing with checkerboards, mesh-dependencies and local minima.
\newblock \emph{Structural optimization}, 16\penalty0 (1):\penalty0 68--75,
  August 1998.
\newblock ISSN 1615-1488.
\newblock \doi{10.1007/BF01214002}.
\newblock URL \url{https://doi.org/10.1007/BF01214002}.

\bibitem[Bruns and Tortorelli(2001)]{bruns_topology_2001}
Tyler~E. Bruns and Daniel~A. Tortorelli.
\newblock Topology optimization of non-linear elastic structures and compliant
  mechanisms.
\newblock \emph{Computer Methods in Applied Mechanics and Engineering},
  190\penalty0 (26):\penalty0 3443--3459, March 2001.
\newblock ISSN 0045-7825.
\newblock \doi{10.1016/S0045-7825(00)00278-4}.
\newblock URL
  \url{https://www.sciencedirect.com/science/article/pii/S0045782500002784}.

\bibitem[Bourdin(2001)]{bourdin_filters_2001}
Blaise Bourdin.
\newblock Filters in topology optimization.
\newblock \emph{International Journal for Numerical Methods in Engineering},
  50\penalty0 (9):\penalty0 2143--2158, March 2001.
\newblock ISSN 0029-5981, 1097-0207.
\newblock \doi{10.1002/nme.116}.
\newblock URL \url{https://onlinelibrary.wiley.com/doi/10.1002/nme.116}.

\bibitem[Aage et~al.(2017)Aage, Andreassen, Lazarov, and
  Sigmund]{aage_giga-voxel_2017}
Niels Aage, Erik Andreassen, Boyan~S. Lazarov, and Ole Sigmund.
\newblock Giga-voxel computational morphogenesis for structural design.
\newblock \emph{Nature}, 550\penalty0 (7674):\penalty0 84--86, October 2017.
\newblock ISSN 1476-4687.
\newblock \doi{10.1038/nature23911}.
\newblock URL \url{https://www.nature.com/articles/nature23911}.

\bibitem[Christiansen et~al.(2015)Christiansen, Lazarov, Jensen, and
  Sigmund]{christiansen_creating_2015}
Rasmus~E. Christiansen, Boyan~S. Lazarov, Jakob~S. Jensen, and Ole Sigmund.
\newblock Creating geometrically robust designs for highly sensitive problems
  using topology optimization: {Acoustic} cavity design.
\newblock \emph{Structural and Multidisciplinary Optimization}, 52\penalty0
  (4):\penalty0 737--754, October 2015.
\newblock ISSN 1615-147X, 1615-1488.
\newblock \doi{10.1007/s00158-015-1265-5}.
\newblock URL \url{http://link.springer.com/10.1007/s00158-015-1265-5}.

\bibitem[Sigmund(2007)]{sigmund_morphology-based_2007}
Ole Sigmund.
\newblock Morphology-based black and white filters for topology optimization.
\newblock \emph{Structural and Multidisciplinary Optimization}, 33\penalty0
  (4):\penalty0 401--424, April 2007.
\newblock ISSN 1615-1488.
\newblock \doi{10.1007/s00158-006-0087-x}.
\newblock URL \url{https://doi.org/10.1007/s00158-006-0087-x}.

\bibitem[Sigmund(2009)]{sigmund_manufacturing_2009}
Ole Sigmund.
\newblock Manufacturing tolerant topology optimization.
\newblock \emph{Acta Mechanica Sinica}, 25\penalty0 (2):\penalty0 227--239,
  April 2009.
\newblock ISSN 1614-3116.
\newblock \doi{10.1007/s10409-009-0240-z}.
\newblock URL \url{https://doi.org/10.1007/s10409-009-0240-z}.

\bibitem[Langtangen and Linge(2017)]{langtangen_finite_2017}
Hans~Petter Langtangen and Svein Linge.
\newblock \emph{Finite {Difference} {Computing} with {PDEs}: {A} {Modern}
  {Software} {Approach}}, volume~16 of \emph{Texts in {Computational} {Science}
  and {Engineering}}.
\newblock Springer International Publishing, Cham, 2017.
\newblock ISBN 978-3-319-55455-6 978-3-319-55456-3.
\newblock \doi{10.1007/978-3-319-55456-3}.
\newblock URL \url{http://link.springer.com/10.1007/978-3-319-55456-3}.

\bibitem[Herrmann et~al.(2024{\natexlab{b}})Herrmann, Bürchner, Dietrich, and
  Kollmannsberger]{herrmann_use_2024}
Leon Herrmann, Tim Bürchner, Felix Dietrich, and Stefan Kollmannsberger.
\newblock On the {Use} of {Neural} {Networks} for {Full} {Waveform} {Inversion}
  [{Software}], August 2024{\natexlab{b}}.
\newblock URL \url{https://data.mendeley.com/datasets/7kps2hnj6g/2}.

\bibitem[Wirgin(2004)]{wirgin_inverse_2004}
Armand Wirgin.
\newblock The inverse crime, January 2004.
\newblock URL \url{http://arxiv.org/abs/math-ph/0401050}.
\newblock arXiv:math-ph/0401050.

\bibitem[Fichtner(2021)]{Fichtner2021}
Andreas Fichtner.
\newblock Lecture notes on inverse theory, 07 2021.
\newblock This content is a preprint and has not been peer-reviewed.

\bibitem[Hug et~al.(2022)Hug, Potten, Stockinger, Thuro, and
  Kollmannsberger]{hug_three-field_2022}
L.~Hug, M.~Potten, G.~Stockinger, K.~Thuro, and S.~Kollmannsberger.
\newblock A three-field phase-field model for mixed-mode fracture in rock based
  on experimental determination of the mode {II} fracture toughness.
\newblock \emph{Engineering with Computers}, 38\penalty0 (6):\penalty0
  5563--5581, December 2022.
\newblock ISSN 0177-0667, 1435-5663.
\newblock \doi{10.1007/s00366-022-01684-9}.
\newblock URL \url{https://link.springer.com/10.1007/s00366-022-01684-9}.

\bibitem[Parvizian et~al.(2007)Parvizian, Düster, and
  Rank]{parvizian_finite_2007}
Jamshid Parvizian, Alexander Düster, and Ernst Rank.
\newblock Finite cell method.
\newblock \emph{Computational Mechanics}, 41\penalty0 (1):\penalty0 121--133,
  December 2007.
\newblock ISSN 1432-0924.
\newblock \doi{10.1007/s00466-007-0173-y}.
\newblock URL \url{https://doi.org/10.1007/s00466-007-0173-y}.

\bibitem[Düster et~al.(2017)Düster, Rank, and Szabó]{duster__2017}
Alexander Düster, Ernst Rank, and Barna Szabó.
\newblock The \textit{ {\textbackslash}textless}span
  style="font-variant:small-caps;"{\textbackslash}textgreaterp{\textbackslash}textless/span{\textbackslash}textgreater
  ‐{Version} of the {Finite} {Element} and {Finite} {Cell} {Methods}.
\newblock In Erwin Stein, René Borst, and Thomas J~R Hughes, editors,
  \emph{Encyclopedia of {Computational} {Mechanics} {Second} {Edition}}, pages
  1--35. Wiley, 1 edition, December 2017.
\newblock ISBN 978-1-119-00379-3 978-1-119-17681-7.
\newblock \doi{10.1002/9781119176817.ecm2003g}.
\newblock URL
  \url{https://onlinelibrary.wiley.com/doi/10.1002/9781119176817.ecm2003g}.

\bibitem[Sigmund(2001)]{sigmund_99_2001}
O.~Sigmund.
\newblock A 99 line topology optimization code written in {Matlab}.
\newblock \emph{Structural and Multidisciplinary Optimization}, 21\penalty0
  (2):\penalty0 120--127, April 2001.
\newblock ISSN 1615-1488.
\newblock \doi{10.1007/s001580050176}.
\newblock URL \url{https://doi.org/10.1007/s001580050176}.

\bibitem[Bezanson et~al.(2017)Bezanson, Edelman, Karpinski, and
  Shah]{bezanson_julia_2017}
Jeff Bezanson, Alan Edelman, Stefan Karpinski, and Viral~B. Shah.
\newblock Julia: {A} {Fresh} {Approach} to {Numerical} {Computing}.
\newblock \emph{SIAM Review}, 59\penalty0 (1):\penalty0 65--98, January 2017.
\newblock ISSN 0036-1445, 1095-7200.
\newblock \doi{10.1137/141000671}.
\newblock URL \url{https://epubs.siam.org/doi/10.1137/141000671}.

\bibitem[Vaswani et~al.(2017)Vaswani, Shazeer, Parmar, Uszkoreit, Jones, Gomez,
  Kaiser, and Polosukhin]{vaswani_attention_2017}
Ashish Vaswani, Noam Shazeer, Niki Parmar, Jakob Uszkoreit, Llion Jones,
  Aidan~N Gomez, Lukasz Kaiser, and Illia Polosukhin.
\newblock Attention is {All} you {Need}.
\newblock In \emph{Advances in {Neural} {Information} {Processing} {Systems}},
  volume~30. Curran Associates, Inc., 2017.
\newblock URL
  \url{https://papers.nips.cc/paper_files/paper/2017/hash/3f5ee243547dee91fbd053c1c4a845aa-Abstract.html}.

\bibitem[Sevilla et~al.(2022)Sevilla, Heim, Ho, Besiroglu, Hobbhahn, and
  Villalobos]{sevilla_compute_2022}
Jaime Sevilla, Lennart Heim, Anson Ho, Tamay Besiroglu, Marius Hobbhahn, and
  Pablo Villalobos.
\newblock Compute {Trends} {Across} {Three} {Eras} of {Machine} {Learning}.
\newblock In \emph{2022 {International} {Joint} {Conference} on {Neural}
  {Networks} ({IJCNN})}, pages 1--8, July 2022.
\newblock \doi{10.1109/IJCNN55064.2022.9891914}.
\newblock URL \url{https://ieeexplore.ieee.org/document/9891914}.
\newblock ISSN: 2161-4407.

\bibitem[Xu and Darve(2019)]{xu_neural_2019}
Kailai Xu and Eric Darve.
\newblock The {Neural} {Network} {Approach} to {Inverse} {Problems} in
  {Differential} {Equations}, January 2019.
\newblock URL \url{http://arxiv.org/abs/1901.07758}.

\bibitem[Berg and Nyström(2021)]{berg_neural_2021}
Jens Berg and Kaj Nyström.
\newblock Neural networks as smooth priors for inverse problems for {PDEs}.
\newblock \emph{Journal of Computational Mathematics and Data Science},
  1:\penalty0 100008, September 2021.
\newblock ISSN 27724158.
\newblock \doi{10.1016/j.jcmds.2021.100008}.
\newblock URL
  \url{https://linkinghub.elsevier.com/retrieve/pii/S2772415821000043}.

\bibitem[Zhu et~al.(2022)Zhu, Xu, Darve, Biondi, and
  Beroza]{zhu_integrating_2022}
Weiqiang Zhu, Kailai Xu, Eric Darve, Biondo Biondi, and Gregory~C. Beroza.
\newblock Integrating deep neural networks with full-waveform inversion:
  {Reparameterization}, regularization, and uncertainty quantification.
\newblock \emph{GEOPHYSICS}, 87\penalty0 (1):\penalty0 R93--R109, January 2022.
\newblock ISSN 0016-8033, 1942-2156.
\newblock \doi{10.1190/geo2020-0933.1}.
\newblock URL \url{https://library.seg.org/doi/10.1190/geo2020-0933.1}.

\bibitem[Jiang et~al.(2024)Jiang, Wang, Ren, Yang, and Li]{jiang_full_2024}
Peng Jiang, Qingyang Wang, Yuxiao Ren, Senlin Yang, and Ningbo Li.
\newblock Full waveform inversion based on inversion network reparameterized
  velocity.
\newblock \emph{Geophysical Prospecting}, 72\penalty0 (1):\penalty0 52--67,
  January 2024.
\newblock ISSN 0016-8025, 1365-2478.
\newblock \doi{10.1111/1365-2478.13292}.
\newblock URL
  \url{https://onlinelibrary.wiley.com/doi/10.1111/1365-2478.13292}.

\bibitem[Singh et~al.(2025)Singh, Herrmann, Sun, Bürchner, Dietrich, and
  Kollmannsberger]{singh_accelerating_2025}
Divya~Shyam Singh, Leon Herrmann, Qing Sun, Tim Bürchner, Felix Dietrich, and
  Stefan Kollmannsberger.
\newblock Accelerating full waveform inversion by transfer learning.
\newblock \emph{Computational Mechanics}, February 2025.
\newblock ISSN 1432-0924.
\newblock \doi{10.1007/s00466-025-02600-w}.
\newblock URL \url{https://doi.org/10.1007/s00466-025-02600-w}.

\bibitem[Liu et~al.(2025)Liu, Li, Su, Huang, and Bai]{liu_deep_2025}
Feng Liu, Yaxing Li, Rui Su, Jianping Huang, and Lei Bai.
\newblock Deep {Reparameterization} for {Full} {Waveform} {Inversion}:
  {Architecture} {Benchmarking}, {Robust} {Inversion}, and {Multiphysics}
  {Extension}, April 2025.
\newblock URL \url{http://arxiv.org/abs/2504.17375}.
\newblock arXiv:2504.17375.

\bibitem[Norder et~al.(2025)Norder, Yin, de~Jong, Stallone, Aydogmus, Sberna,
  Bessa, and Norte]{norder_pentagonal_2025}
Lucas Norder, Shunyu Yin, Matthijs H.~J. de~Jong, Francesco Stallone, Hande
  Aydogmus, Paolo~M. Sberna, Miguel~A. Bessa, and Richard~A. Norte.
\newblock Pentagonal photonic crystal mirrors: scalable lightsails with
  enhanced acceleration via neural topology optimization.
\newblock \emph{Nature Communications}, 16\penalty0 (1):\penalty0 2753, March
  2025.
\newblock ISSN 2041-1723.
\newblock \doi{10.1038/s41467-025-57749-y}.
\newblock URL \url{https://www.nature.com/articles/s41467-025-57749-y}.

\bibitem[Hoyer et~al.(2019)Hoyer, Sohl-Dickstein, and
  Greydanus]{hoyer_neural_2019}
Stephan Hoyer, Jascha Sohl-Dickstein, and Sam Greydanus.
\newblock Neural reparameterization improves structural optimization, September
  2019.
\newblock URL \url{http://arxiv.org/abs/1909.04240}.

\bibitem[Deng and To(2020)]{deng_topology_2020}
Hao Deng and Albert~C. To.
\newblock Topology optimization based on deep representation learning ({DRL})
  for compliance and stress-constrained design.
\newblock \emph{Computational Mechanics}, 66\penalty0 (2):\penalty0 449--469,
  August 2020.
\newblock ISSN 1432-0924.
\newblock \doi{10.1007/s00466-020-01859-5}.
\newblock URL \url{https://doi.org/10.1007/s00466-020-01859-5}.

\bibitem[Chen and Shen(2021)]{chen_new_2021}
Liang Chen and Mo-How~Herman Shen.
\newblock A {New} {Topology} {Optimization} {Approach} by {Physics}-{Informed}
  {Deep} {Learning} {Process}.
\newblock \emph{Advances in Science, Technology and Engineering Systems
  Journal}, 6\penalty0 (4):\penalty0 233--240, July 2021.
\newblock ISSN 24156698, 24156698.
\newblock \doi{10.25046/aj060427}.
\newblock URL \url{https://astesj.com/v06/i04/p27/}.

\bibitem[Chandrasekhar and
  Suresh(2021{\natexlab{a}})]{chandrasekhar_tounn_2021}
Aaditya Chandrasekhar and Krishnan Suresh.
\newblock {TOuNN}: {Topology} {Optimization} using {Neural} {Networks}.
\newblock \emph{Structural and Multidisciplinary Optimization}, 63\penalty0
  (3):\penalty0 1135--1149, March 2021{\natexlab{a}}.
\newblock ISSN 1615-147X, 1615-1488.
\newblock \doi{10.1007/s00158-020-02748-4}.
\newblock URL \url{http://link.springer.com/10.1007/s00158-020-02748-4}.

\bibitem[Chandrasekhar and
  Suresh(2021{\natexlab{b}})]{chandrasekhar_multi-material_2021}
Aaditya Chandrasekhar and Krishnan Suresh.
\newblock Multi-{Material} {Topology} {Optimization} {Using} {Neural}
  {Networks}.
\newblock \emph{Computer-Aided Design}, 136:\penalty0 103017, July
  2021{\natexlab{b}}.
\newblock ISSN 0010-4485.
\newblock \doi{10.1016/j.cad.2021.103017}.
\newblock URL
  \url{https://www.sciencedirect.com/science/article/pii/S0010448521000282}.

\bibitem[Chandrasekhar and Suresh(2022)]{chandrasekhar_approximate_2022}
Aaditya Chandrasekhar and Krishnan Suresh.
\newblock Approximate {Length} {Scale} {Filter} in {Topology} {Optimization}
  using {Fourier} {Enhanced} {Neural} {Networks}.
\newblock \emph{Computer-Aided Design}, 150:\penalty0 103277, September 2022.
\newblock ISSN 0010-4485.
\newblock \doi{10.1016/j.cad.2022.103277}.
\newblock URL
  \url{https://www.sciencedirect.com/science/article/pii/S0010448522000574}.

\bibitem[Mallon et~al.(2024)Mallon, Thornton, Hill, and
  Badia]{mallon_neural_2024}
Connor~N. Mallon, Aaron~W. Thornton, Matthew~R. Hill, and Santiago Badia.
\newblock Neural {Level} {Set} {Topology} {Optimization} {Using} {Unfitted}
  {Finite} {Elements}, February 2024.
\newblock URL \url{http://arxiv.org/abs/2303.13672}.

\bibitem[Sanu et~al.(2024)Sanu, Aragon, and Bessa]{sanu_neural_2024}
Suryanarayanan~Manoj Sanu, Alejandro~M. Aragon, and Miguel~A. Bessa.
\newblock Neural topology optimization: the good, the bad, and the ugly, July
  2024.
\newblock URL \url{http://arxiv.org/abs/2407.13954}.

\bibitem[Kuszczak et~al.(2025)Kuszczak, Kus, Bosi, and
  Bessa]{kuszczak_meta-neural_2025}
Igor Kuszczak, Gawel Kus, Federico Bosi, and Miguel~A. Bessa.
\newblock Meta-neural {Topology} {Optimization}: {Knowledge} {Infusion} with
  {Meta}-learning, February 2025.
\newblock URL \url{http://arxiv.org/abs/2502.01830}.
\newblock arXiv:2502.01830.

\end{thebibliography}

\end{document}